\documentclass{aa}

\usepackage{graphics}
\usepackage{epsfig}
\begin{document}

\thesaurus{05(10.03.1,10.07.2,10.19.3)}

\title{The metal-rich globular clusters of the Milky Way}

\author{F. Heitsch\inst{1} \and T. Richtler\inst{2}}

\offprints{F. Heitsch}

\institute{Max-Planck-Institut f\"ur Astronomie, K\"onigstuhl 17, 63119 Heidelberg, Germany
           \and
           Sternwarte der Universit\"at Bonn, Auf dem H\"ugel 71, 53115 Bonn, Germany}

\date{Recieved / Accepted}

\maketitle
\begin{abstract}
We present new (V,V-I)-photometry of the metal-rich globular clusters \object{NGC~5927}, 6316, 6342, 
6441 and 6760. The clusters show differential reddening up to $\delta E_{V-I}=0.32$ mag, for which
the CMDs are corrected via extinction maps. There are hints of a variation in the extinction
law. Two different ways to determine the parameters metallicity, reddening and distance lead to 
consistent results. 
The metallicities of the clusters range between $-0.7 \le [\mbox{M}/\mbox{H}] \le 0.0$ dex and
the absolute reddening between $0.43 \le E_{V-I} \le 0.76$ mag. Taking the differential reddening into
account leads to slightly increased distances. From the resulting parameters we conclude that the usual
halo-disk-distinction in the system of globular clusters seems questionable.

\end{abstract}

\section{Introduction}
The globular cluster system of the inner Milky Way is still not well understood. This is
particularly true for the clusters' classification with respect of the galactic population 
structure. Because reliably determined parameters, e.g. metallicity, reddening, distance and
age, are the basic requirement of any discussion, we present new photometry in (V, I) of
the metal-rich globular clusters (GC's) \object{NGC~5927}, 6316, 6342, 6441 and 6760. 
We also re-discuss \object{NGC~6528} and \object{NGC~6553}, where the data have already been published 
(Richtler et al. \cite{RTL98}, Sagar et al. \cite{SAG98}). 

As there has been evidence for a correlation between metallicity and spatial distribution of
the GC's since the late 1950's, Zinn (\cite{ZIN85}) classified the clusters via their
kinematics, spatial distribution and metallicity into two subsystems: The disk-system with
clusters of metallicity $[\mbox{M}/\mbox{H}] \ge -0.8$ dex and the halo-system with 
$[\mbox{M}/\mbox{H}] \le -0.8$ dex.
The disk-system shows a high rotational velocity and a small velocity dispersion, the
halo-system vice versa. Armandroff (\cite{ARM89}) derived a scale height of $1.1$ kpc
for the disk-system, which he identified with the galactic thick disk
via rotational velocities and velocity dispersions. By comparing the
metal-rich GC's of the inner $3$ kpc with the underlying stellar population,
Minniti (\cite{MIN95}) assigned these objects to the bulge rather than to the disk.
Burkert \& Smith (\cite{BUR97}) used kinematical arguments and the masses of the clusters to 
divide the metal-rich subsystem of Zinn (\cite{ZIN85}) into a bulge, a bar and a disk-group.

The thing these subdivisions have in common is, that they refer to the entire system of clusters
and they try to formulate their criteria by identifying subsystems within the whole system.
For the other way around, i.e. to classify observed objects with any of these subgroups, accurate 
parameters are needed. The halo clusters are well discernible from any other subsystem, but
the metal-rich clusters near the galactic center are not. The determination of their para\-meters 
encounters observational difficulties, as their low galactic latitudes lead to strong contamination
with field stars and to strong (differential) reddening. These effects have to be taken care of. 

There is a variety of photometry existing for the program clusters. Recent studies on \object{NGC~5927} were
done by Fullton et al. (\cite{FUL96}) and Samus et al. (\cite{SAM96}). Armandroff (\cite{ARM88}) presented
and discussed CMDs including \object{NGC~6316}, 6342 and 6760. There is a photometry of \object{NGC~6441} in (B,V) of Hesser \&
Hartwick (\cite{HES76}) and a more recent one in Rich et al. (\cite{RIC97}). CMDs of \object{NGC~6528} have been 
discussed by Ortolani et al. (\cite{ORT90}) and Richtler et al. (\cite{RTL98}). Guarnieri et al. (\cite{GUA98})
as well as Sagar et al. (\cite{SAG98}) present (V,I)-photometry of \object{NGC~6553}. Zinn (\cite{ZIN85}), Armandroff
(\cite{ARM89}), Richtler et al. (\cite{RTL94}), Minniti (\cite{MIN95}) and Burkert \& Smith (\cite{BUR97}) discuss 
the subdivision of the GC-system into a halo, disk and/or bulge component. 

As the data and their reduction shall be published in a forthcoming paper, section 2 deals
only briefly with this subject. In section 3 and 4, the derived CMDs are presented and the effects of
differential reddening are discussed and removed. Section 5 contains the methods and results of 
the parameter determination, and in section 6 we will discuss the resulting classification
and its problems.

\section{Observations and reduction}
The observations in V and I were carried out at La Silla/Chile between July 16th and 19th 1993.
We used the 2.2m with CCD ESO \#19, which covers with $1024 \times 1024 $ pix an area of $5.7' \times 5.7'$ 
on the sky. The seeing was $1.1''$. In addition to the ground-based data, we used data of the 
Hubble-Space-Telescope (HST) for \object{NGC~5927}. These data have already been published by Fullton et al. 
(\cite{FUL96}). 

To reduce the data, we used the DAOPHOT-package (Stetson \cite{STE87}, \cite{STE92}), together with the
ESO-MIDAS-system (version 1996). The calibrating equations (\ref{equCalib}), determined via Landolt 
standard stars (Landolt 1992), are
\begin{eqnarray}
V_{st} = V_{inst} - (1.38 \pm 0.05) + (0.057 \pm 0.003)(V-I)_{st}\nonumber \\
I_{st} = I_{inst} - (2.62 \pm 0.05) - (0.060 \pm 0.003)(V-I)_{st}
\label{equCalib}
\end{eqnarray}
Together with the error of the photometry and of the PSF-aperture-shift, we get an absolute error for
a single measurement of $\pm 0.06$ mag. For a more extensive treatment of the data and their reduction,
see Heitsch \& Richtler (\cite{HEI99}).
To calibrate the HST-data, we used the relations and coefficients as described by Holtzman (\cite{HOL95}).
In agreement with Guarnieri et al. (\cite{GUA98}), we detected a systematic shift between calibrated 
ground-based and HST-magnitudes of about $0.2$ mag with the HST-magnitudes being fainter. This difference 
might be due to crowding influences on the calibration stars in the ESO-frame, as explained by 
Guarnieri et al. 

\section{Colour-Magnitude-Diagrams}

\subsection{\object{NGC~5927}}
The unselected CMD for \object{NGC~5927} is shown in Fig. \ref{dia5927all}.
The cluster's HB and RGB are clearly distinguishable, with the HB overlapping the RGB, as well as 
the stars of some field population to the blue of the cluster's structures covering the TOP-region. 
Some $0.5$ mag below the HB, the RGB-bump is discernible. The elongation
of the HB and the broadening of the RGB are due to differential reddening, as we now argue. 
As the HB of metal-rich GCs generally is rather clumped and the HB-stars all have the same luminosity, 
differential reddening should cause an elongation of the HB parallel to the reddening vector. 
\begin{figure}[h]
  \begin{picture}(6.0,3.0)
  \put(0.0,0.2){\makebox(5.0,3.0){\epsfig{file=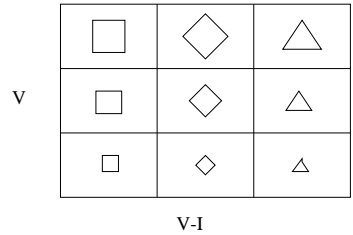,scale=0.45
                                           ,bbllx=4.7cm,bblly=4.0cm}}}
  \end{picture}
  \caption{Symbols for brightness and colour of the HB-stars in diagrams \ref{dia5927HB} to 
           \ref{dia6760HB}. With the box covering the HB-region of a cluster, all the stars in one
           of the nine subfields are denoted by the corresponding symbol. Increasing symbol size
           denotes increasing brightness. Increasing colour-index is represented by a change from
           squares to lozenges to triangles.\label{diaHBexplic}}
\end{figure}
\begin{figure*}[h]
  \begin{picture}(18.0,8.8)
    \put(0.5,0.2)
    {
      \begin{picture}(0.0,0.0)
      \put(0.0,0.0){\makebox(7.0,7.0){\epsfig{file=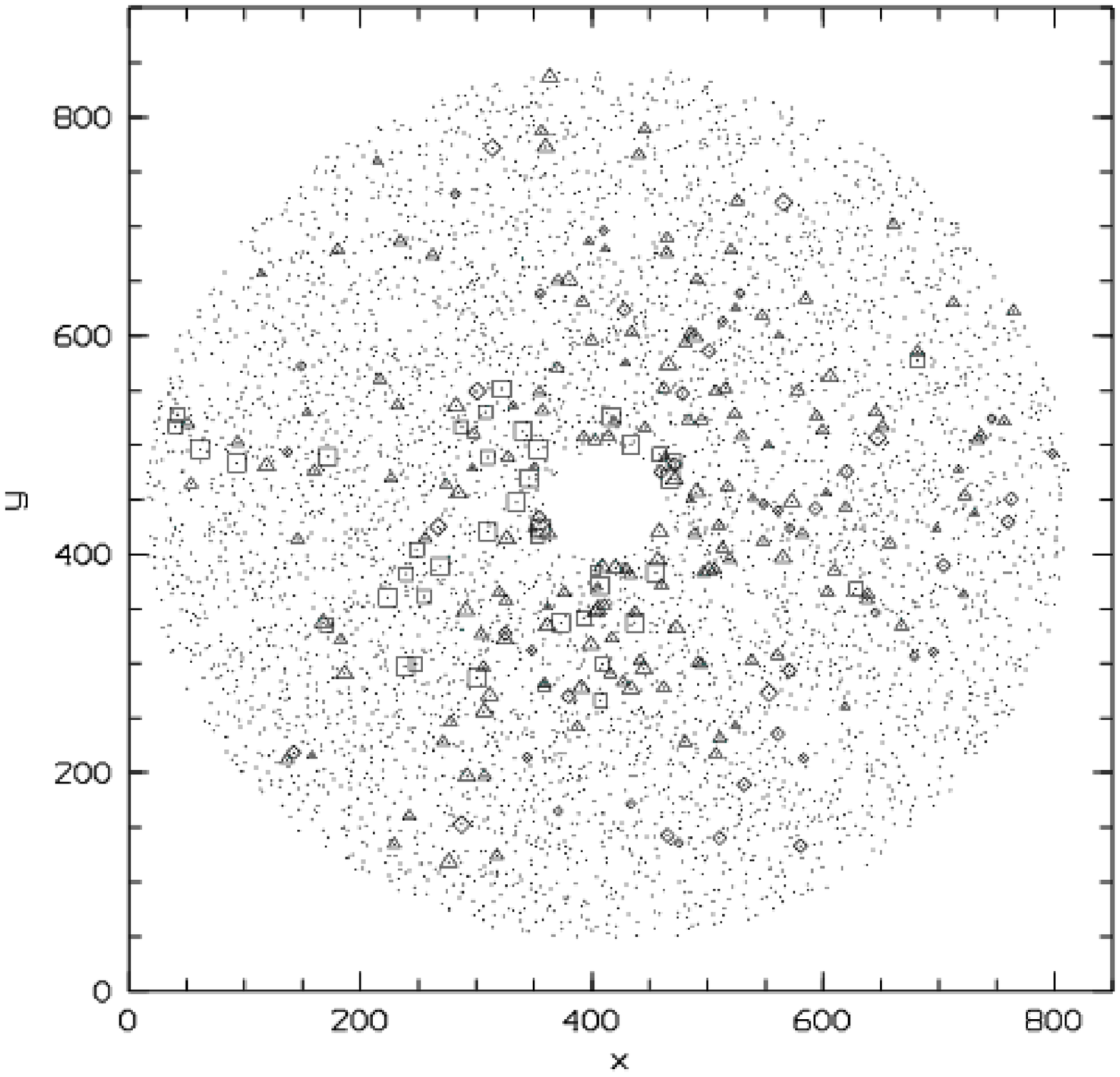,scale=0.48
                                               ,bbllx=20.3cm,bblly=18.0cm}}}
      \end{picture}
    }
    \put(4.4,8.2){\makebox(0.6,0.6){\bf N}}
    \put(8.4,4.4){\makebox(0.4,0.6){\bf E}}

    \put(9.7,0.2)
    {
      \begin{picture}(0.0,0.0)
      \put(0.0,0.0){\makebox(7.0,7.0){\epsfig{file=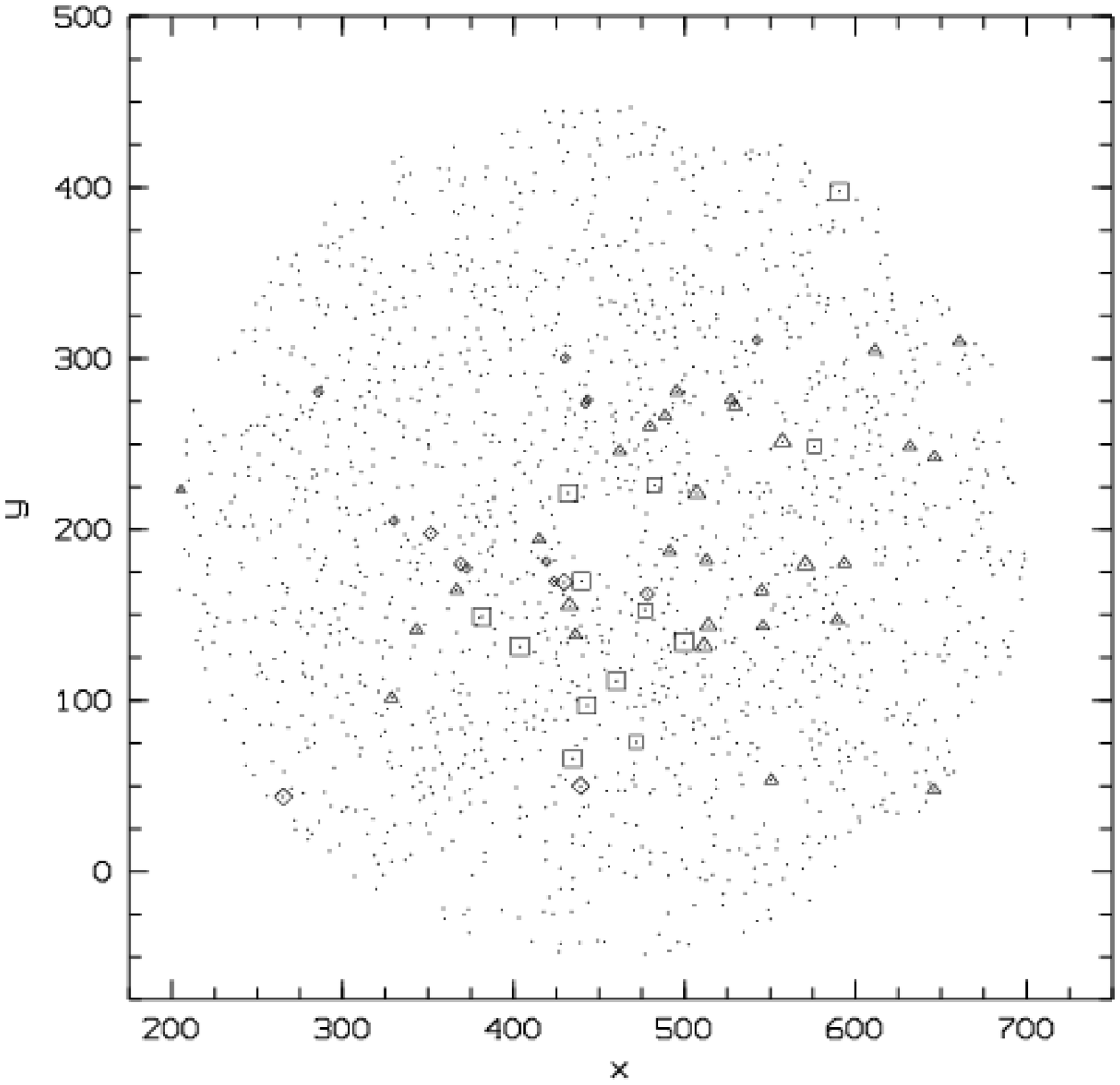,scale=0.48
                                               ,bbllx=20.3cm,bblly=18.0cm}}}
      \end{picture}
    }
    \put(13.6,8.2){\makebox(0.6,0.6){\bf N}}
    \put(17.6,4.4){\makebox(0.4,0.6){\bf E}}
  \end{picture}
  \hfill
  \parbox{8.8cm}{
  \caption{Coordinates of all the stars in \object{NGC~5927} between $50 \le r \le
           400$ pix, r being the distance to the cluster's center in pixel. The HB-stars
           are marked according to Fig. \ref{diaHBexplic}. Blue bright stars are to be found 
           in an area $0 \le x \le 500$ pix and $250 \le y \le 575$ pix. The stars are selected
           for photometric errors $\le 0.03$ mag.\label{dia5927HB}}}
  \hfill
  \parbox{8.8cm}{
  \caption{Coordinates of all the stars in \object{NGC~6342} between $30 \le r \le 250$ pix. The HB-stars
           are marked according to Fig. \ref{diaHBexplic}. Blue bright stars concentrate in
           an area to the south of the cluster's center. The stars are selected for photometric
           errors $\le 0.03$ mag.\label{dia6342HB}}}
\end{figure*}

\begin{figure*}[h]
  \begin{picture}(18.0,8.8)
    \put(0.5,0.2)
    {
      \begin{picture}(0.0,0.0)
      \put(0.0,0.0){\makebox(7.0,7.0){\epsfig{file=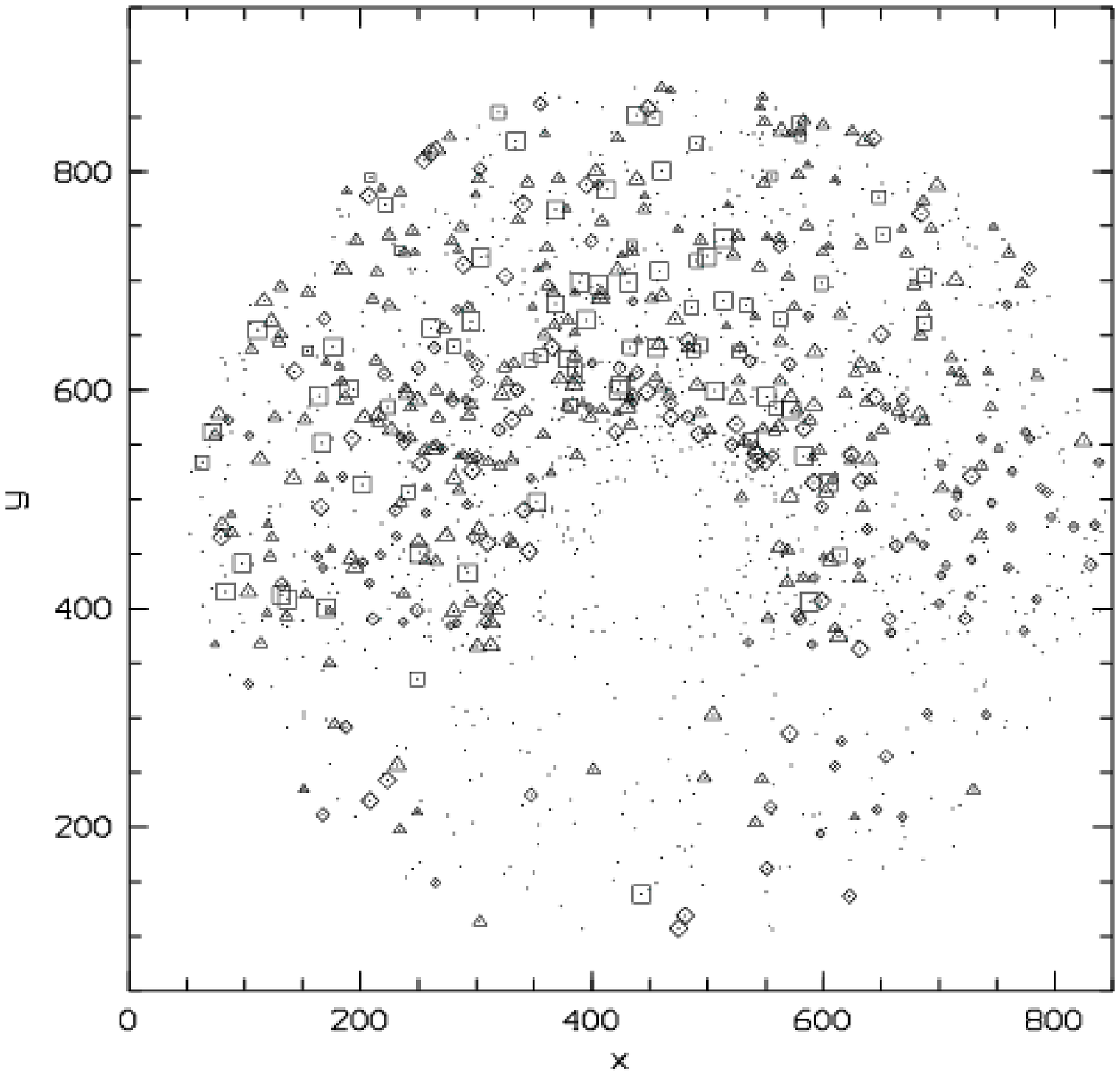,scale=0.48
                                               ,bbllx=20.3cm,bblly=18.0cm}}}
      \end{picture}
    }
    \put(4.4,8.2){\makebox(0.6,0.6){\bf N}}
    \put(8.4,4.4){\makebox(0.4,0.6){\bf E}}

    \put(9.7,0.2)
    {
      \begin{picture}(0.0,0.0)
      \put(0.0,0.0){\makebox(7.0,7.0){\epsfig{file=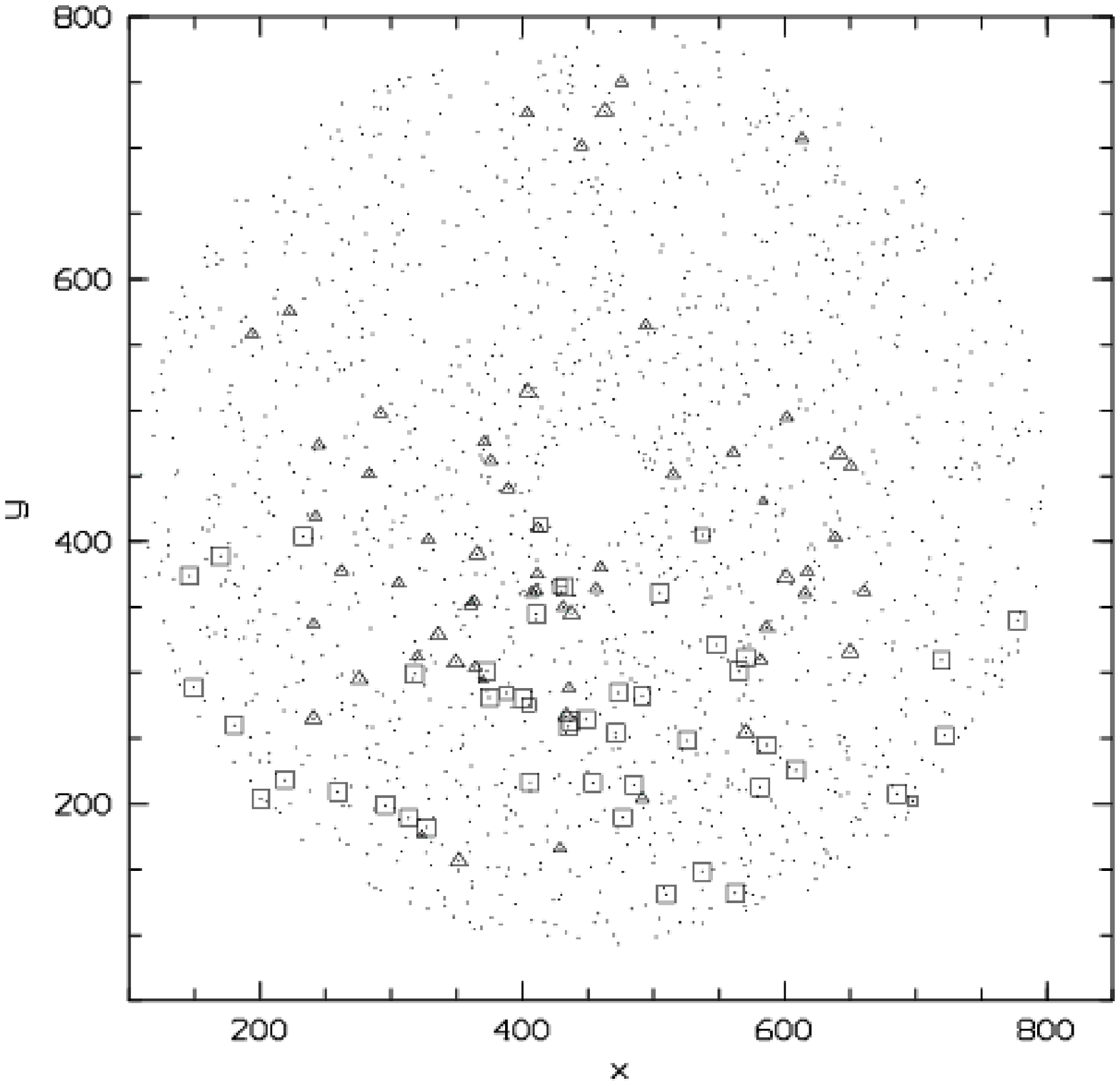,scale=0.48
                                               ,bbllx=20.3cm,bblly=18.0cm}}}
      \end{picture}
    }
    \put(13.6,8.2){\makebox(0.6,0.6){\bf N}}
    \put(17.6,4.4){\makebox(0.4,0.6){\bf E}}
  \end{picture}
  \hfill
  \parbox{8.8cm}{
  \caption{Coordinates of all the stars in \object{NGC~6441} between $30 \le r \le 400$ pix. The HB-stars
           are marked according to Fig. \ref{diaHBexplic}. The lower star density for $y \le 360$
           pix is due to a coordinate shift of the calibration frames. The stars are selected for 
           photometric errors $\le 0.03$ mag.\label{dia6441HB}}} 
  \hfill
  \parbox{8.8cm}{
  \caption{Coordinates of all the stars in \object{NGC~6760} between $40 \le r \le 350$ pix. The HB-stars
           are marked according to Fig. \ref{diaHBexplic}. The blue bright stars are found
           to the south of the cluster's center. The stars are selected for photometric
           errors $\le 0.03$ mag.\label{dia6760HB}}} 
\end{figure*}

Fig. \ref{dia5927HB} shows the coordinates of the radially selected cluster 
stars with special markings for the HB-stars as given by Fig. \ref{diaHBexplic}.
If differential reddening is indeed responsible for the observed elongation, we do 
not expect to find any red faint stars in areas where blue bright stars are to be found, unless
the reddening is very (!) patchy. In the case of \object{NGC~5927}, we note (Fig. \ref{dia5927HB}) that 
the blue bright stars are located in an area west of the cluster's center, and thus differential
reddening is indeed responsible for the elongated HB structure. 

In Fig. \ref{dia5927HST} we present the calibrated HST-CMDs of \object{NGC~5927}. They all show a slightly
broadened lower RGB as well as a slightly tilted HB. The TOP is well resolved. However, the CMDs
do not extend to the bright stars of the AGB/RGB due to pixel overflow on the exposures. As 
mentioned above, the HST-CMDs are shifted with respect to the ESO-CMDs to fainter magnitudes. 
Richtler et al. (\cite{RTL98}) argue that the ground-based calibration is not erroneous. 
Thus, we will use the ground-based calibrationed data for the further analysis. For a  detailed 
discussion see Heitsch \& Richtler (\cite{HEI99}).

\subsection{\object{NGC~6316}}
The unselected CMD (Fig. \ref{dia6316all}) shows beside the cluster a strong contribution from the field population.
The field main sequence is striking. The cluster RGB is broadened, but since the clumpy HB indicates only small 
differential reddening, the RGB width is probably to a large part due to the field contamination.  
Determining a correlation between HB-stars and coordinates as in Fig. \ref{dia5927HB} led to no convincing results,
because the field does not contain enough stars. 

\subsection{\object{NGC~6342}}
\object{NGC~6342} (Fig. \ref{dia6342all}) shows a sparsely populated AGB/RGB due to the
small size of the cluster. The TOP-region and upper MS are reached.
Fig. \ref{dia6342HB} gives the location of HB-stars as in Fig. \ref{dia5927HB}. Blue, bright HB-stars are to 
be found in an area to the south of the cluster's center. 

\subsection{\object{NGC~6441}}
\object{NGC~6441} (Fig. \ref{dia6441all}) is located behind a dense field population, the stars of which can be found 
between $1.0 \le V-I \le 1.5$ mag. This population covers the lower part of the RGB of \object{NGC~6441} as well.
Its TOP is not reached. We mention some special features. 
First, we find some stars between 
$1.8 \le V-I \le 2.0$ mag and above the HB of \object{NGC~6441}. These could be HB-stars of a population which is
similar to \object{NGC~6441} and is located between the cluster and ourselves, as the stars are shifted in V only.
In this case, we would have to assume that the absolute reddening is caused by some cloud between this 
population and the observer. Otherwise it would have to be shifted in $V-I$ as well. 
Second, we find some stars to the blue of the clumpy, tilted HB of \object{NGC~6441}. These stars seemingly belong
to the cluster, as they are still visible when selecting for small radii. Probably they belong to the blue 
HB of \object{NGC~6441}, which has been discovered by Rich et al. (\cite{RIC97}). 
The difference in star density (Fig. \ref{dia6441HB}) is due to the fact that the calibration exposures
were shifted by around 360 pix to the south. Taking this into account, we find that the blue bright stars
are mostly found in the western two thirds of the cluster. 

\subsection{\object{NGC~6760}}
\begin{figure*}[hp]
  \begin{picture}(18.0,8.8)
    \put(0.5,0.2)
    {
      \begin{picture}(0.0,0.0)
      \put(0.0,0.0){\makebox(7.0,7.0){\epsfig{file=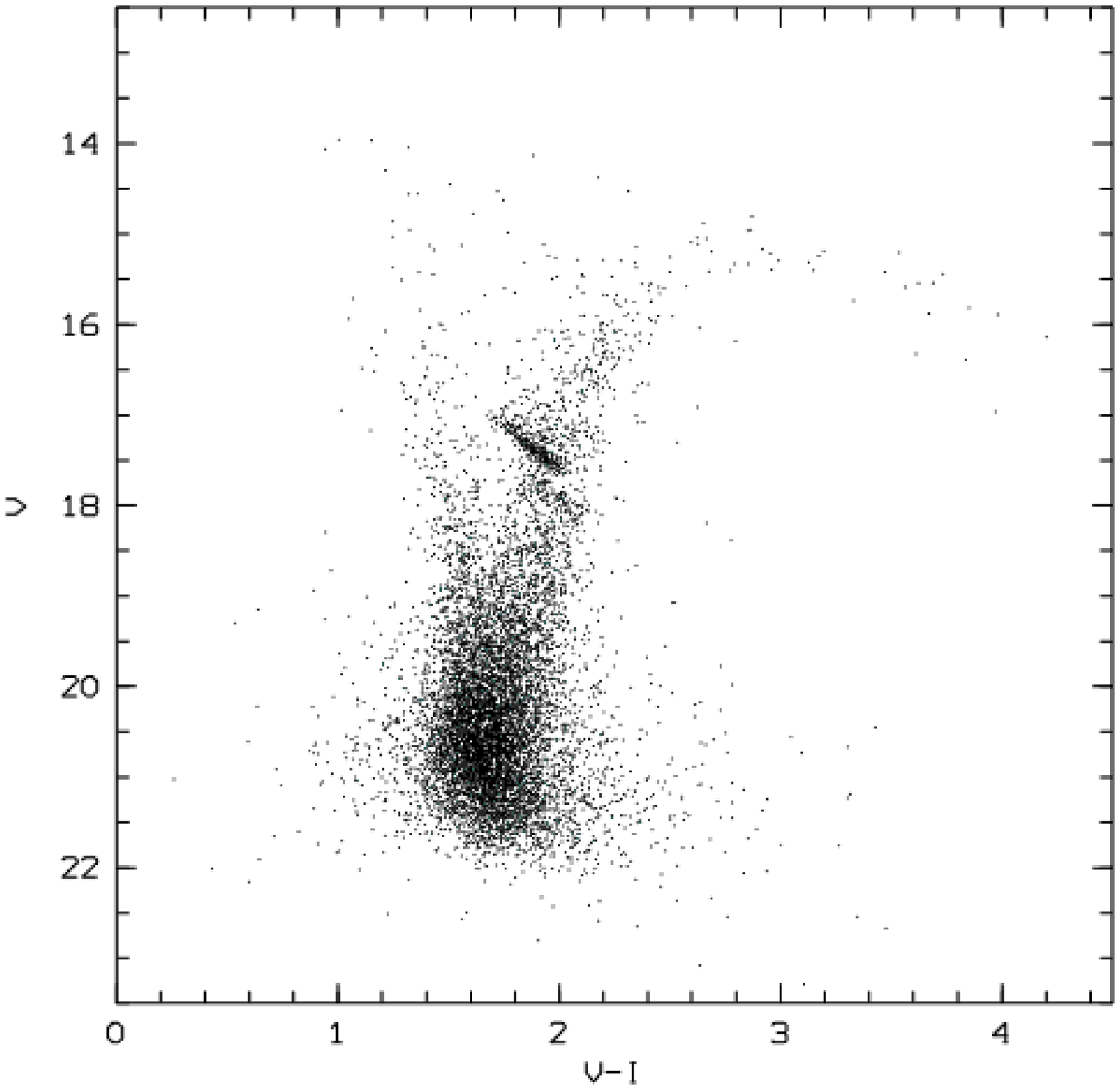,scale=0.48
                                               ,bbllx=20.5cm,bblly=18.0cm}}}
      \end{picture}
    }

    \put(9.7,0.2)
    {
      \begin{picture}(0.0,0.0)
      \put(0.0,0.0){\makebox(7.0,7.0){\epsfig{file=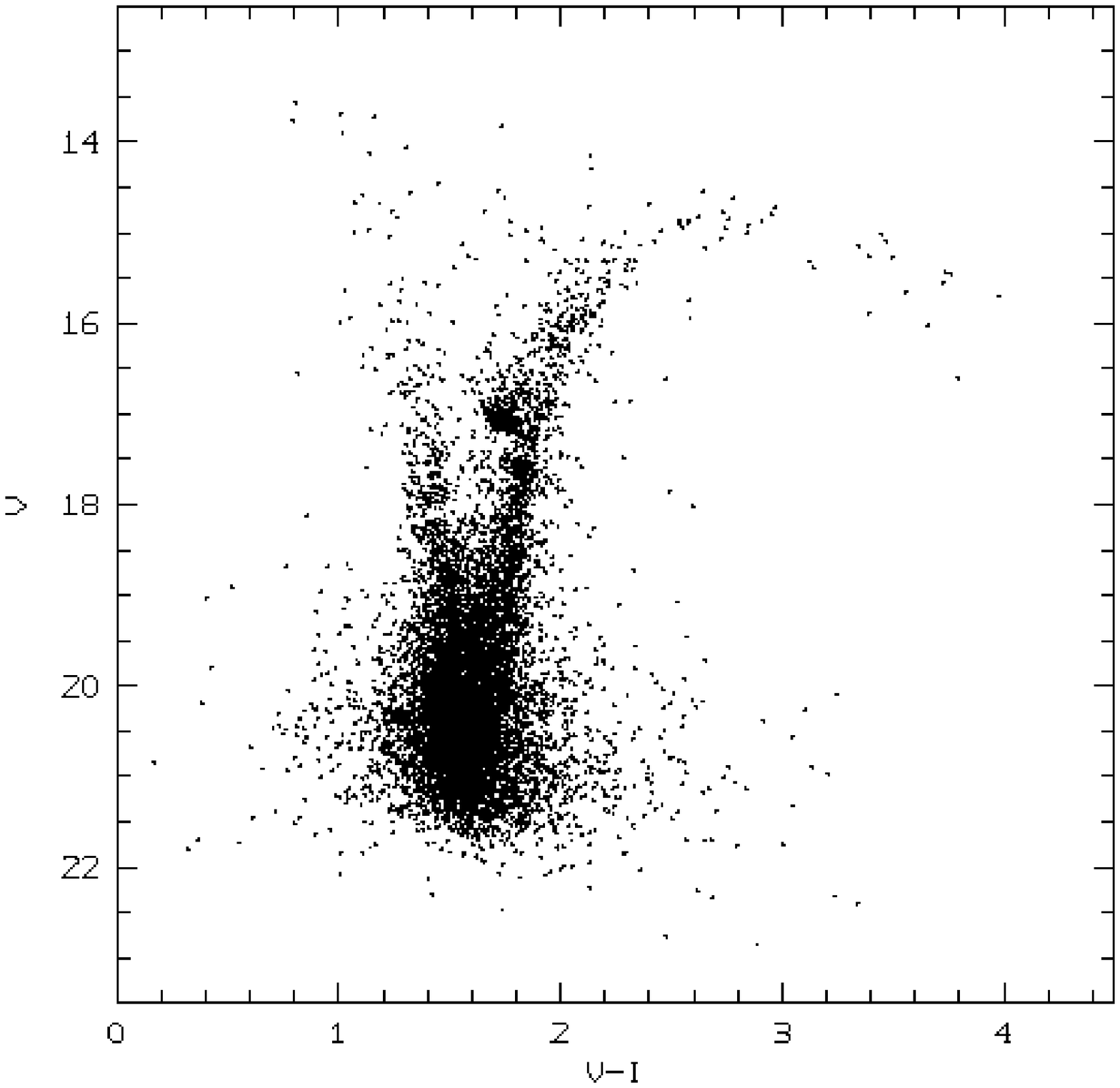,scale=0.48
                                               ,bbllx=20.5cm,bblly=18.0cm}}}
      \end{picture}
    }
  \end{picture}
  \hfill
  \parbox{8.8cm}{
  \caption{Unselected CMD for \object{NGC~6760}. The effect of differential reddening is clearly visible.
           \label{dia6760all}}}
  \hfill
  \parbox{8.8cm}{
  \caption{Differentially dereddened CMD for \object{NGC~6760}. The HB lies well to the blue of the RGB, the
           RGB-bump clearly on the RGB below the HB. 
           \label{dia6760dc}}}
\end{figure*}

Fig. \ref{dia6760all} not only shows the already discussed structures such as HB, RGB and field population,
but it shows the RGB-bump below the HB as well. The RGB is rather broadened. The HB and RGB-bumps are 
elongated and tilted with the same slope. The HB-stars, marked according to colour and brightness, are 
found in Fig. \ref{dia6760HB}. 

\subsection{\object{NGC~6528} and \object{NGC~6553}}
\begin{figure*}[hp]
  \begin{picture}(18.0,8.8)
    \put(0.5,0.2)
    {
      \begin{picture}(0.0,0.0)
      \put(0.0,0.0){\makebox(7.0,7.0){\epsfig{file=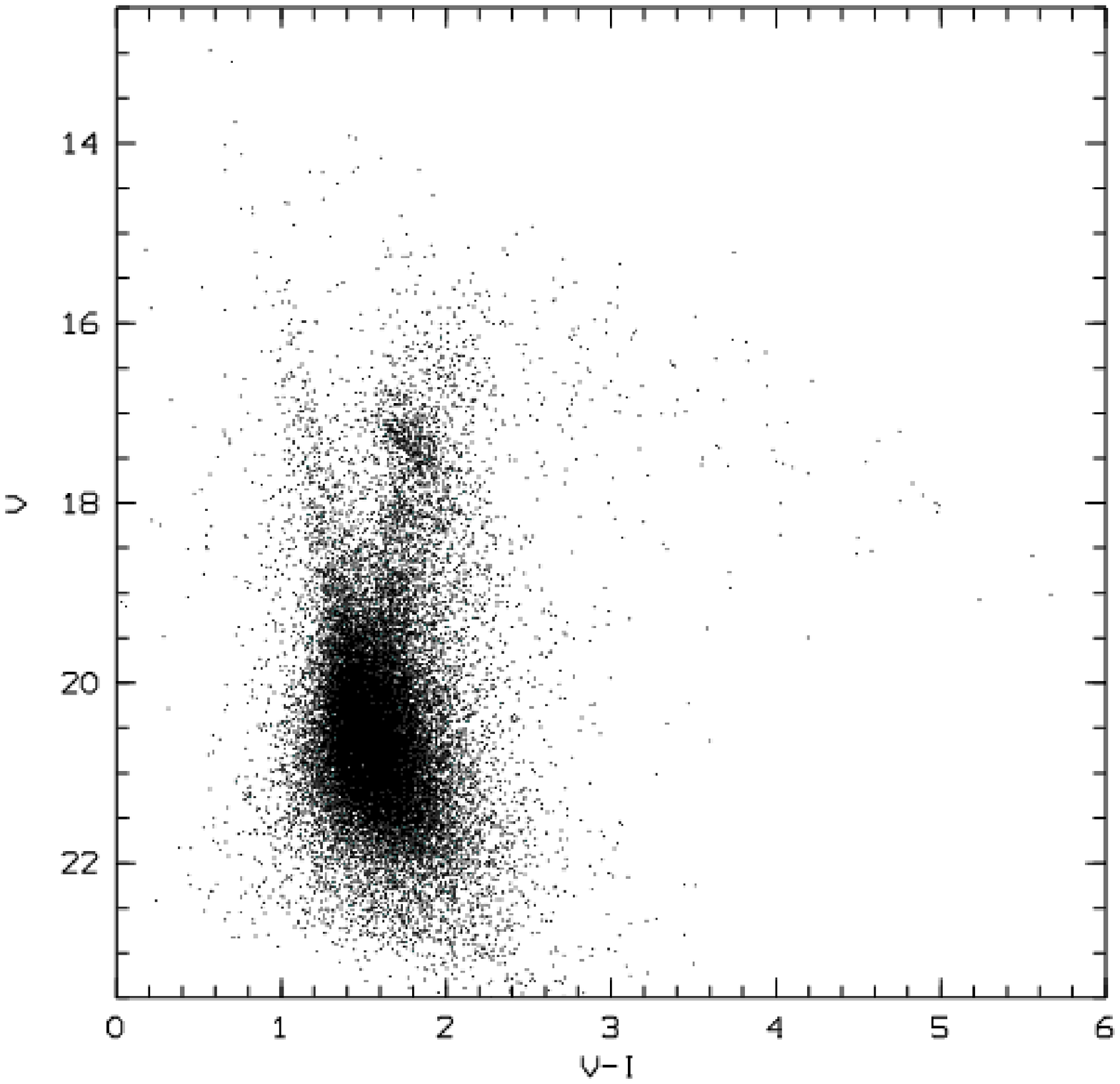,scale=0.48
                                               ,bbllx=20.5cm,bblly=18.0cm}}}
      \end{picture}
    }

    \put(9.7,0.2)
    {
      \begin{picture}(0.0,0.0)
      \put(0.0,0.0){\makebox(7.0,7.0){\epsfig{file=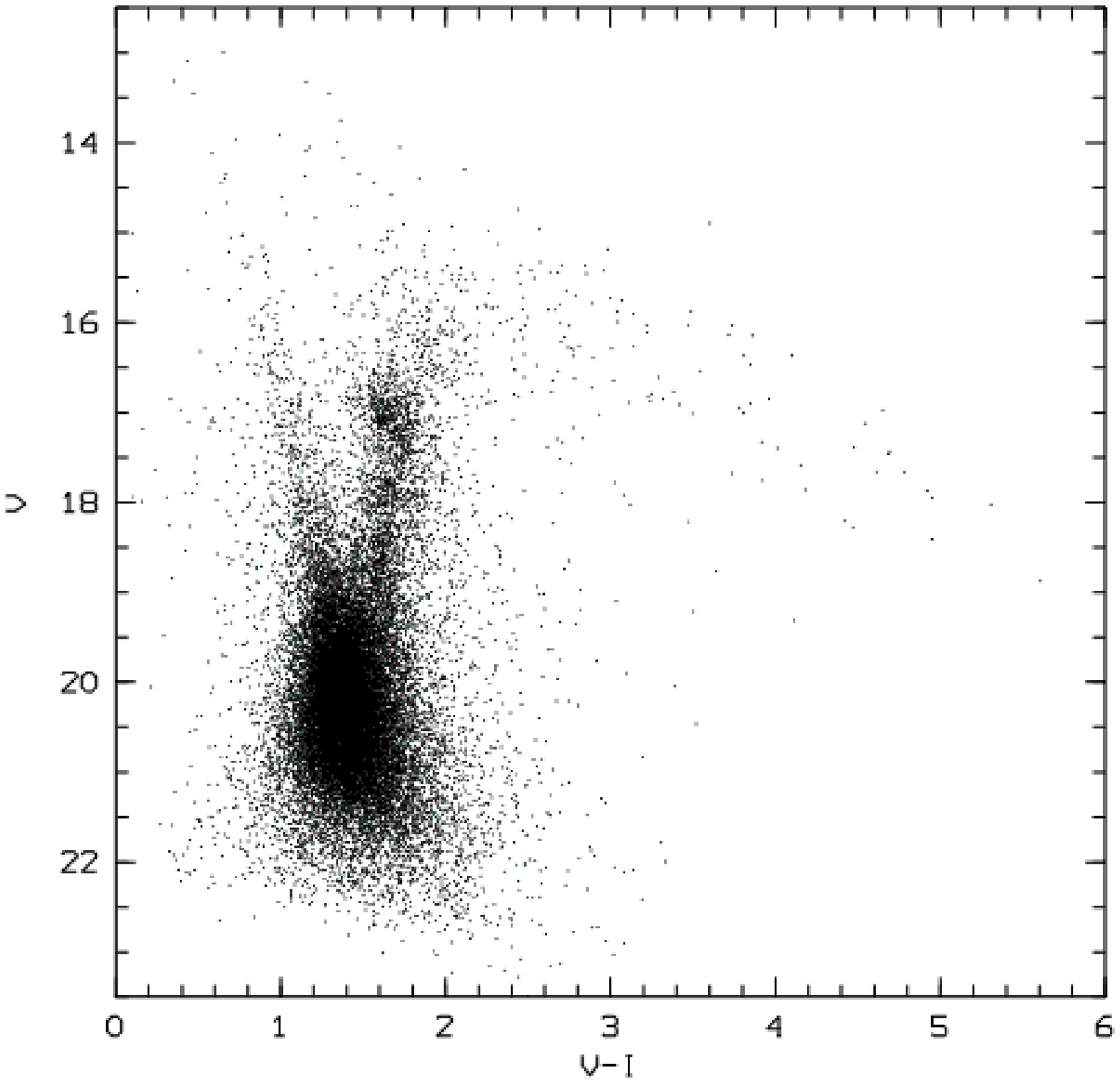,scale=0.48
                                               ,bbllx=20.5cm,bblly=18.0cm}}}
      \end{picture}
    }
  \end{picture}
  \hfill
  \parbox{8.8cm}{
  \caption{Unselected CMD for \object{NGC~6528}. Below the AGB/RGB there are traces of the background population.
           \label{dia6528all}}}
  \hfill
  \parbox{8.8cm}{
  \caption{Differentially dereddened CMD for \object{NGC~6528}. The HB still overlaps strongly with the RGB. Thus, 
           the correction was not completely successful due to the strong contamination by field stars. 
           However, the RGB narrowed perceptibly, and the RGB-bump is now visible below the HB. 
           \label{dia6528dc}}}
\end{figure*}

\begin{figure*}[hp]
  \begin{picture}(18.0,8.8)
    \put(0.5,0.2)
    {
      \begin{picture}(0.0,0.0)
      \put(0.0,0.0){\makebox(7.0,7.0){\epsfig{file=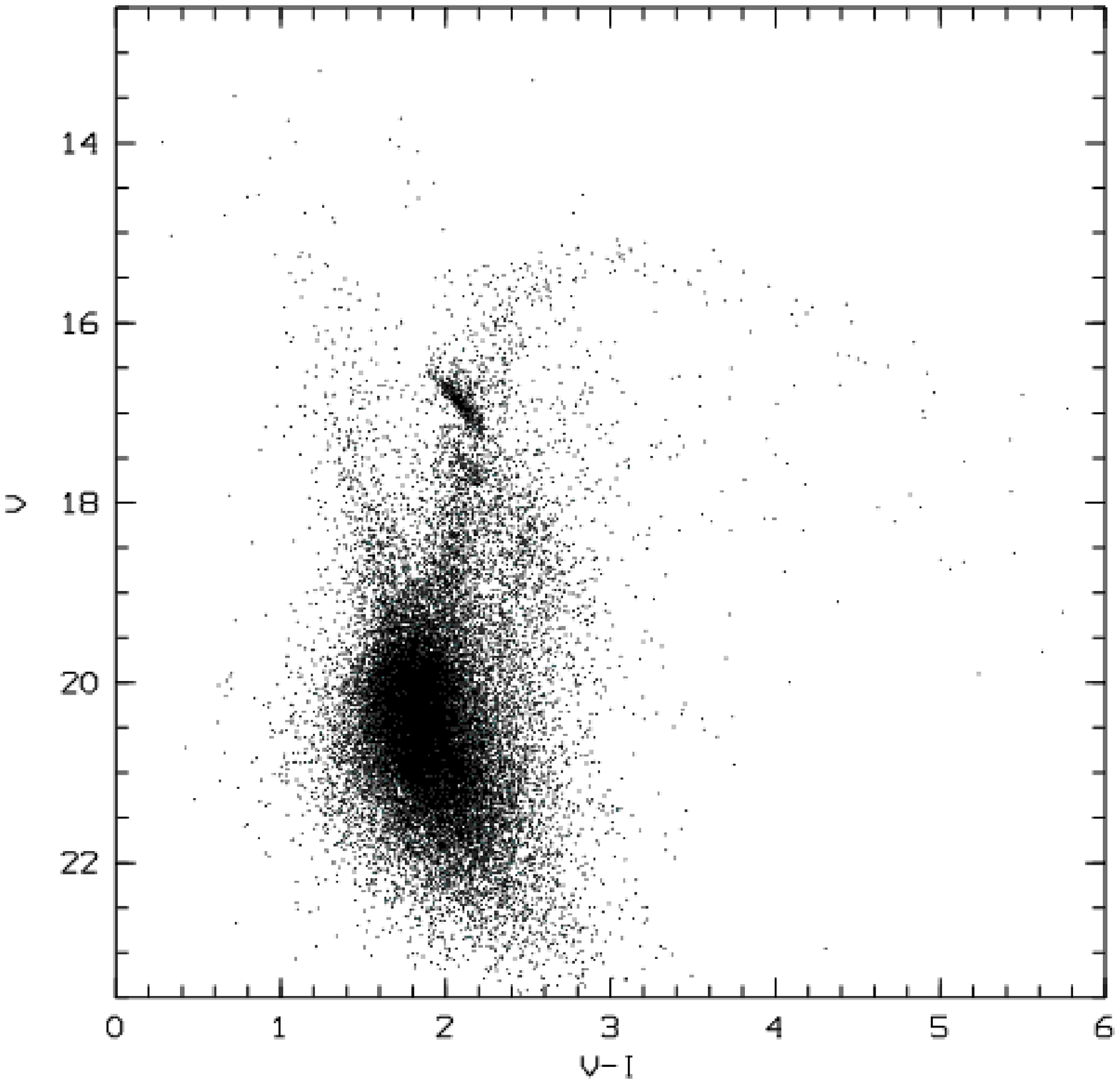,scale=0.48
                                               ,bbllx=20.5cm,bblly=18.0cm}}}
      \end{picture}
    }

    \put(9.7,0.2)
    {
      \begin{picture}(0.0,0.0)
      \put(0.0,0.0){\makebox(7.0,7.0){\epsfig{file=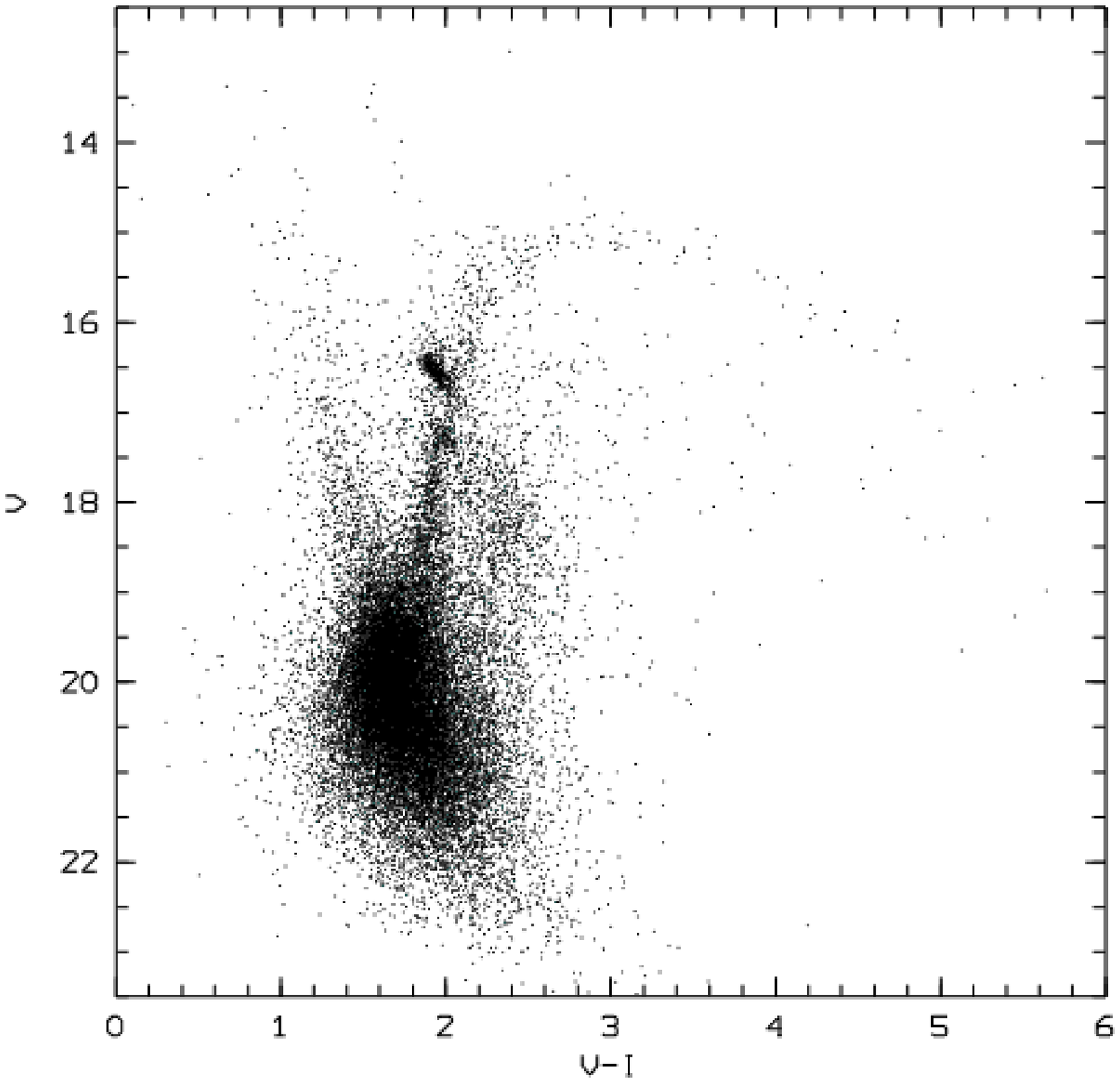,scale=0.48
                                               ,bbllx=20.5cm,bblly=18.0cm}}}
      \end{picture}
    }
  \end{picture}
  \hfill
  \parbox{8.8cm}{
  \caption{Unselected CMD for \object{NGC~6553}. Strong contamination by the field population is visible.
           \label{dia6553all}}}
  \hfill
  \parbox{8.8cm}{
  \caption{Differentially dereddened CMD for \object{NGC~6553}. RGB and RGB-bump are clearly defined now, and
           the HB lies well to the blue of the RGB. 
           \label{dia6553dc}}}
\end{figure*}

The CMDs for \object{NGC~6528} and 6553 (Fig. \ref{dia6528all}, \ref{dia6553all}) have already been published 
(Richtler et al. \cite{RTL98}, Sagar et al. \cite{SAG98}). As described in their papers, \object{NGC~6528} not 
only shows a broadened RGB and a tilted and elongated HB, but it shows also some background population below 
the AGB/RGB. The field population covers the TOP-region of the cluster. Moreover, the RGB-bump of \object{NGC~6528} is 
clearly visible some $0.5$ mag below the HB. The CMD of \object{NGC~6553} shows the same characteristics as \object{NGC~6528}, 
however they are even more distinct. Here we clearly see the background population with its RGB and AGB/RGB 
strongly differentially reddened. 
\begin{figure*}[hp]
  \begin{picture}(18.0,8.8)
    \put(0.5,0.2)
    {
      \begin{picture}(0.0,0.0)
      \put(0.0,0.0){\makebox(7.0,7.0){\epsfig{file=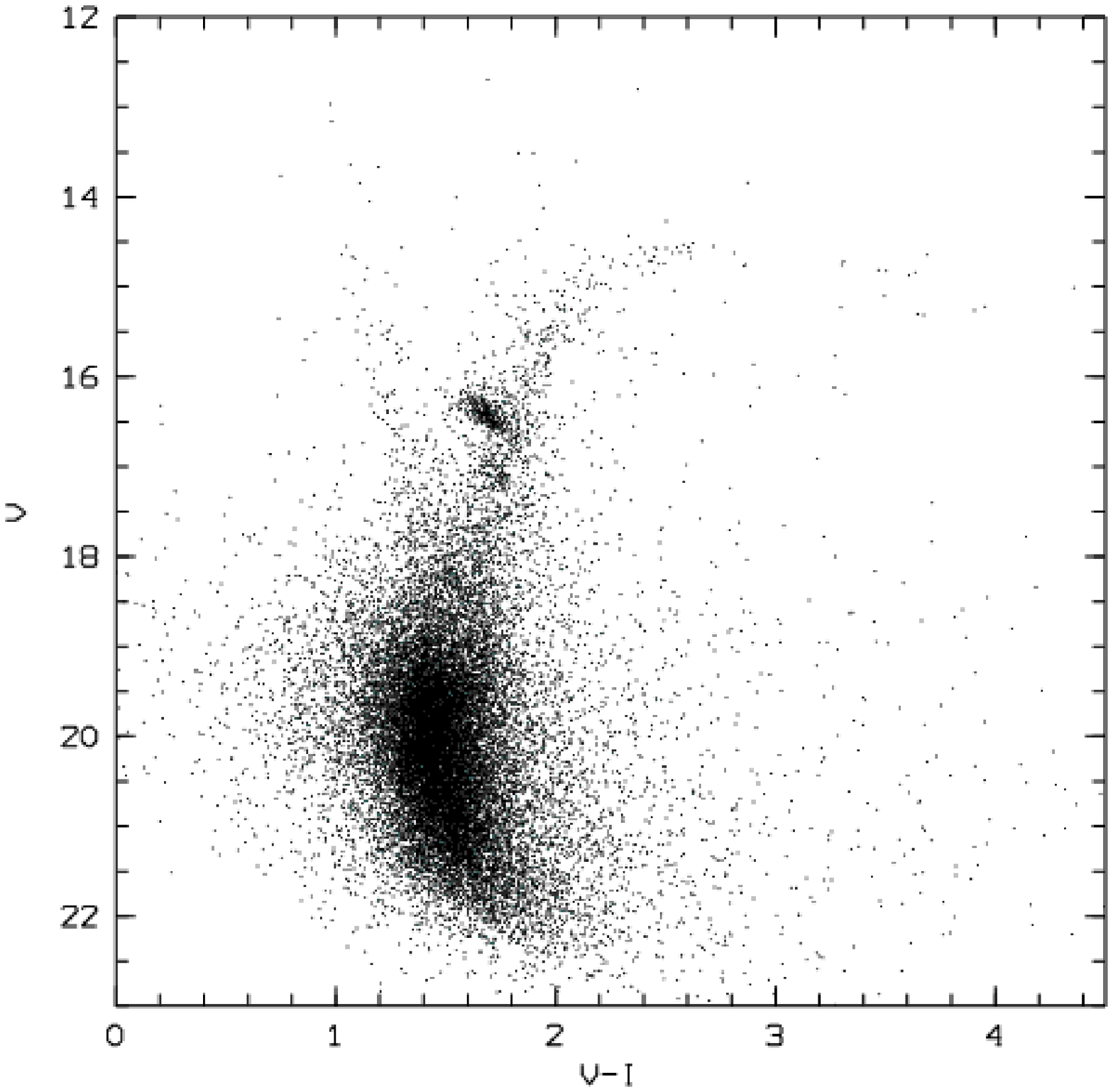,scale=0.48
                                               ,bbllx=20.5cm,bblly=18.0cm}}}
      \end{picture}
    }

    \put(9.7,0.2)
    {
      \begin{picture}(0.0,0.0)
      \put(0.0,0.0){\makebox(7.0,7.0){\epsfig{file=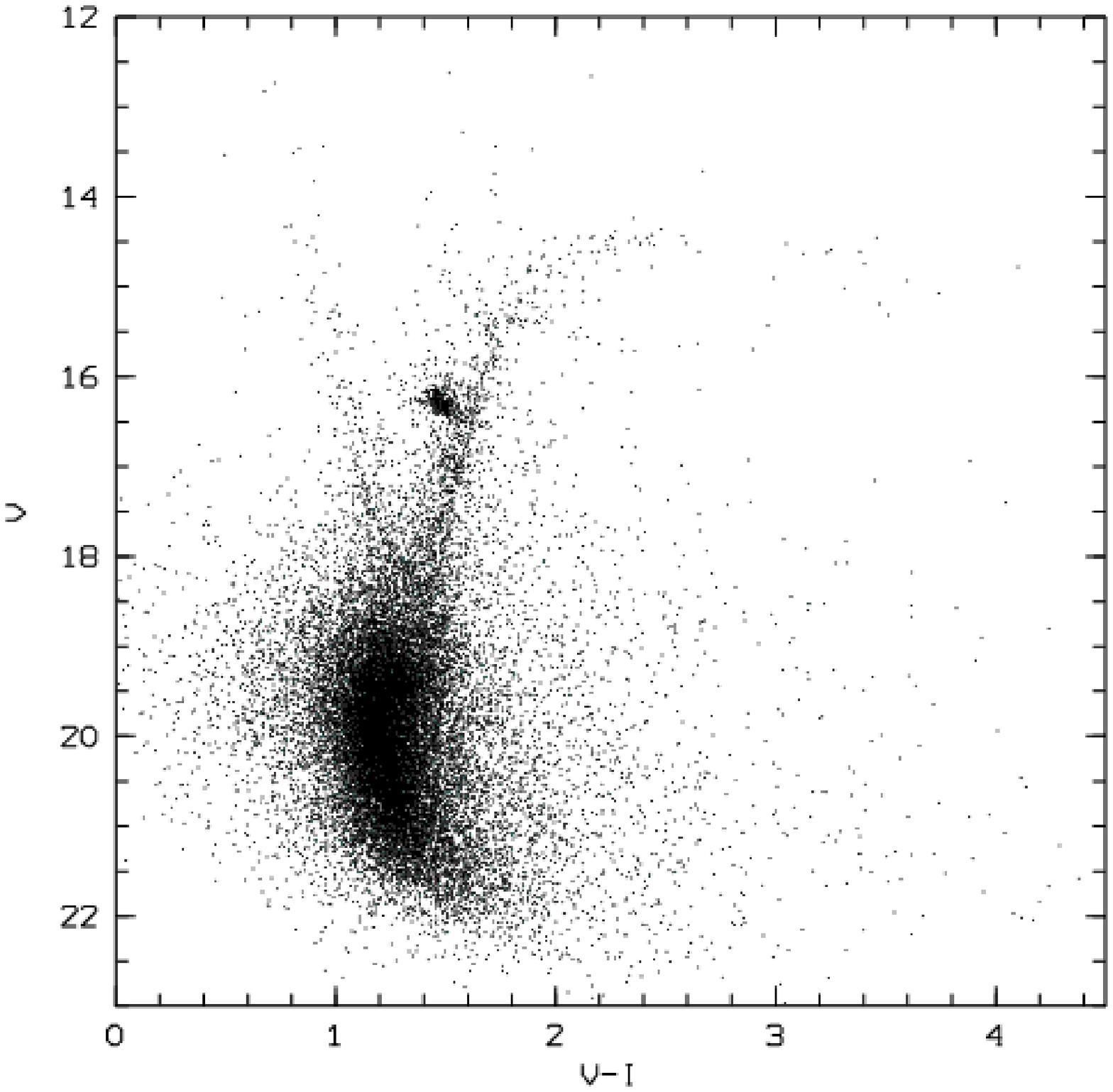,scale=0.48
                                               ,bbllx=20.5cm,bblly=18.0cm}}}
      \end{picture}
    }
  \end{picture}
  \hfill
  \parbox{8.8cm}{
  \caption{Unselected CMD for \object{NGC~5927}. RGB and HB are clearly visible as
           well as some stars of the field population.\label{dia5927all}}}
  \hfill
  \parbox{8.8cm}{
  \caption{Differentially dereddened CMD for \object{NGC~5927}. The RGB-bump is well discernible now.
           \label{dia5927dc}}}
\end{figure*}

\begin{figure*}[hp]
  \begin{picture}(18.0,8.8)
    \put(0.5,0.2)
    {
      \begin{picture}(0.0,0.0)
      \put(0.0,0.0){\makebox(7.0,7.0){\epsfig{file=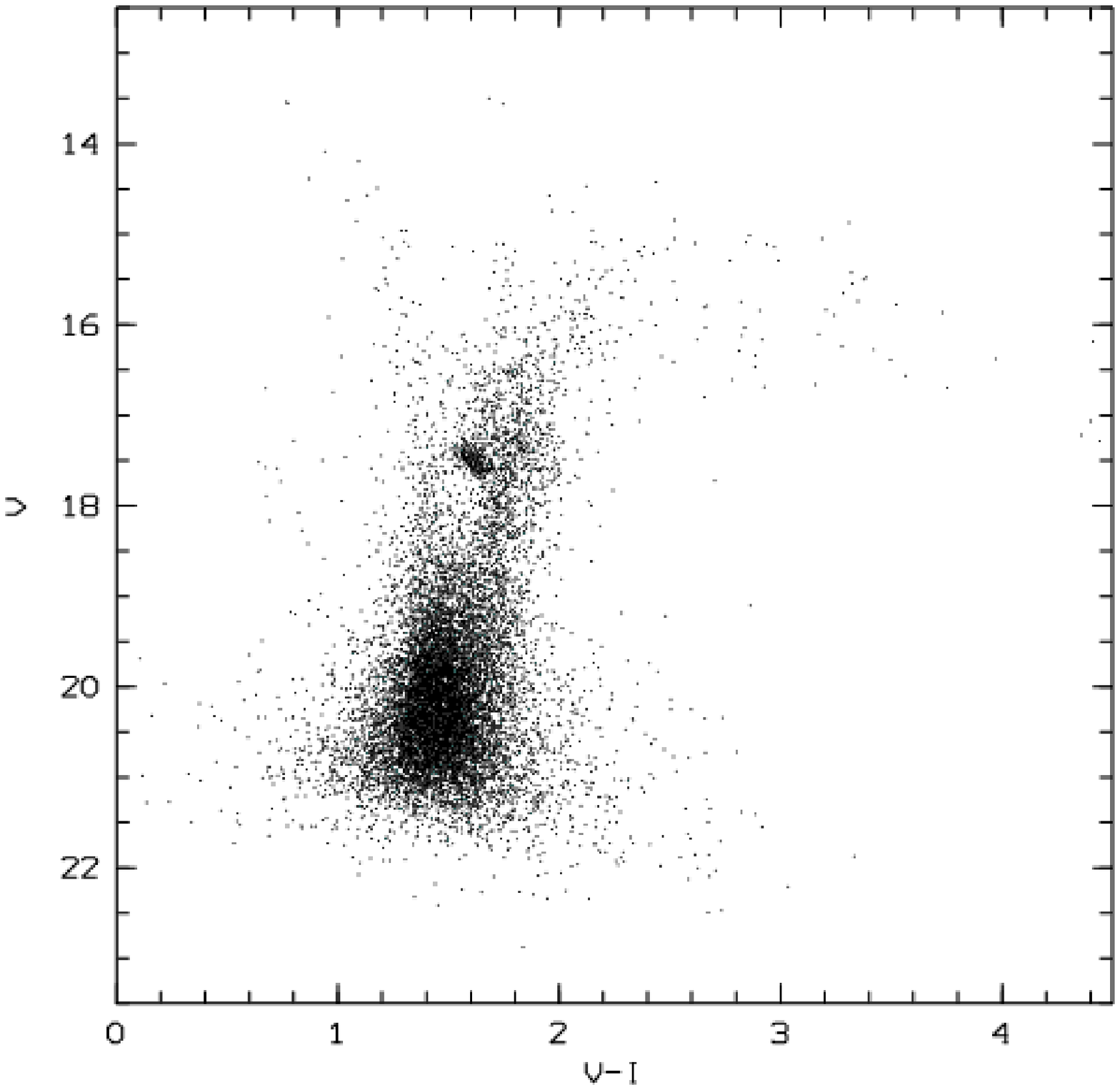,scale=0.48
                                               ,bbllx=20.5cm,bblly=18.0cm}}}
      \end{picture}
    }

    \put(9.7,0.2)
    {
      \begin{picture}(0.0,0.0)
      \put(0.0,0.0){\makebox(7.0,7.0){\epsfig{file=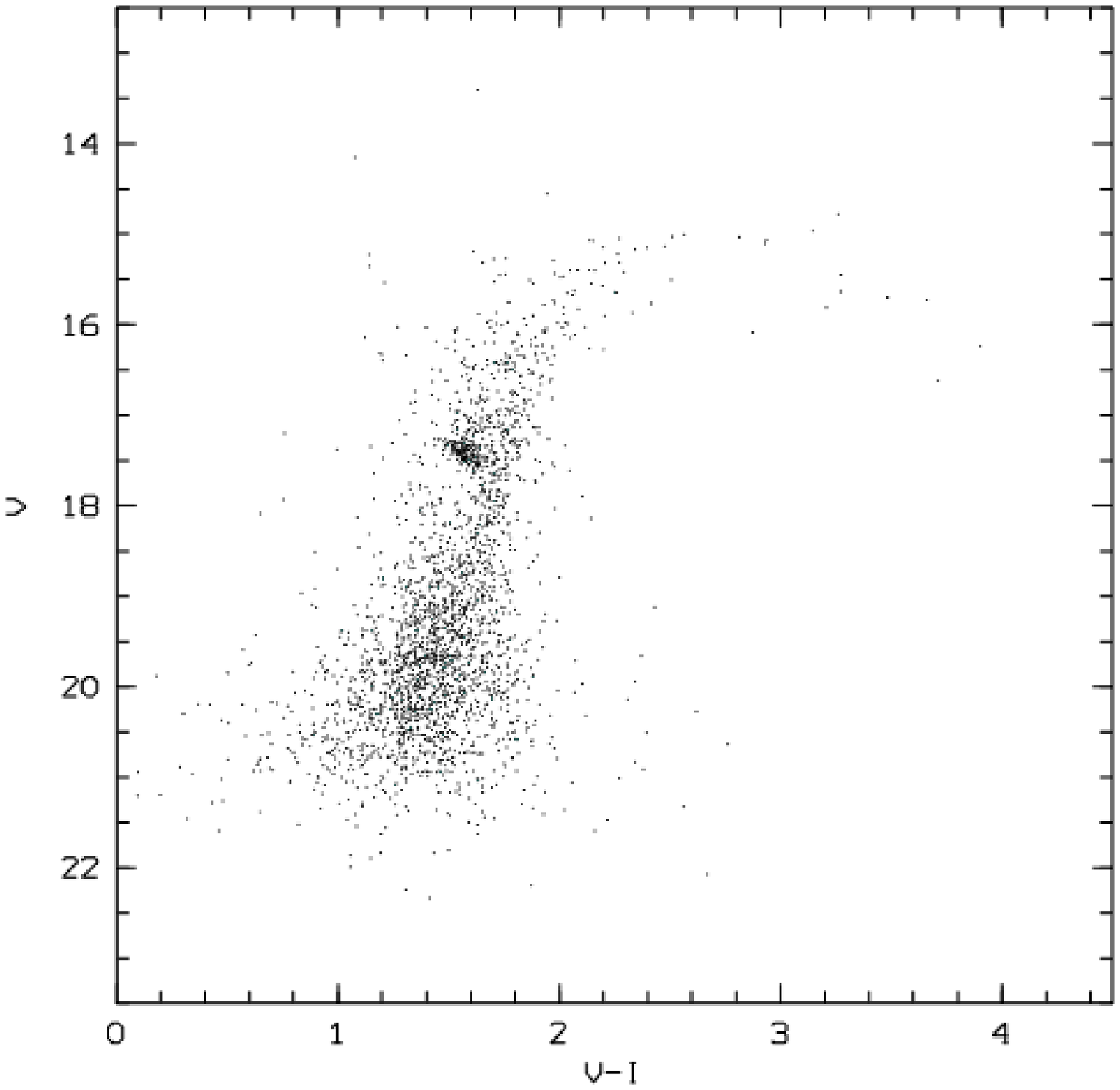,scale=0.48
                                               ,bbllx=20.5cm,bblly=18.0cm}}}
      \end{picture}
    }
  \end{picture}
  \hfill
  \parbox{8.8cm}{
  \caption{Unselected CMD for \object{NGC~6316}. The AGB/RGB of the field population is well discernible.
           \label{dia6316all}}}
  \hfill
  \parbox{8.8cm}{
  \caption{Differentially dereddened CMD for NGC 6316.
           Due to the only slight differential reddening, the effect of the correction is not as
           distinct as in \object{NGC~5927}, for example. Moreover, the dereddened CMD contains stars of the inner
           $2.24'\times 2.24'$ around the cluster's center only (See extinction maps in paragraph
           \ref{ssecCorDifRed}).\label{dia6316dc}}}

\end{figure*}
\begin{figure*}[hp]
  \begin{picture}(18.0,8.8)
    \put(0.5,0.2)
    {
      \begin{picture}(0.0,0.0)
      \put(0.0,0.0){\makebox(7.0,7.0){\epsfig{file=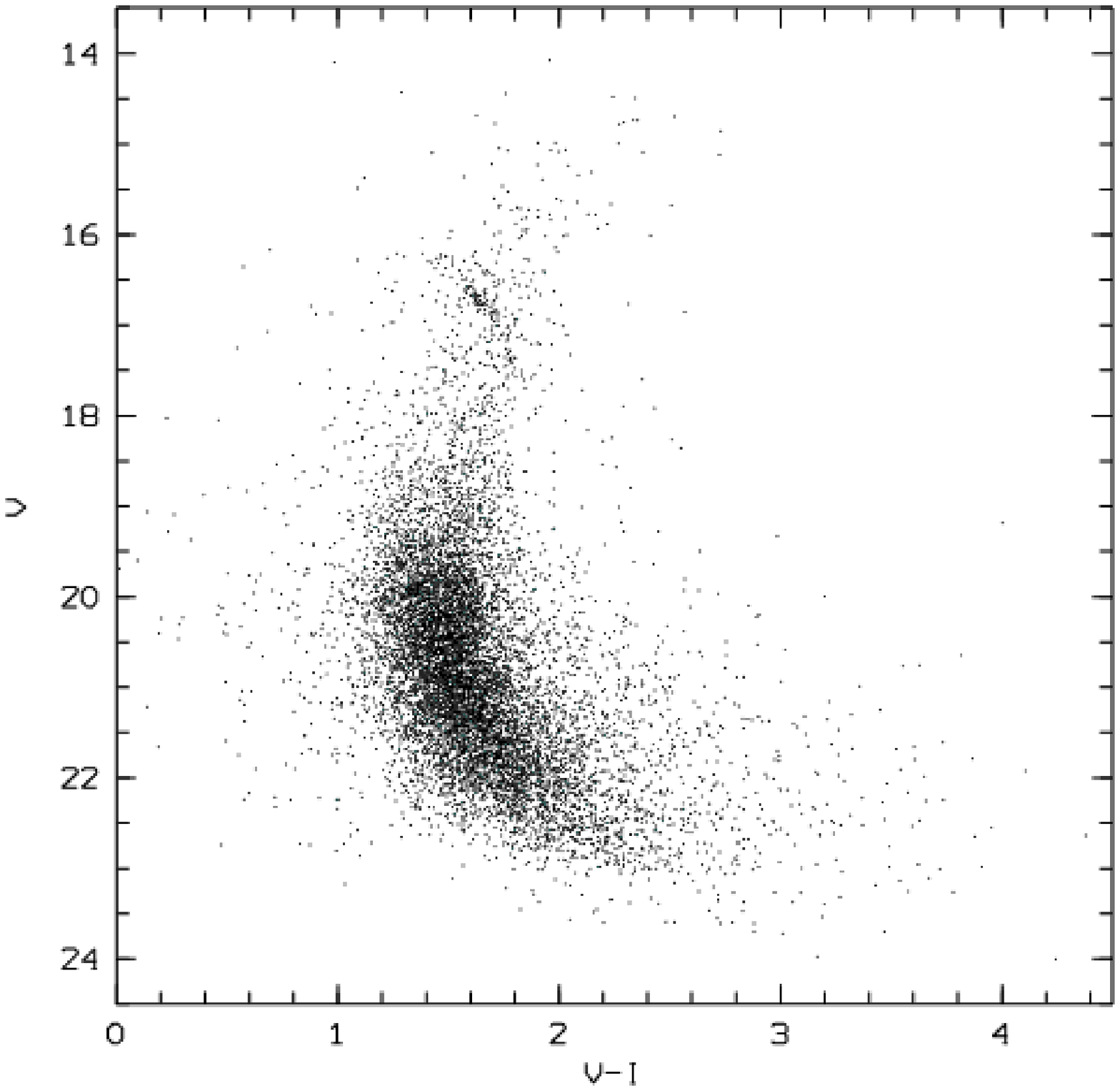,scale=0.48
                                               ,bbllx=20.5cm,bblly=18.0cm}}}
      \end{picture}
    }

    \put(9.7,0.2)
    {
      \begin{picture}(0.0,0.0)
      \put(0.0,0.0){\makebox(7.0,7.0){\epsfig{file=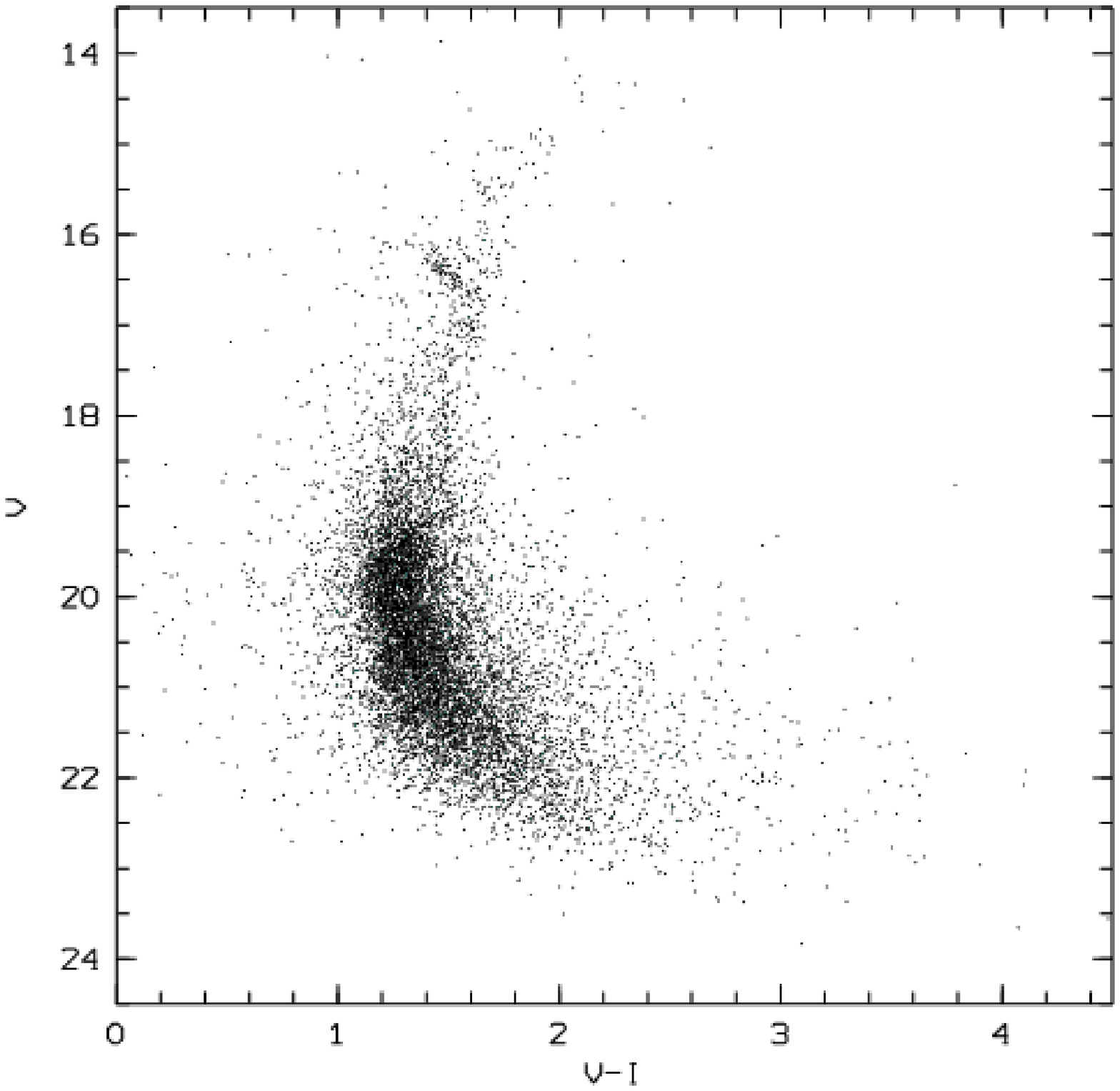,scale=0.48
                                               ,bbllx=20.5cm,bblly=18.0cm}}}
      \end{picture}
    }
  \end{picture}
  \hfill
  \parbox{8.8cm}{
  \caption{Unselected CMD for \object{NGC~6342}. TOP and upper MS are reached.
           \label{dia6342all}}}
  \hfill
  \parbox{8.8cm}{
  \caption{Differentially dereddened CMD for \object{NGC~6342}. The effect of the correction is best
           seen at the TOP-region. \label{dia6342dc}}}
\end{figure*}
\begin{figure*}[hp]
  \begin{picture}(18.0,8.8)
    \put(0.5,0.2)
    {
      \begin{picture}(0.0,0.0)
      \put(0.0,0.0){\makebox(7.0,7.0){\epsfig{file=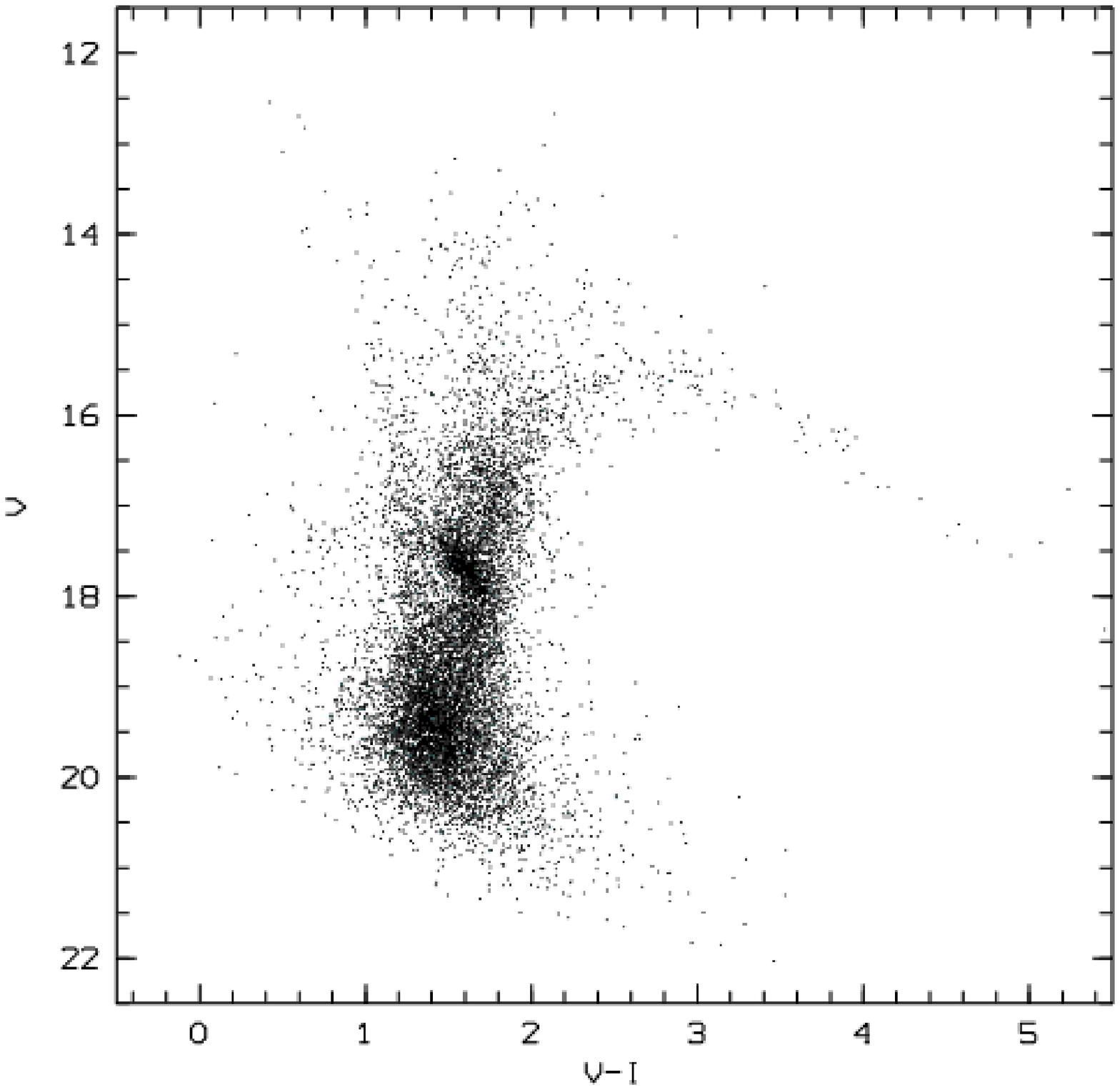,scale=0.48
                                               ,bbllx=20.5cm,bblly=18.0cm}}}
      \end{picture}
    }

    \put(9.7,0.2)
    {
      \begin{picture}(0.0,0.0)
      \put(0.0,0.0){\makebox(7.0,7.0){\epsfig{file=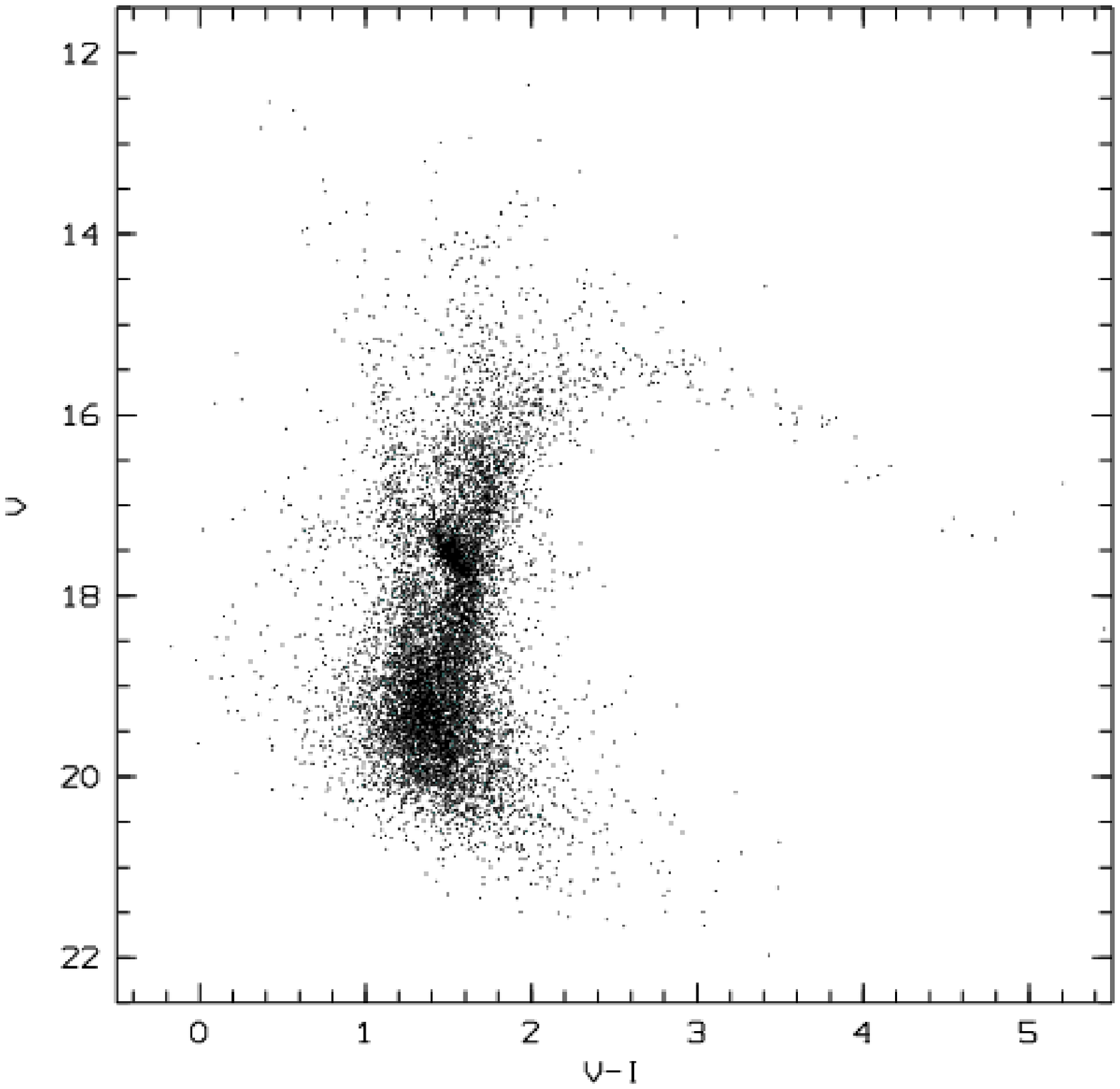,scale=0.48
                                               ,bbllx=20.5cm,bblly=18.0cm}}}
      \end{picture}
    }
  \end{picture}
  \hfill
  \parbox{8.8cm}{
  \caption{Unselected CMD for \object{NGC~6441}. Strong contamination by the field population.\label{dia6441all}}}
  \hfill
  \parbox{8.8cm}{
  \caption{Differentially dereddened CMD for \object{NGC~6441}. The effect of the correction shows best in the 
           narrower lower RGB. The HB still partly overlaps with the RGB, which means, that the 
           correction was not completely succesful. This is mostly due to the strong field population.
           \label{dia6441dc}}}
\end{figure*}

\begin{figure*}[hp]
  \begin{picture}(17.6,9.3)
    \put(0.0,4.8)
    {
      \begin{picture}(0.0,0.0)
      \put(0.0,0.0){\makebox(4.0,4.0){\epsfig{file=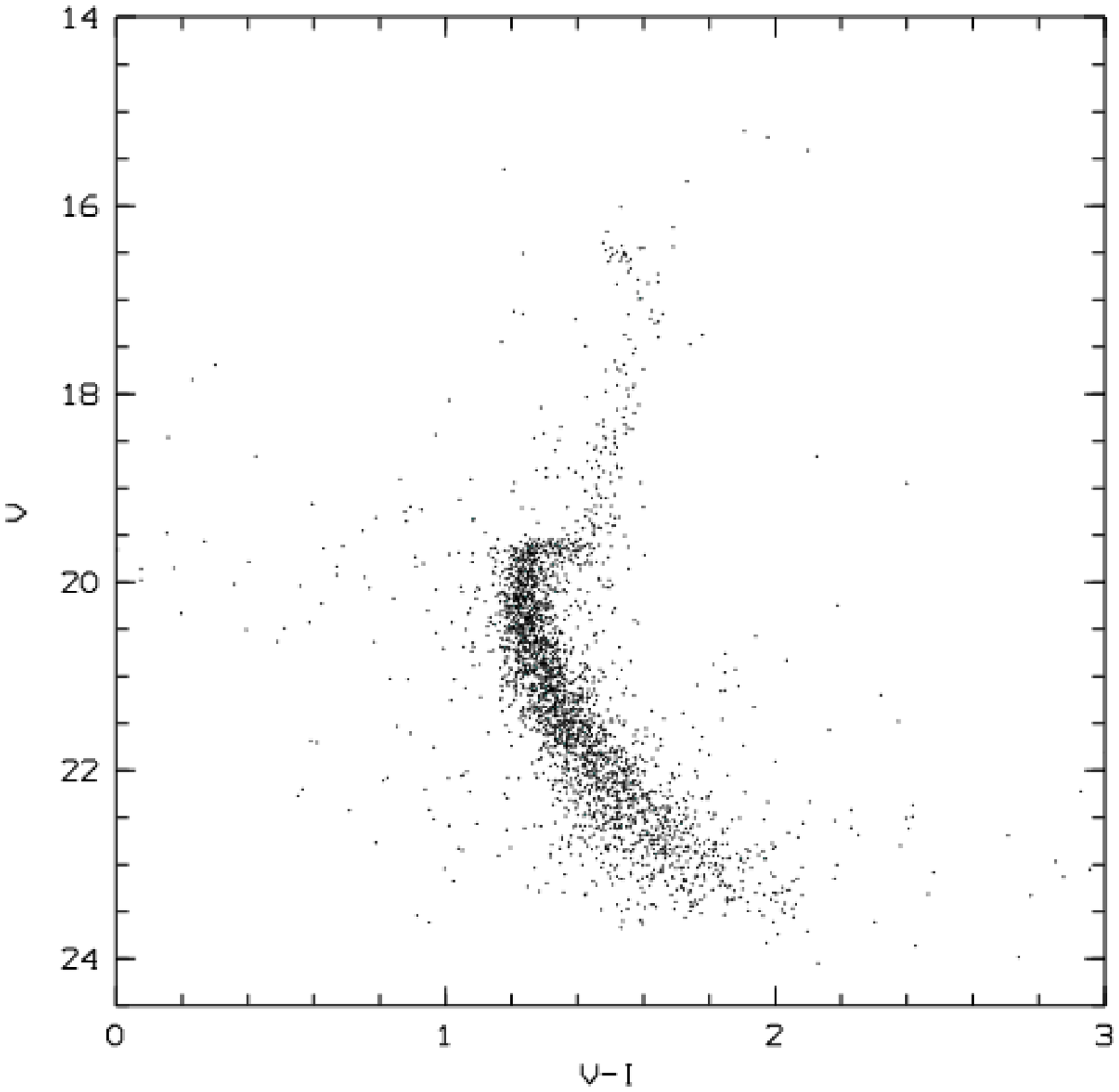,scale=0.24
                                               ,bbllx=20.5cm,bblly=18.0cm}}}
      \end{picture}
    }

    \put(4.6,4.8)
    {
      \begin{picture}(0.0,0.0)
      \put(0.0,0.0){\makebox(4.0,4.0){\epsfig{file=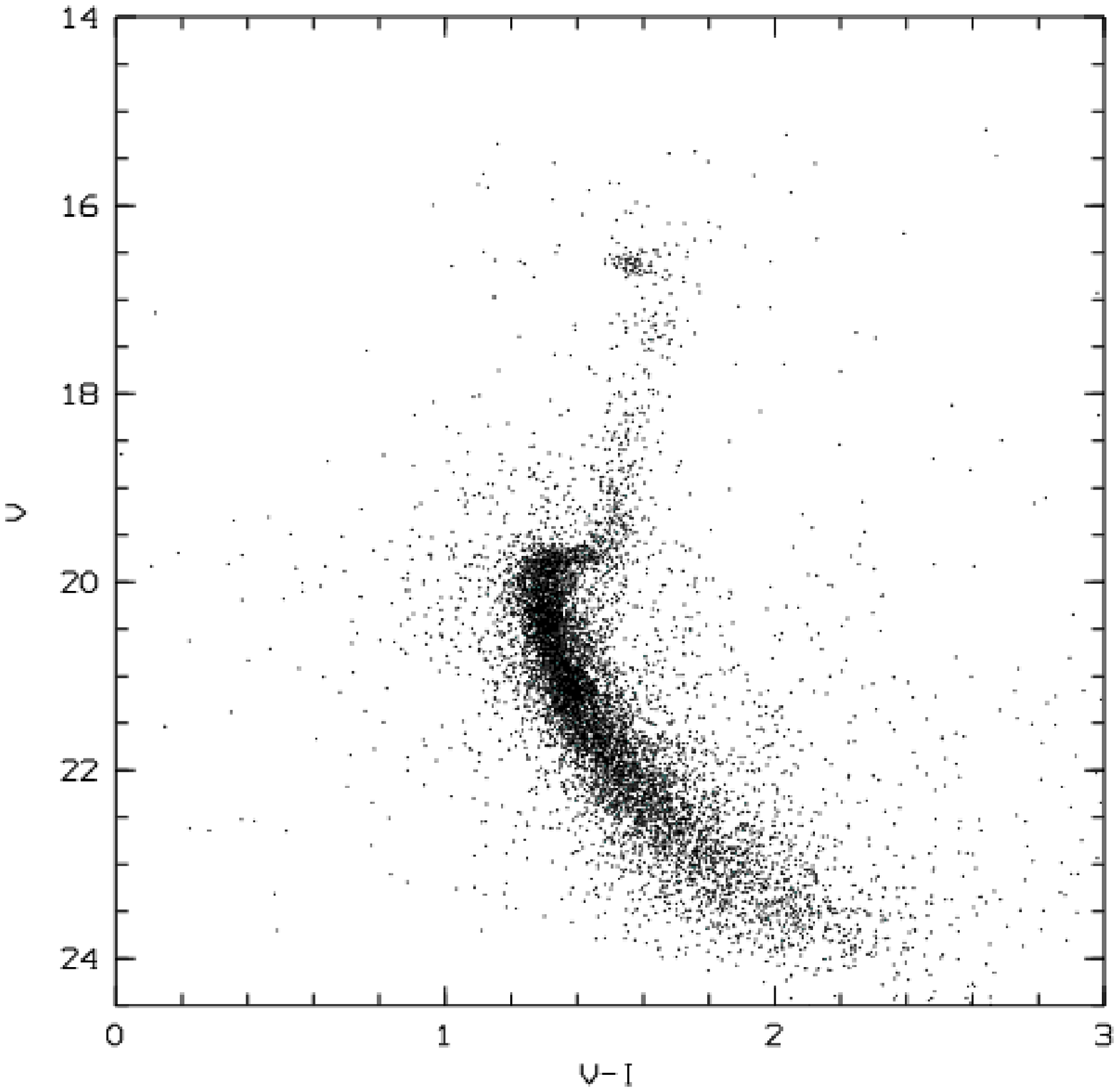,scale=0.24
                                               ,bbllx=20.5cm,bblly=18.0cm}}}
      \end{picture}
    }
    \put(0.0,0.0)
    {
      \begin{picture}(0.0,0.0)
      \put(0.0,0.0){\makebox(4.0,4.0){\epsfig{file=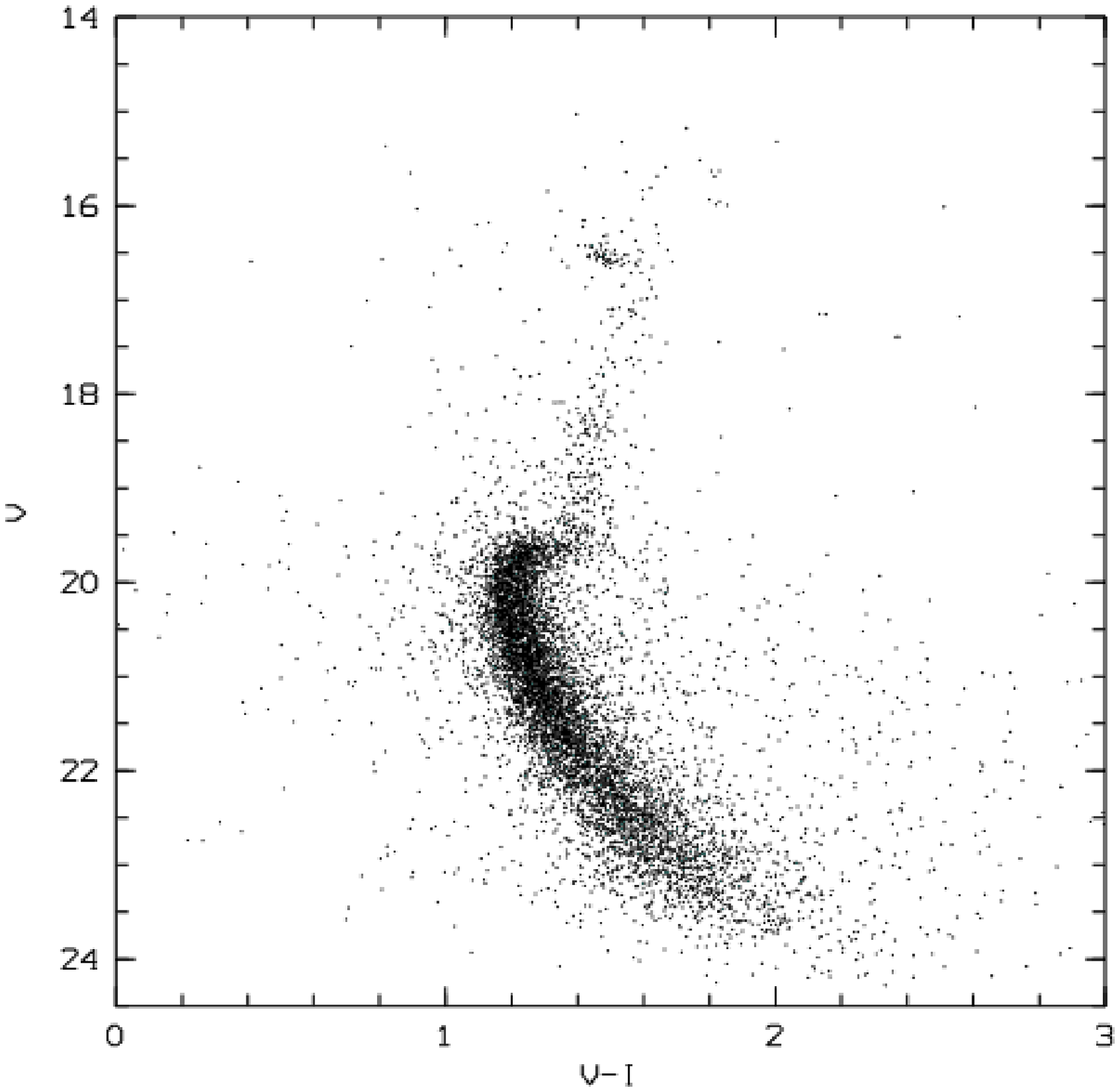,scale=0.24
                                               ,bbllx=20.5cm,bblly=18.0cm}}}
      \end{picture}
    }

    \put(4.6,0.0)
    {
      \begin{picture}(0.0,0.0)
      \put(0.0,0.0){\makebox(4.0,4.0){\epsfig{file=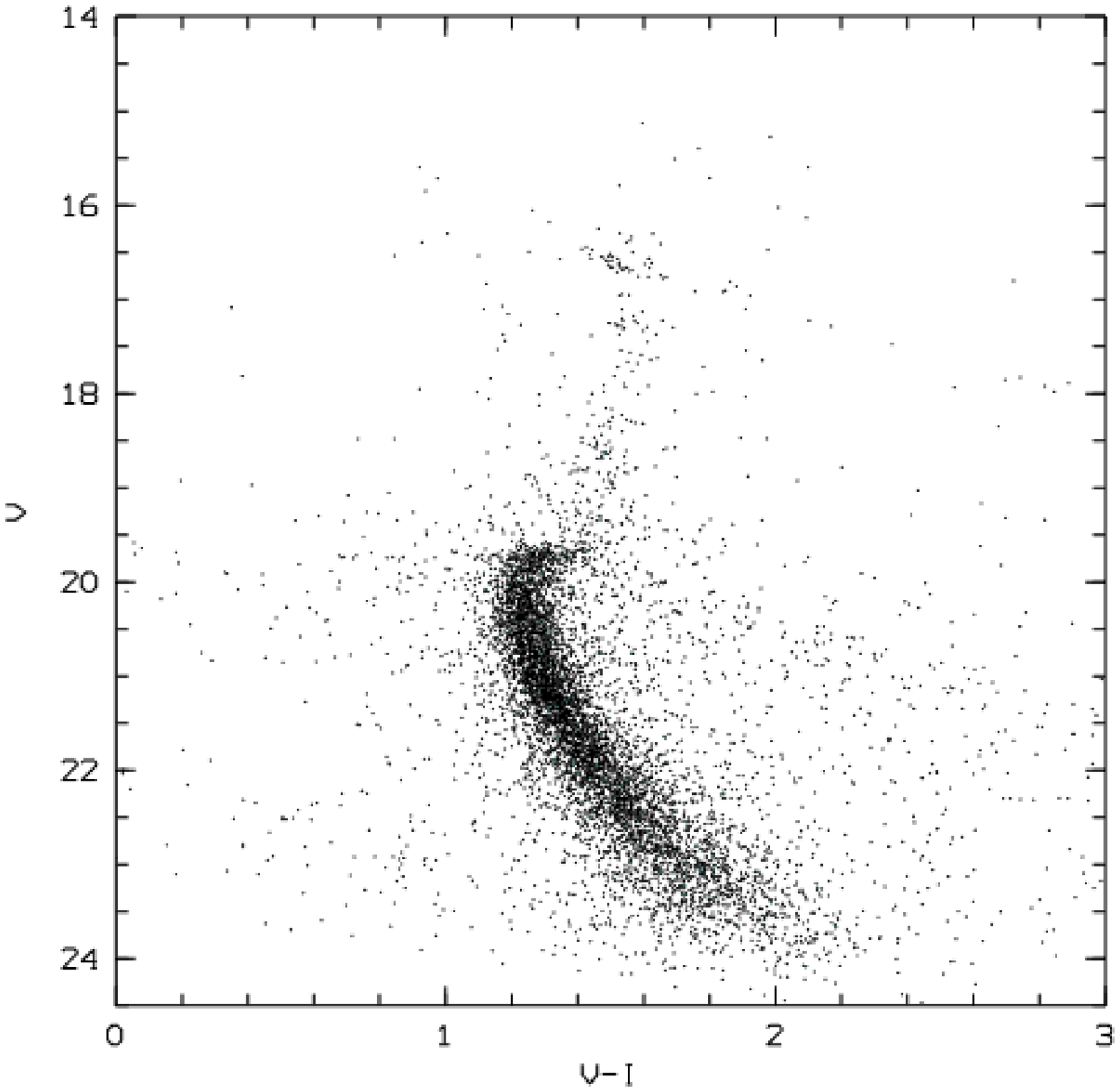,scale=0.24
                                               ,bbllx=20.5cm,bblly=18.0cm}}}
      \end{picture}
    }

    \put(9.7,0.2)
    {
      \begin{picture}(0.0,0.0)
      \put(0.0,0.0){\makebox(8.0,8.0){\epsfig{file=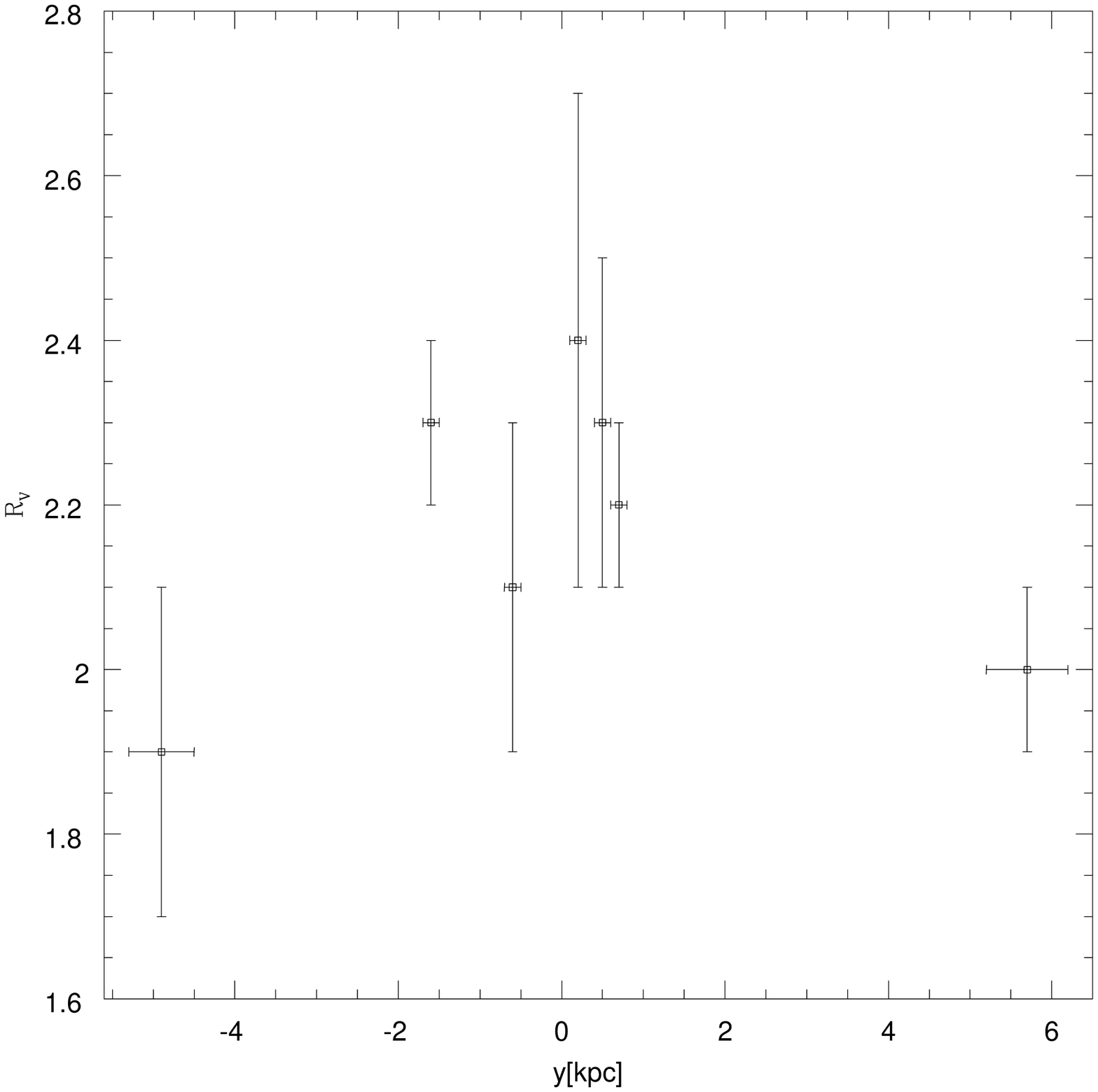,scale=0.43
                                               ,bbllx=21.0cm,bblly=15.0cm}}}
      \end{picture}
    }
  \end{picture}
  \hfill
  \parbox{8.8cm}
  {
    \caption{\object{NGC~5927}: Unselected HST-CMDs (PC, WF2, WF3 and WF4). The
             TOP is well resolved.
             \label{dia5927HST}}
  }
  \hfill
  \parbox{8.8cm}
  {
    \caption{Slopes of the HBs, i.e. the reddening vectors against galactic longitude (here in cartesian 
             coordinates).\label{diaExtVar}}
  }
\end{figure*}

\section{Correction for differential reddening\label{ssecCorDifRed}}
\begin{figure*}[h]
  \begin{picture}(18.0,19.0)
    \put(0.0,12.2)
    {
      \begin{picture}(0.0,0.0)
      \put(0.0,0.0){\makebox(5.5,5.5){\epsfig{file=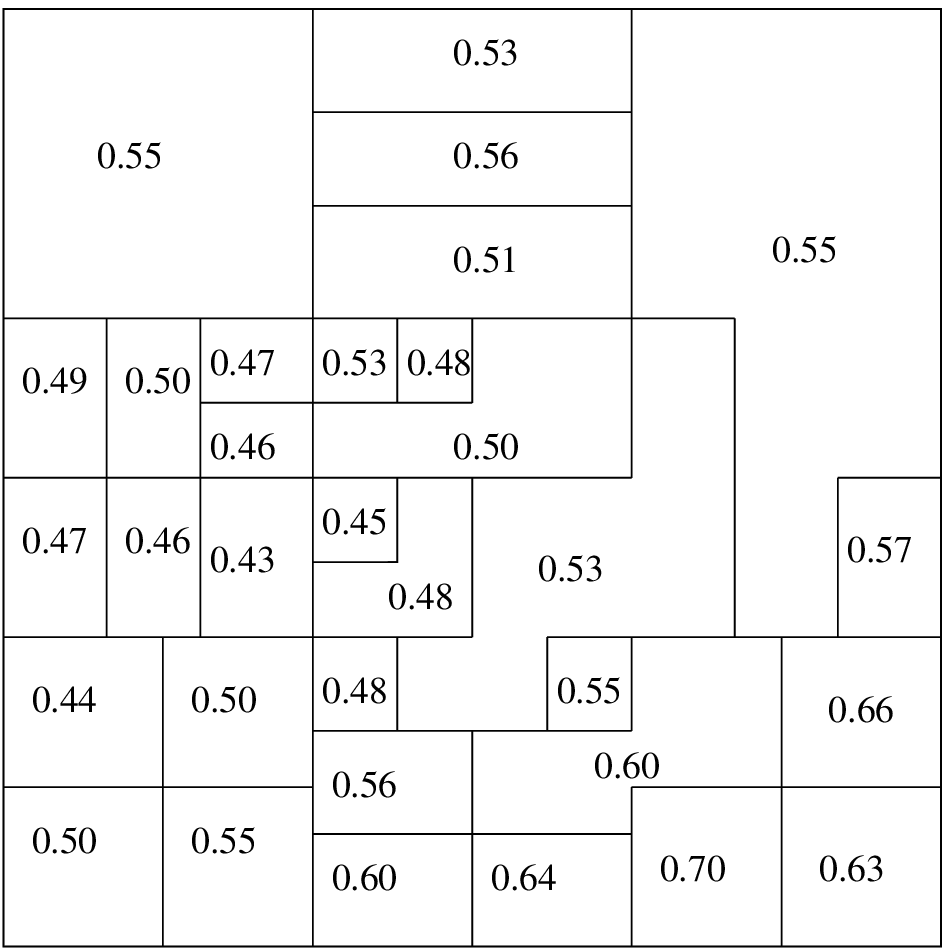,scale=0.52
                                               ,bbllx=10.5cm,bblly=5.3cm}}}
      \end{picture}
    }
    \put(0.5,17.2){\makebox(1.,0.4){\footnotesize \object{NGC~5927}}}
    \put(2.5,17.2){\makebox(0.6,0.6){\bf N}}
    \put(5.1,14.7){\makebox(0.4,0.6){\bf E}}

    \put(5.8,12.2)
    {
      \begin{picture}(0.0,0.0)
      \put(0.0,0.0){\makebox(5.5,5.5){\epsfig{file=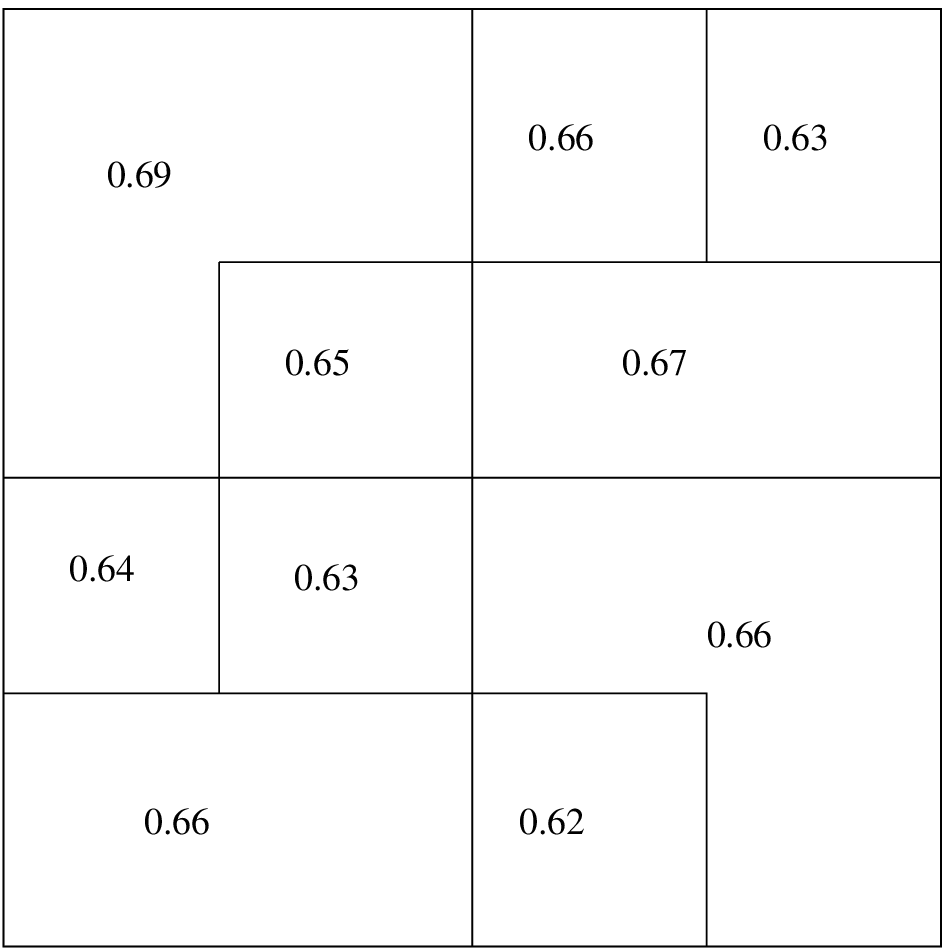,scale=0.52
                                               ,bbllx=10.5cm,bblly=5.3cm}}}
      \end{picture}
    }
    \put(6.3,17.2){\makebox(1.0,0.4){\footnotesize \object{NGC~6316}}}
    \put(8.3,17.2){\makebox(0.6,0.6){\bf N}}
    \put(10.9,14.7){\makebox(0.4,0.6){\bf E}}

    \put(11.7,12.2)
    {
      \begin{picture}(0.0,0.0)
      \put(0.0,0.0){\makebox(5.5,5.5){\epsfig{file=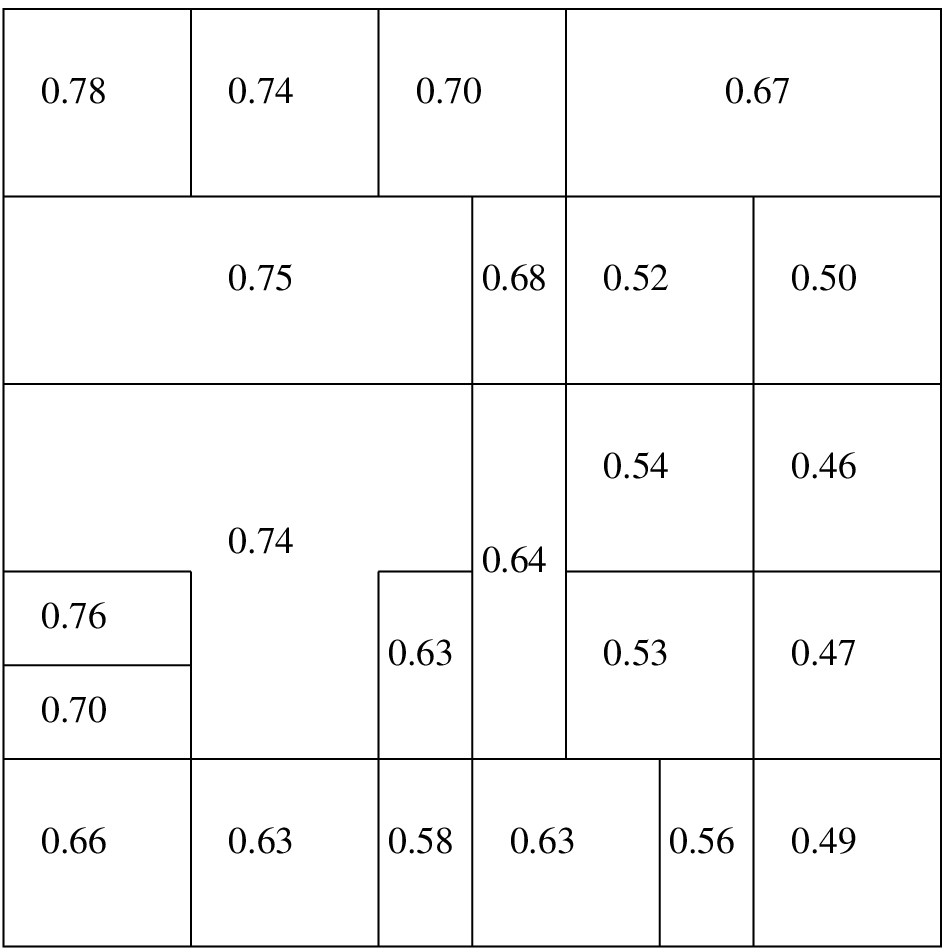,scale=0.52
                                               ,bbllx=10.5cm,bblly=5.3cm}}}
      \end{picture}
    }
    \put(12.2,17.2){\makebox(1.0,0.4){\footnotesize \object{NGC~6342}}}
    \put(14.2,17.2){\makebox(0.6,0.6){\bf N}}
    \put(16.8,14.7){\makebox(0.4,0.6){\bf E}}

    \put(0.0,6.2)
    {
      \begin{picture}(0.0,0.0)
      \put(0.0,0.0){\makebox(5.5,5.5){\epsfig{file=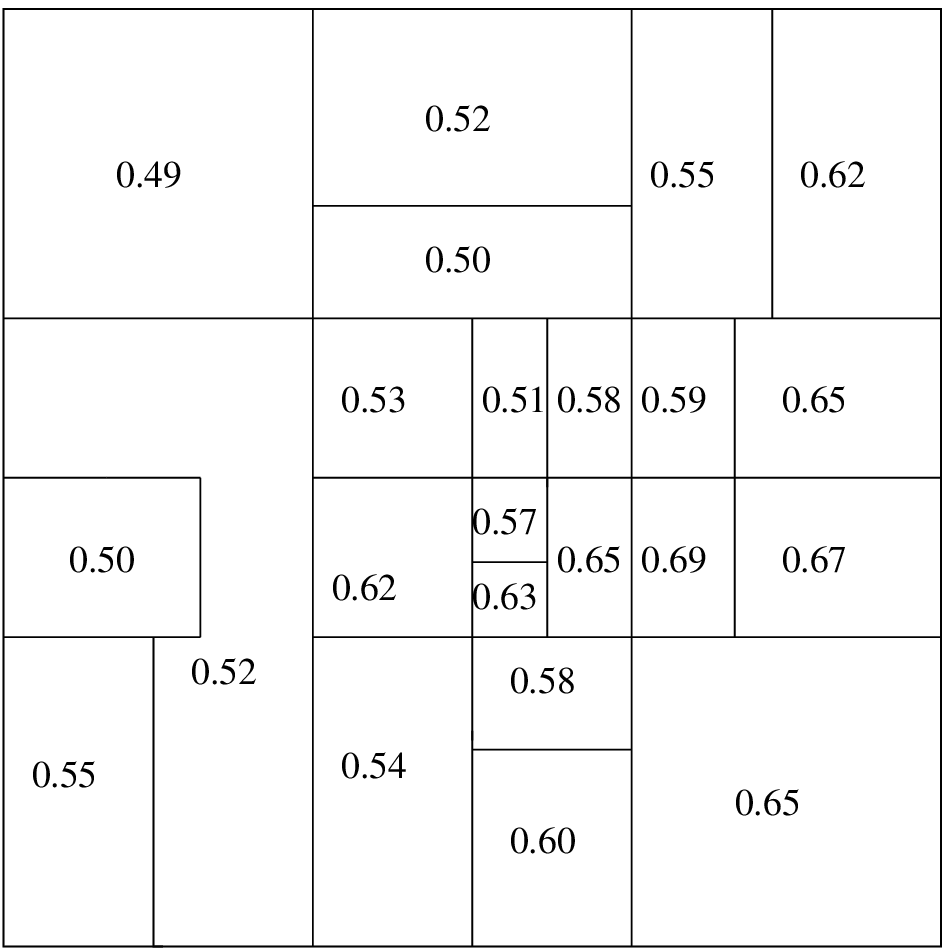,scale=0.52
                                               ,bbllx=10.5cm,bblly=5.3cm}}}
      \end{picture}
    }
    \put(0.5,11.2){\makebox(1.0,0.4){\footnotesize \object{NGC~6441}}}
    \put(2.5,11.2){\makebox(0.6,0.6){\bf N}}
    \put(5.1,8.7){\makebox(0.4,0.6){\bf E}}

    \put(5.8,6.2)
    {
      \begin{picture}(0.0,0.0)
      \put(0.0,0.0){\makebox(5.5,5.5){\epsfig{file=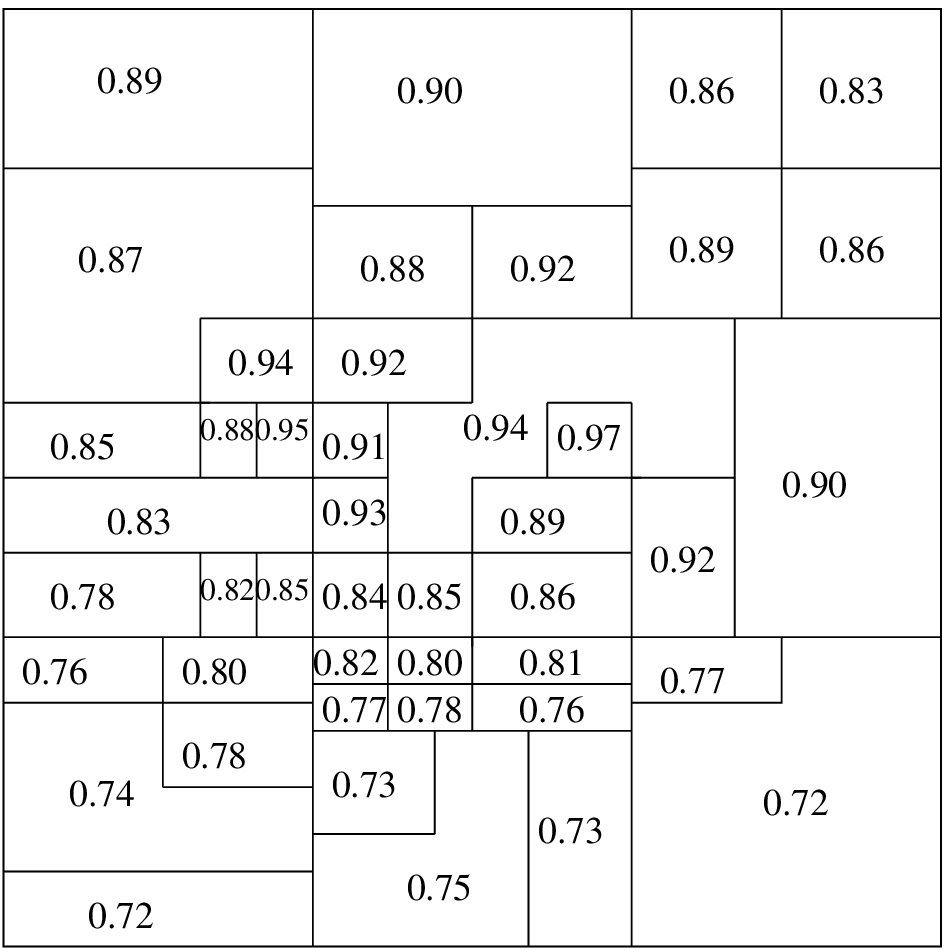,scale=0.52
                                               ,bbllx=10.5cm,bblly=5.3cm}}}
      \end{picture}
    }
    \put(6.3,11.2){\makebox(1.0,0.4){\footnotesize \object{NGC~6760}}}
    \put(8.3,11.2){\makebox(0.6,0.6){\bf N}}
    \put(10.9,8.7){\makebox(0.4,0.6){\bf E}}

    \put(11.7,6.2)
    {
      \begin{picture}(0.0,0.0)
      \put(0.0,0.0){\makebox(5.5,5.5){\epsfig{file=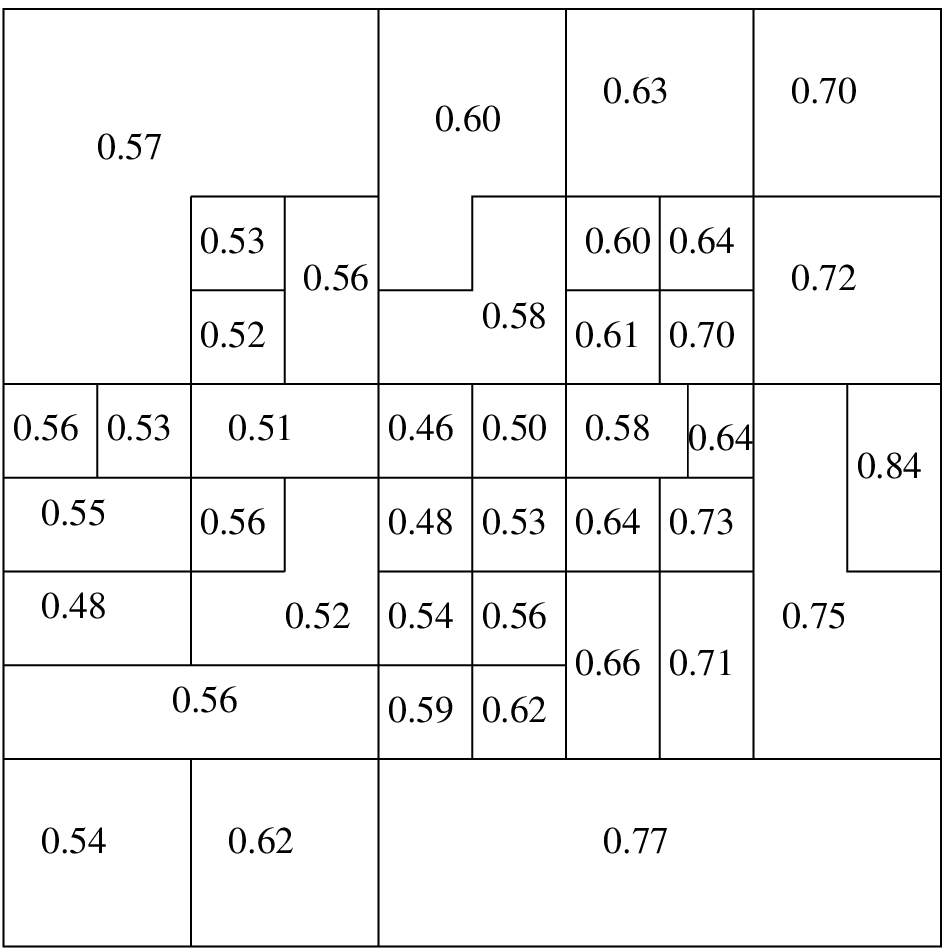,scale=0.52
                                               ,bbllx=10.5cm,bblly=5.3cm}}}
      \end{picture}
    }
    \put(12.2,11.2){\makebox(1.0,0.4){\footnotesize \object{NGC~6528}}}
    \put(14.2,11.2){\makebox(0.6,0.6){\bf N}}
    \put(16.8,8.7){\makebox(0.4,0.6){\bf E}}

    \put(5.8,0.2)
    {
      \begin{picture}(0.0,0.0)
      \put(0.0,0.0){\makebox(5.5,5.5){\epsfig{file=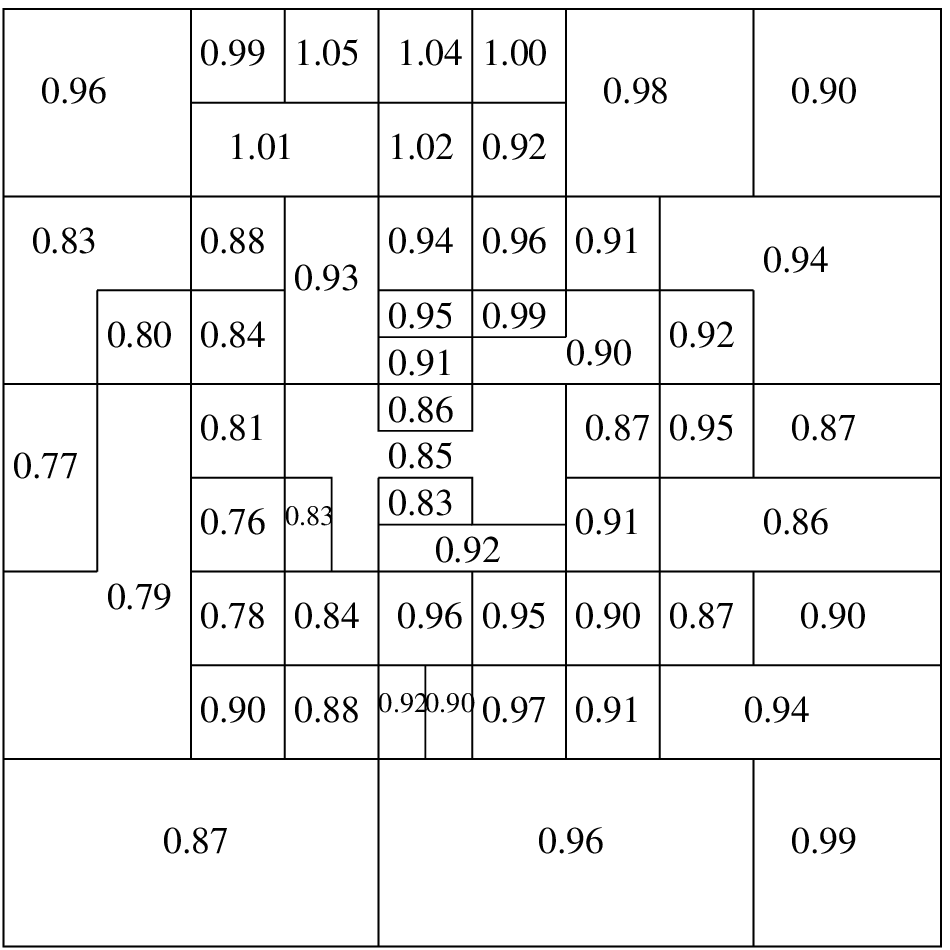,scale=0.52
                                               ,bbllx=10.5cm,bblly=5.3cm}}}
      \end{picture}
    }
    \put(6.3,5.2){\makebox(1.0,0.4){\footnotesize \object{NGC~6553}}}
    \put(8.3,5.2){\makebox(0.6,0.6){\bf N}}
    \put(10.9,2.7){\makebox(0.4,0.6){\bf E}}
  \end{picture}
  \caption{Extinction maps for \object{NGC~5927} to \object{NGC~6553}, derived from
           plotting CMDs for each of the areas. The numbers in the
           subfields give the absolute reddening derived via isochrone
           fitting. The differential reddening was determined using
           the minimal absolute reddening as a point of reference. As
           the derived scale of reddening variation strongly depends
           on the number of stars, the resulting scales can only be
           estimates for upper boundaries. The maps cover $5.7' \times 5.7'$,
           i.e. they cover the whole area of the original frames. An 
           exception is made with NGC 6136, where we constrained the map
           to the inner $2.24' \times 2.24'$, because of the clusters small
           size.
           \label{diaAllClusExt}}
\end{figure*}

In order to correct the CMDs for differential reddening, we used a
refined version of the method described by Grebel et al. (\cite{GRE95}). The entire 
frame is divided into subframes. These are determined by covering the whole frame with
a regular subgrid and dividing the grid cells further until
the number of stars in one cell becomes too small to define the CMD structure.
The CMDs will be shifted according to the reddening vector (described below)
with respect to CMDs from neighbouring cells. The shift
in colour supplies the differential reddening. If two neighbouring subframes have the same
reddening, these subframes are merged. There are two problems with this method:  
First, one has to be careful to use the HB as a means for comparing two CMDs, as the HB 
may be intrinsically elongated. Useful results can only be achieved by comparing the RGBs and TOPs, as 
far as they are accessible. Second, the size of the subfields must be large enough to render meaningful CMDs.
Fig. \ref{diaAllClusExt} shows the resulting extinction maps for the seven clusters. The 
smallest subfields have a size of about $28'' \times 28''$. But as some of them still 
showed differential reddening, the scale of the structures responsible for the differential reddening 
is be expected to be even smaller. The smallest scales we got from
a comparison of coordinates of stars with different reddening amounted to values of $4''$. 

\begin{figure*}[hp]
  \begin{picture}(18.0,8.8)
    \put(0.5,0.2)
    {
      \begin{picture}(0.0,0.0)
      \put(0.0,0.0){\makebox(7.0,7.0){\epsfig{file=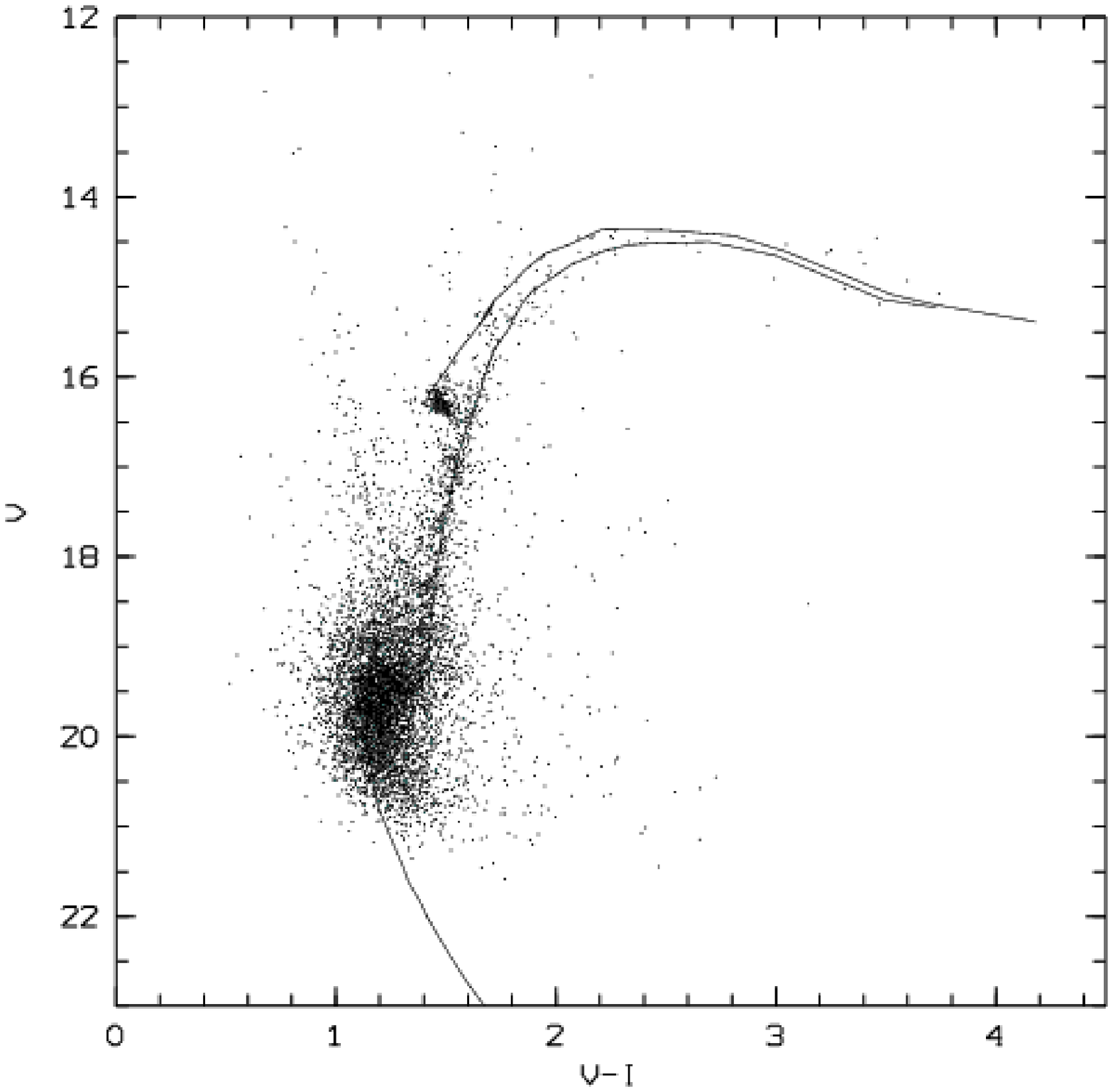,scale=0.48
                                               ,bbllx=20.5cm,bblly=18.0cm}}}
      \end{picture}
    }

    \put(9.7,0.2)
    {
      \begin{picture}(0.0,0.0)
      \put(0.0,0.0){\makebox(7.0,7.0){\epsfig{file=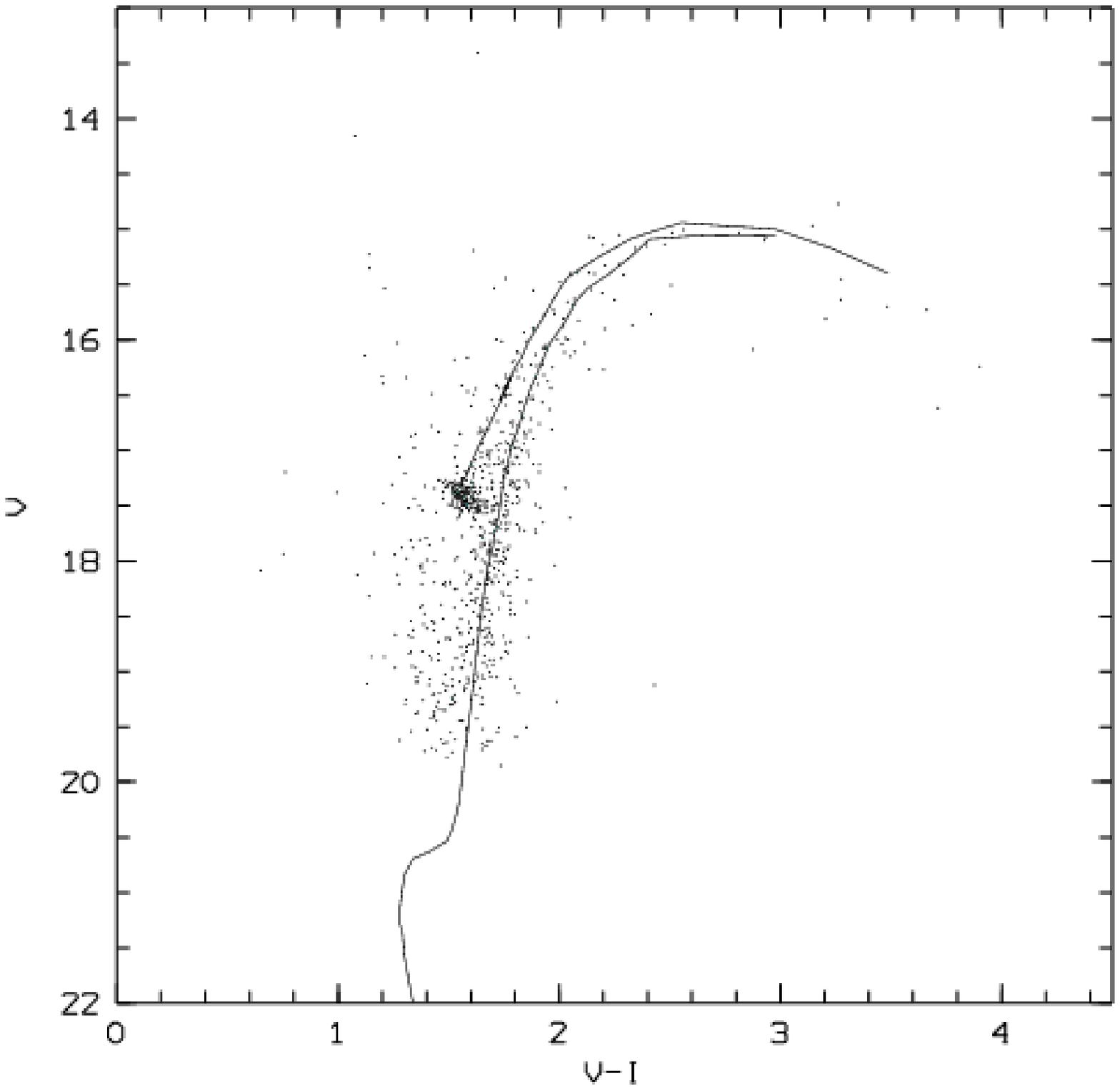,scale=0.48
                                               ,bbllx=20.5cm,bblly=18.0cm}}}
      \end{picture}
    }
  \end{picture}
  \hfill
  \parbox{8.8cm}{
  \caption{Isochrone fitting for \object{NGC~5927}. $[\mbox{M}/\mbox{H}]=-0.40$ dex, $t=14.5$ gyr. The CMD is
           selected for photometric errors $\le 0.04$ mag and for radii $(50 \le r \le 400)$ pix. 
           The first selection gives preference to brighter stars, i.e. to stars which not necessarily 
           are cluster members. This effect is counterbalanced by the second selection.  
           \label{dia5927dciso}}}
  \hfill
  \parbox{8.8cm}{
  \caption{\object{NGC~6316}, selected for photometric errors $\le 0.04$ mag and with an isochrone 
           $[\mbox{M}/\mbox{H}]=-0.70$, $t=14.5$ gyr. The radial selection is due to the
           correction for differential reddening (see Figs. \ref{dia6316dc} and \ref{diaAllClusExt}).
           \label{dia6316dciso}}}
\end{figure*}

\begin{figure*}[h]
  \begin{picture}(18.0,8.8)
    \put(0.5,0.2)
    {
      \begin{picture}(0.0,0.0)
      \put(0.0,0.0){\makebox(7.0,7.0){\epsfig{file=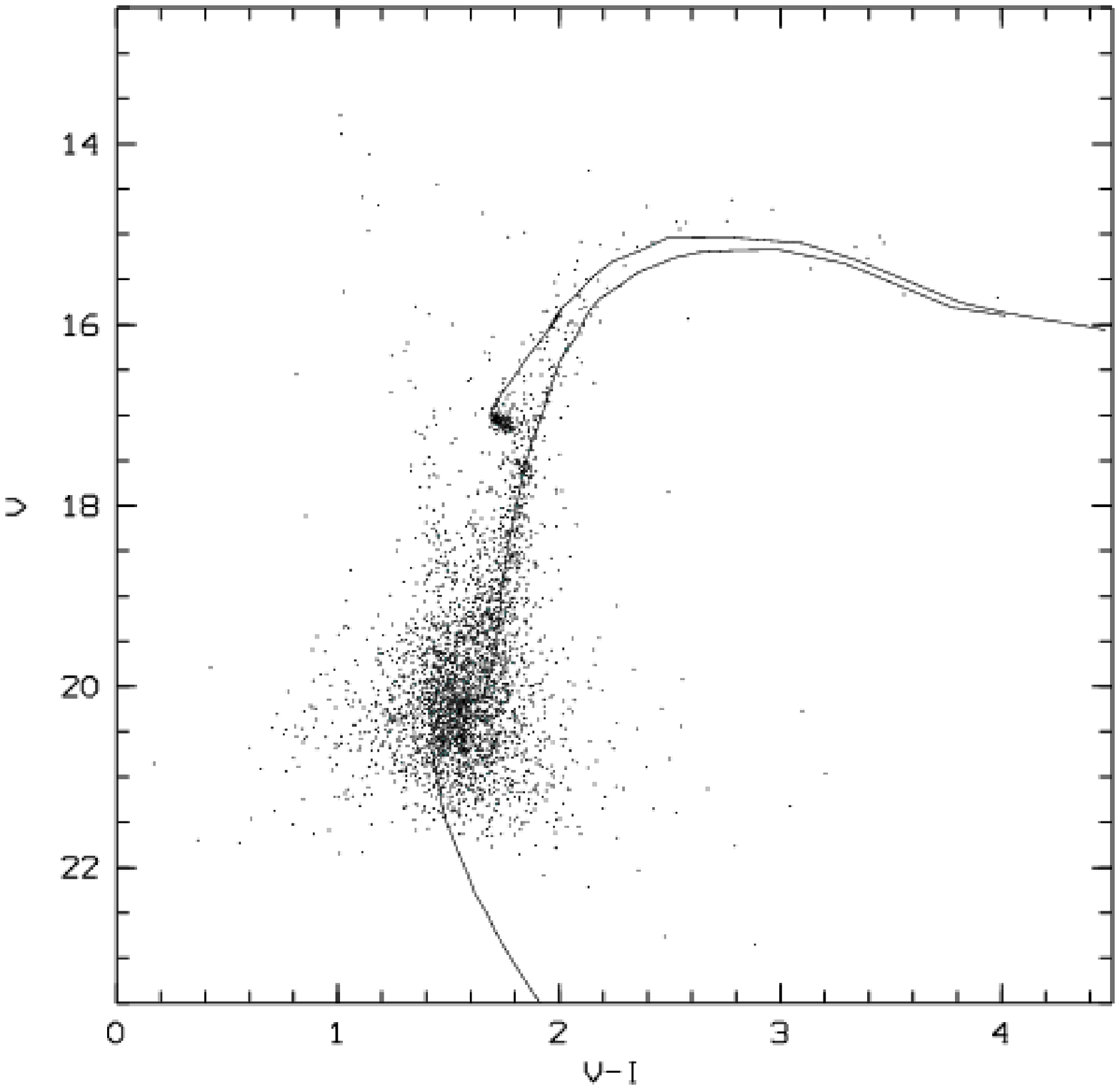,scale=0.48
                                               ,bbllx=20.5cm,bblly=18.0cm}}}
      \end{picture}
    }

    \put(9.7,0.2)
    {
      \begin{picture}(0.0,0.0)
      \put(0.0,0.0){\makebox(7.0,7.0){\epsfig{file=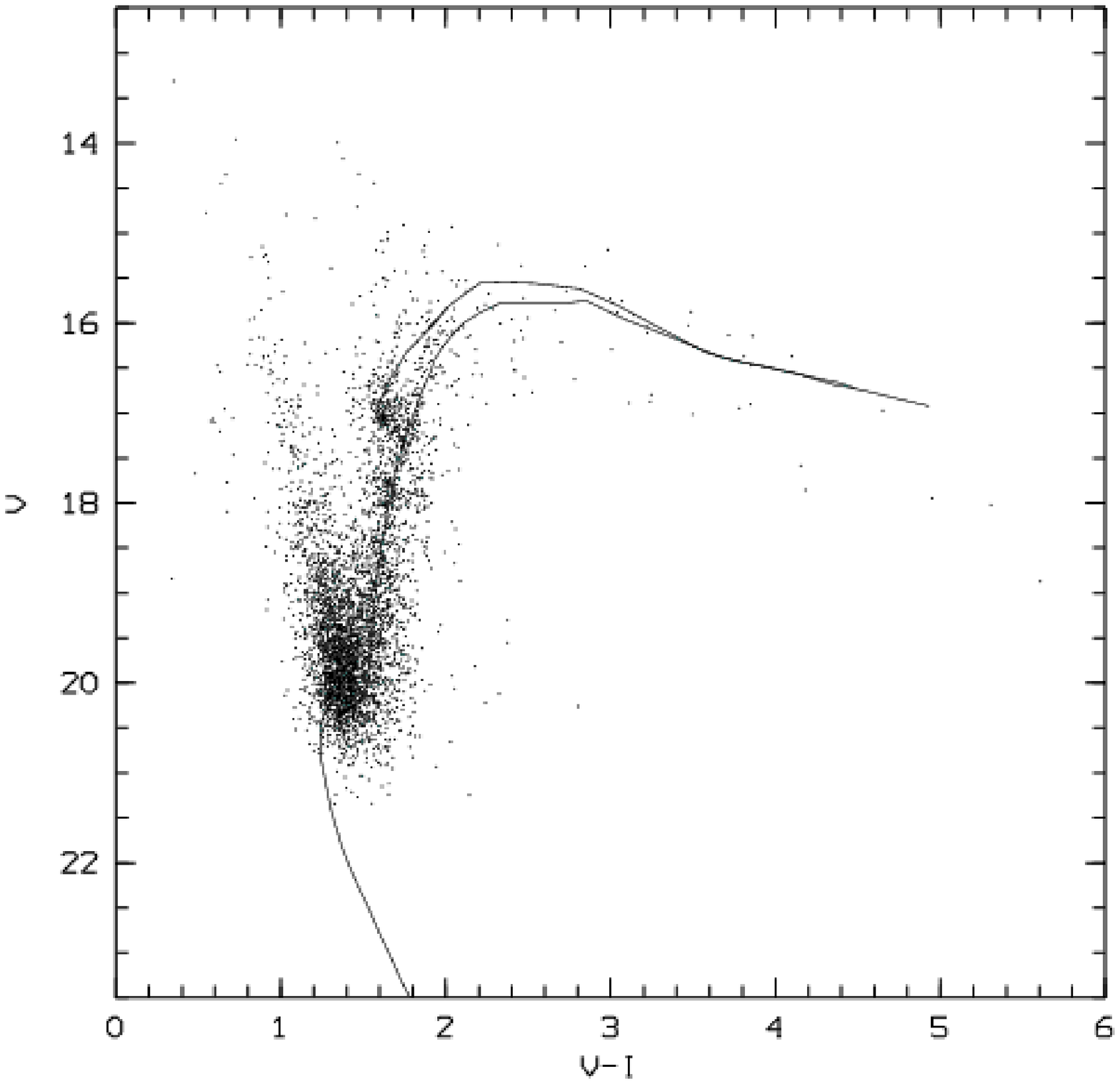,scale=0.48
                                               ,bbllx=20.5cm,bblly=18.0cm}}}
      \end{picture}
    }
  \end{picture}
  \hfill
  \parbox{8.8cm}{
  \caption{The isochrone for \object{NGC~6760} with $[\mbox{M}/\mbox{H}]=-0.40$ dex and $t=14.5$ gyr 
           is slightly too
           metal-rich, as the AGB/RGB arcs above the model. This result is corroborated by the metallicity-
           estimates given below. The CMD is radially selected for $100 \le r \le 300$ pix.
           \label{dia6760dciso}}}
  \hfill
  \parbox{8.8cm}{
  \caption{The radially selected CMD of \object{NGC~6528} clearly shows the two AGB/RGB's of the cluster's and the
           background population. The isochrone ($[\mbox{M}/\mbox{H}]=0.0$ dex, $t=14.5$ gyr) is slightly too
           metal-rich as well.  
           \label{dia6528dciso}}}
\end{figure*}

\begin{figure*}[h]
  \begin{picture}(18.0,8.8)
    \put(0.5,0.2)
    {
      \begin{picture}(0.0,0.0)
      \put(0.0,0.0){\makebox(7.0,7.0){\epsfig{file=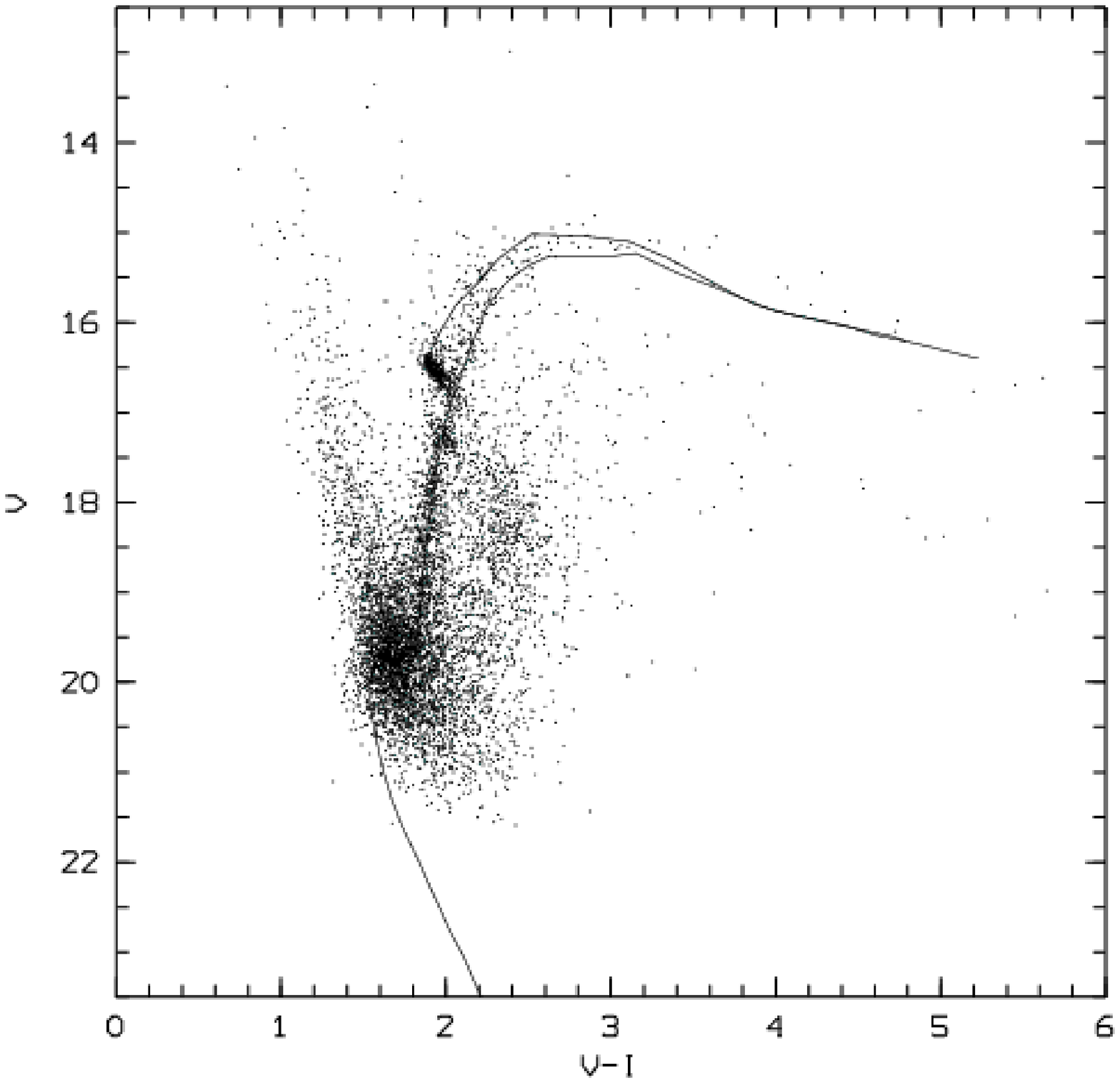,scale=0.48
                                               ,bbllx=20.5cm,bblly=18.0cm}}}
      \end{picture}
    }
  \end{picture}
  \hfill
  \parbox{8.8cm}{
  \caption{Radially selected CMD ($50 \le r \le 400$ pix) of \object{NGC~6553} with an isochrone of 
           $[\mbox{M}/\mbox{H}]=0.0$ dex and $t=14.5$ gyr. The strongly differentially reddened field
           population is now clearly separated from the cluster (to the red of the lower RGB). The CMD
           is also selected for photometrical errors $\le 0.05$ mag.
           \label{dia6553dciso}}}
\end{figure*}

For the correction of the CMDs we need the extinction
\begin{equation}
A_V=R^B_V E_{B-V}=R^I_V E_{V-I}\mbox{.}
\label{equExtLaw}
\end{equation}
However, assuming a uniform reddening law led to CMDs which in some cases showed the corrected HBs having
larger or smaller slopes than the uncorrected HBs. Moreover, as there is some uncertainty in the literature 
regarding $R^B_V$, with values varying between $R^B_V=3.1$ (Savage \& Mathis \cite{SAV79}) and $R^B_V=3.6$ 
(Grebel \& Roberts \cite{GRR95}), we determined the slope of the reddening vector via the tilted HBs of our 
CMDs. This leads to reasonable results only if the HBs are intrinsically clumpy. This assumption is 
corroborated by the fact that the well dereddened CMDs (Fig. \ref{dia5927dciso}, \ref{dia6760dciso}, 
\ref{dia6553dciso}) have clumpy HBs indeed.  
Table \ref{tabDifRed} shows the slopes $R^I_V$ for each cluster. In Fig. \ref{diaExtVar}, the slopes are 
plotted against the galactic longitude. These variations, although at the margin of the errors,
confirm earlier observations by Meyer \& Savage (\cite{MEY81}) and Turner (\cite{TUR94}). Meyer \& Savage 
determined via two-color-diagrams the deviation of single stars in the extinction behaviour from the galactic 
mean extinction law. Turner demonstrated the inapplicability of a mean galactic reddening law for objects
lying close to the galactic plane. 
\begin{table}[h]
\footnotesize
\begin{center}
\begin{tabular}{lcc}
\hline     
NGC   & $\Delta E_{V-I}^{max}$ & $R_V^I$      \\ \hline
5927  & $0.27$                 & $1.9\pm 0.2$ \\
6316  & $0.07$                 & $2.1\pm 0.2$ \\
6342  & $0.32$                 & $2.2\pm 0.1$ \\
6441  & $0.20$                 & $2.3\pm 0.1$ \\
6760  & $0.25$                 & $2.0\pm 0.1$ \\
6528  & $0.31$                 & $2.4\pm 0.3$ \\
6553  & $0.29$                 & $2.3\pm 0.2$ \\
\hline
\end{tabular}
\end{center}
\normalsize
\caption{Maximum differential reddening and slope of the reddening vector.
         \label{tabDifRed}}
\end{table}    
To correct the diagrams for differential reddening, we referred all sub-CMDs of one cluster to the one
with detected minimal reddening and we shifted all other sub-CMDs onto that. As we thus use the minimal 
absolute reddening as a point of reference, the absolute reddening determined later on will be smaller than 
value given in the literature.
The differentially dereddened CMDs are shown in Figs. \ref{dia5927dc} through \ref{dia6553dc}. As the
correction led to a clearly improved appearance for all clusters, the corrected versions of the CMDs 
will be used for further investigation.

\begin{figure*}[h]
  \begin{picture}(18.0,8.8)
    \put(0.5,0.2)
    {
      \begin{picture}(0.0,0.0)
      \put(0.0,0.0){\makebox(7.0,7.0){\epsfig{file=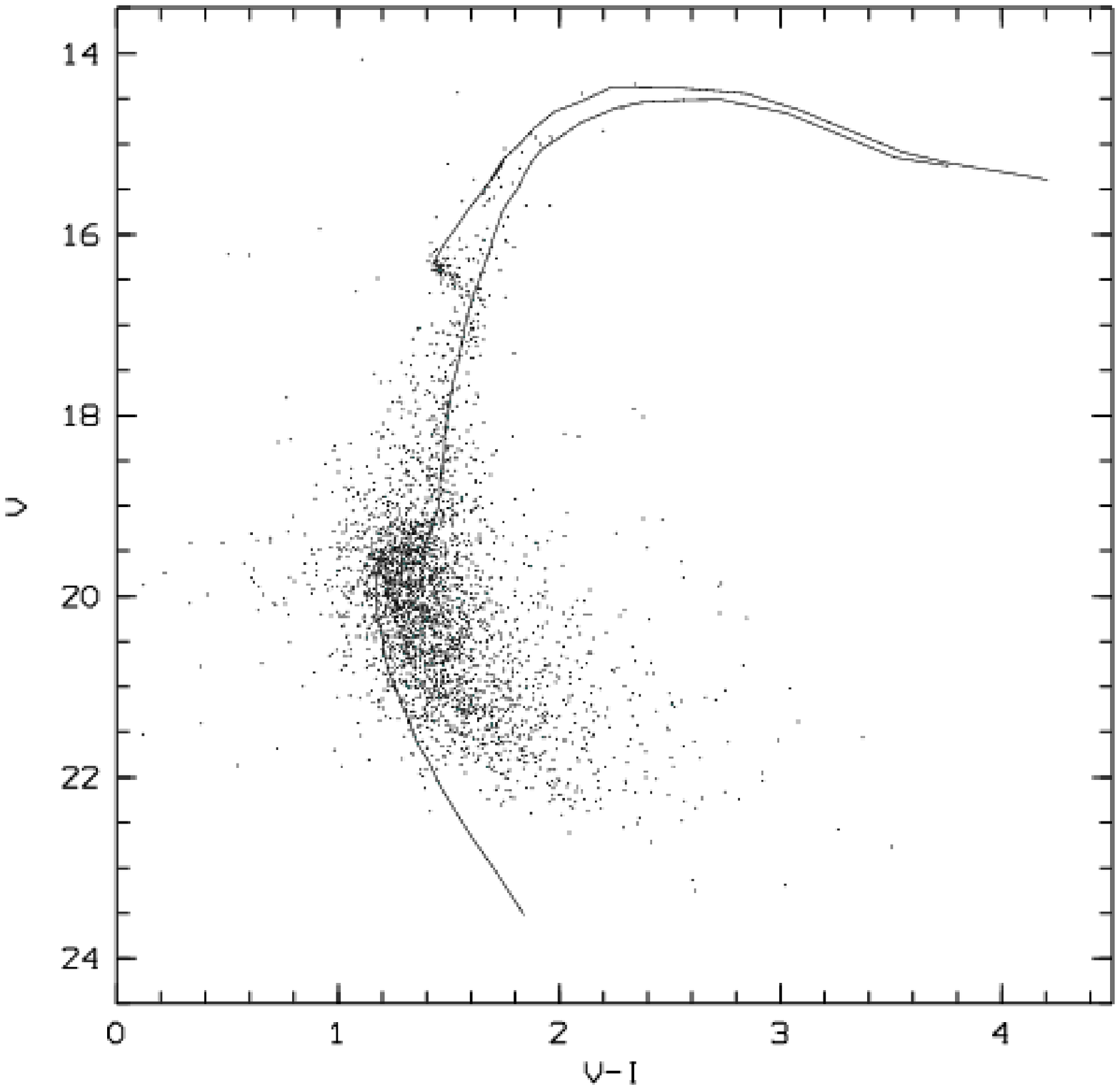,scale=0.48
                                               ,bbllx=20.5cm,bblly=18.0cm}}}
      \end{picture}
    }

    \put(9.7,0.2)
    {
      \begin{picture}(0.0,0.0)
      \put(0.0,0.0){\makebox(7.0,7.0){\epsfig{file=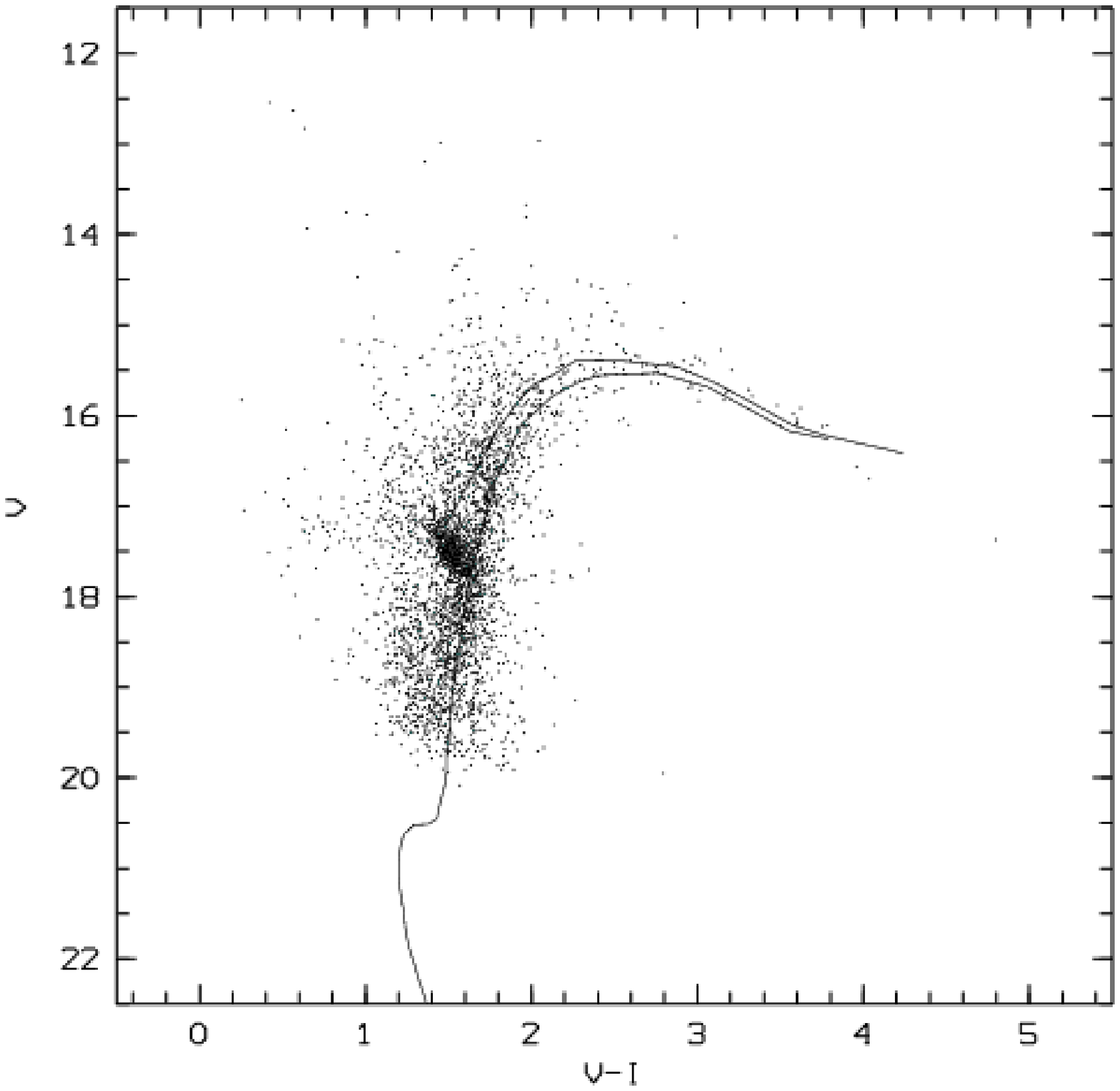,scale=0.48
                                               ,bbllx=20.5cm,bblly=18.0cm}}}
      \end{picture}
    }
  \end{picture}
  \hfill
  \parbox{8.8cm}{
  \caption{Isochrone for \object{NGC~6342} ($[\mbox{M}/\mbox{H}]=-0.40$ dex, $t=14.5$ gyr). The CMD is selected
           for radii $\le 200$ pix.
           \label{dia6342dciso}}}
  \hfill
  \parbox{8.8cm}{
  \caption{CMD of \object{NGC~6441} with isochrone $[\mbox{M}/\mbox{H}]=-0.40$ dex, $t=14.5$ gyr. The CMD is 
           radially
           selected $(50 \le r \le 400)$ pix. The stars to the blue of the HB still are visible.
           \label{dia6441dciso}}}
\end{figure*}
\clearpage
\section{The Globular Cluster Parameters}
This section deals with the determination of the metallicity, reddening and distance 
using the {\it differentially dereddened} CMDs. There are two possible ways to achieve the goal. In the 
first, theoretical models are compared with the CMDs, in the second, empirical relations between parameters 
and loci in the CMDs are used. 

\subsection{Isochrone fitting}
To derive metallicity, distance and absolute reddening via isochrone fitting, we used the Padova-tracks
(Bertelli et al. \cite{BER94}) with a fixed age of $14.5$ Gyr ($\log(age)=10.160$).
Isochrones with different ages ($10.120 \le \log(age) \le 10.200$) led to identical results.
To avoid systematic errors, we used the middle of the broadened structures to fit the isochrones by eye.
These loci are easily determined for the ascending part of the RGB, as it runs more or less
perpendicular to the reddening vector. Regarding the upper part of the RGB, we take into account that
we cannot distinguish between the AGB and the RGB in our diagrams. Hence, the densest regions of the AGB/RGB
lie between the model's tracks. We additionally used the HB and the lower part of the RGB, as far as 
they were accessible. The parameters resulting from the isochrone fit are given in Table \ref{tabIso}.
\begin{table}[h]
  \footnotesize
  \begin{center}
  \begin{tabular}{lccc}
  \hline
  NGC   & $(m-M)_V$       & $E_{V-I}$       & $[\mbox{M}/\mbox{H}]$\\ \hline
  5927  & $15.45\pm 0.03$ & $0.43\pm 0.02$  & $-0.40$		   \\  
  6316  & $16.76\pm 0.04$ & $0.62\pm 0.03$  & $-0.70$		   \\
  6342  & $15.36\pm 0.04$ & $0.46\pm 0.03$  & $-0.40$		   \\
  6441  & $16.48\pm 0.05$ & $0.49\pm 0.03$  & $-0.40$		   \\
  6760  & $16.18\pm 0.04$ & $0.72\pm 0.03$  & $-0.40$		   \\
  6528  & $15.94\pm 0.05$ & $0.46\pm 0.03$  & $0.00$		   \\
  6553  & $15.42\pm 0.04$ & $0.76\pm 0.03$  & $0.00$		   \\ 
  \hline
  \end{tabular}
  \end{center}
  \normalsize
  \caption{Distance modulus $(m-M)_V$, total reddening $E_{V-I}$ and metallicity 
           $[\mbox{M}/\mbox{H}]$ of all the sample's clusters derived by isochrone fitting.
           The errors are eye-estimates of how accurately we could place the isochrones. 
           Note that the isochrones are fitted to the differentially dereddened CMDs.\label{tabIso}}
\end{table}

Figures \ref{dia5927dciso} to \ref{dia6553dciso} show the differentially dereddened CMDs with the fitted
isochrones. For a discussion and comparison of these parameters with the literature, see paragraph 
\ref{ssecComp}.

\subsection{Metallicity and reddening: relations }
\subsubsection{Metallicity}

The luminosity difference between HB and the turn over of the AGB/RGB in $(V,V-I)$-CMDs is very sensitive to
metallicity in the metal-rich domain (e.g. Ortolani et al. \cite{ORT97}). Moreover, it is a differential 
metallicity indicator, thus it is independent of absolute colour or luminosity, in contrast to the 
$[\mbox{M}/\mbox{H}]-(V-I)_{0,g}$-method (see e.g. Sarajedini \cite{SAR94}). We present a
preliminary linear calibration of this method, 
\begin{equation}
[\mbox{M}/\mbox{H}]=a(V_{HB}-V_{RGB}^{max}) + b,
\label{equFeHdV}
\end{equation}
as there has not been any so far. Because there still are only very few $(V,V-I)$-CMDs which clearly show 
both the HB and turn over of the AGB/RGB and which have reliable metallicity determinations, we used the 
Padova-isochrones and a CMD of NGC 6791 (Garnavich et al. \cite{GAR94}) to set up a calibration. NGC 6791 
is one of the richest old open clusters with a good metallicity determination and it is therefore suitable 
to serve as a zero-point check.

\begin{figure}
  \resizebox{\hsize}{!}{\includegraphics{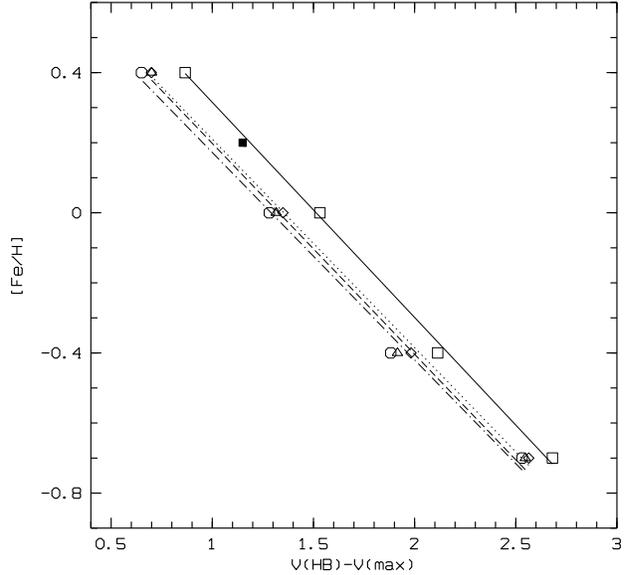}}
  \caption{$[\mbox{M}/\mbox{H}]-\Delta V$-relation for different ages. The open symbols
           stand for the isochrone-values. Squares, lines: $9.8$ Gyr, diamonds, dotted: $13.2$ Gyr,
           triangles, dashed: $14.5$ Gyr, circles, dash-dotted: $15.8$ Gyr.                
           The filled square represents the value for NGC 6791 (Garnavich et al. \cite{GAR94}). 
           \label{diaCalibdV}}
\end{figure}

As the form of the RGB depends slightly on age as well (e.g. Stetson et al. \cite{STE96}), we have to check 
this dependence before applying our calibration. Fig. \ref{diaCalibdV} shows the linear relation between 
$[\mbox{M}/\mbox{H}]$ and $\Delta V \equiv V_{HB} - V_{max}$ for four GC-ages. Table \ref{tabCalibdV} 
contains the respective coefficients. As the metal-poorest isochrones of the Padova-sample 
($[\mbox{M}/\mbox{H}] = -1.70, 1.30$ dex) do not show a maximum of the AGB/RGB, they have not been used.  
Fig. \ref{diaCalibdV} makes clear that the age has only a minor influence on the resulting metallicity. 
To be consistent with the isochrone-fit, we used the relation for $\log(age)=10.160$.

To estimate the metallicities of our clusters, we now only have to measure the relevant luminosities.
The results are given in Table \ref{tabFeHdV}.
The value for \object{NGC~5927} given in column $[M/H]_2$ relates to a single star (Fran\c{c}ois \cite{FRA91});
\object{NGC~6528} and \object{NGC~6553} are from Richtler et al. (\cite{RTL98}) and Sagar et al. (\cite{SAG98}). 
\begin{table}
  \begin{center}
  \begin{tabular}{lcc}
  \hline
  $\log(age)$ & a [dex/mag]          & b [dex]         \\ \hline
  $9.990$     & $-0.614 \pm 0.020$ & $0.929 \pm 0.026$ \\
  $10.120$    & $-0.595 \pm 0.017$ & $0.805 \pm 0.024$ \\
  $10.160$    & $-0.602 \pm 0.032$ & $0.800 \pm 0.043$ \\
  $10.200$    & $-0.592 \pm 0.033$ & $0.764 \pm 0.045$ \\ \hline
  \end{tabular}
  \end{center}
  \caption{Calibration coefficients of the $[M/H]-\Delta V$-relation
           for different cluster ages. The calibration equation is
           $[M/H]$ $=a\Delta V + b$. \label{tabCalibdV}}
\end{table}
\begin{table*}
\begin{center}
\begin{tabular}{lcccccc}
\hline
NGC &$V_{HB}$       &$V_{RGB}^{max}$&$\Delta V$&$[M/H]$[dex]  &$[M/H]_1$&$[M/H]_2$\\\hline
5927&$16.30\pm 0.03$&$14.49\pm 0.04$&$1.81$    &$-0.29\pm 0.08$&$-0.37$   &$-1.08$ \\
6316&$17.42\pm 0.05$&$15.09\pm 0.04$&$2.33$    &$-0.60\pm 0.09$&$-0.55$   & \\
6342&$16.47\pm 0.05$&$14.44\pm 0.06$&$2.03$    &$-0.42\pm 0.09$&$-0.65$   & \\
6441&$17.51\pm 0.07$&$15.53\pm 0.07$&$1.98$    &$-0.39\pm 0.10$&$-0.53$   & \\
6760&$17.08\pm 0.02$&$14.81\pm 0.04$&$2.27$    &$-0.57\pm 0.09$&$-0.52$   & \\
6528&$17.07\pm 0.06$&$15.74\pm 0.08$&$1.33$    &$ 0.00\pm 0.09$&$-0.17$   &$-0.15$ \\
6553&$16.52\pm 0.07$&$15.13\pm 0.06$&$1.39$    &$-0.04\pm 0.08$&$-0.25$   &$-0.10$ \\
\hline
\end{tabular}
\end{center}
\caption{Metallicities of all GCs via the differentially dereddened CMDs. Column $[M/H]$ contains the
         values derived by the $[M/H]-\Delta V$-relation. The values of
         column $[\mbox{M}/\mbox{H}]_1$ have been taken from Harris (\cite{HAR96}). Column $[M/H]_2$ 
         gives additional values as discussed in the text. The errors only take account of the
         uncertainties of the luminosities and the calibration errors of Table \ref{tabCalibdV}.
         \label{tabFeHdV}}
\end{table*}

\subsubsection{Reddening}
It should be remembered, that we used the differentially dereddened CMDs to determine the parameters. Thus,
the given reddenings are minimal ones.

As mentioned above, the absolute colour of the RGB at the level of the HB can be used to estimate the 
metallicity. Conversely (Armandroff \cite{ARM88}), if we know the metallicity, we can determine the absolute
colour $(V-I)_{0,g}$ and thus the absolute reddening of the cluster. 

\begin{figure}
  \resizebox{\hsize}{!}{\includegraphics{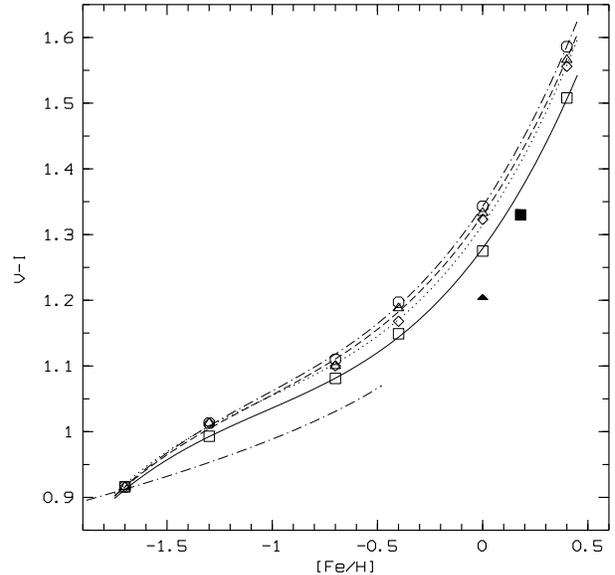}}
  \caption{Calibration of the non-linear $(V-I)_{0,g}-[\mbox{M}/\mbox{H}]$-relation
           for different ages. The key for the symbols is the same as used in Fig. \ref{diaCalibdV},
           except for the filled triangle denoting the $(V-I)_{0,g}-[\mbox{M}/\mbox{H}]$ pair of M67.
           In addition, the quadratic relation of Caretta \& Bragaglia (\cite{CAA98}) is plotted in
           dash-dotted line. It intersects our relation at low $[Fe/H]$ but shows a difference of 
           $\Delta (V-I)=0.4$ mag in the more metal-rich regime.
           \label{diaVIcalib}}
\end{figure}

These relations between colour $(V-I)_{0,g}$ and metallicity are well calibrated for the metal-poor to
intermediate regime. However, it is difficult to set up a calibration for the metal-rich regime of our clusters.
Linear calibrations have been provided by e.g. Sarajedini (\cite{SAR94}). A more recent calibration by 
Caretta \& Bragaglia (\cite{CAA98}) uses a 2nd order polynomial.
To set up a calibration for the metal-rich regime we used again the Padova-tracks together with NGC 6791 
to derive the coefficients for a relation of the form
\begin{equation}
(V-I)_{0,g}=a + b \cdot [\mbox{M}/\mbox{H}] + c \cdot [\mbox{M}/\mbox{H}]^2 + d \cdot [\mbox{M}/\mbox{H}]^3
\label{equVIFeHCalib}
\end{equation}
In addition, we used the $[\mbox{M}/\mbox{H}]$ and $(V-I)_{0,g}$ values for M67 given by Montgomery et al. 
(\cite{MON93}) to check the zero point. Taking into accound that M67 is even younger than NGC 6791 by 3 to 
5 Gyrs, the measured quantities fit reasonably well.
Table \ref{tabVIcalib} contains the calibration coefficients, Fig. \ref{diaVIcalib} the graphic relations,
again for different ages. As above, we used the relation for $\log(age)=10.160$. 
\begin{table*}[hb]
  \footnotesize
  \begin{center}
  \begin{tabular}{lcccc}
  \hline
  $\log(age)$ & a                 & b                  & c                 & d         \\ \hline
  $9.990$     &$1.279\pm 0.002$ &$0.438 \pm 0.004$ &$0.287\pm 0.010$ &$0.092\pm 0.005$ \\
  $10.120$    &$1.315\pm 0.003$ &$0.468 \pm 0.007$ &$0.307\pm 0.016$ &$0.099\pm 0.008$ \\
  $10.160$    &$1.330\pm 0.004$ &$0.467 \pm 0.008$ &$0.281\pm 0.021$ &$0.088\pm 0.011$ \\
  $10.200$    &$1.343\pm 0.006$ &$0.479 \pm 0.006$ &$0.289\pm 0.014$ &$0.091\pm 0.008$ \\ \hline
  \end{tabular}
  \end{center}
  \normalsize
  \caption{Calibration coefficients for the $(V-I)_{0,g}-[\mbox{M}/\mbox{H}]$-relation (equation 
           \ref{equVIFeHCalib}).\label{tabVIcalib}}
\end{table*}

Using the metallicities listed in Table \ref{tabFeHdV}, column $[\mbox{M}/\mbox{H}]$, we get the absolute 
reddening as given in Table \ref{tabReddening}. Metallicities as well as reddenings fit very well with the 
values derived via isochrone-fitting, but are significantly lower than the values given in the literature. 
This is partly explained by the fact that we take the minimal reddening from the reddening map. 
Another part of the explanation may be that previous isochrone fits tend to use the red ridge of the RGB 
and thus overestimate the reddening. 
\begin{table*}[hb]
\begin{center}
\begin{tabular}{lcccccc}
\hline
NGC  &$E_{V-I}^{iso}$ &$(V-I)_{HB}$   &$(V-I)_{0,g} $ &$E_{V-I}^{rel}$ &$E_{V-I}^{lit}$ &literature \\\hline
5927 &$0.43\pm 0.02$  &$1.63\pm 0.02$ &$1.22\pm 0.04$ &$0.41 \pm 0.05$ & $0.66$     &Sarajedini \& Norris 
(\cite{SAN94})\\
6316 &$0.62\pm 0.03$  &$1.76\pm 0.03$ &$1.13\pm 0.06$ &$0.63 \pm 0.06$ & $0.61$     &Davidge et al.
(\cite{DAV92})\\
6342 &$0.46\pm 0.03$  &$1.65\pm 0.04$ &$1.18\pm 0.05$ &$0.47 \pm 0.06$ & $0.65$     &Armandroff \& Zinn
(\cite{ARZ88})\\
6441 &$0.49\pm 0.03$  &$1.64\pm 0.04$ &$1.19\pm 0.05$ &$0.46 \pm 0.06$ & $0.64$     &Deutsch et al.
(\cite{DEU96})\\
6760 &$0.72\pm 0.03$  &$1.88\pm 0.02$ &$1.14\pm 0.05$ &$0.74 \pm 0.06$ & $1.07$     &Armandroff \& Zinn 
(\cite{ARZ88})\\
6528 &$0.46\pm 0.03$  &$1.79\pm 0.04$ &$1.33\pm 0.06$ &$0.46 \pm 0.06$ & $0.70$     &Richtler et al.
(\cite{RTL98})\\
6553 &$0.76\pm 0.03$  &$2.08\pm 0.04$ &$1.31\pm 0.06$ &$0.77 \pm 0.06$ & $0.95$     &Sagar et al.
(\cite{SAG98})\\
\hline
\end{tabular}
\end{center}
\caption{Absolute reddening for all GCs. Column $E_{V-I}^{iso}$ contains values derived via 
         isochrone-fitting, column $E_{V-I}^{rel}$ values via $(V-I)_{0,g}-[\mbox{M}/\mbox{H}]$-relation.
         $(V-I)_{g}$ gives the colour of the RGB at the level of $V_{HB}$, $(V-I)_{0,g}$ 
         the corresponding dereddened colour, calculated via the 
         $(V-I)_{0,g}-[\mbox{M}/\mbox{H}]$-relation. 
         In the last column, we cited values form literature and their sources, which, of course,
         cannot be more than a selection.
         Any measurement of colours was done in the differentially dereddened CMDs, which provides the
         explanation for the difference between our values and that taken from other works.
         \label{tabReddening}}
\end{table*}

\begin{figure}
  \resizebox{\hsize}{!}{\includegraphics{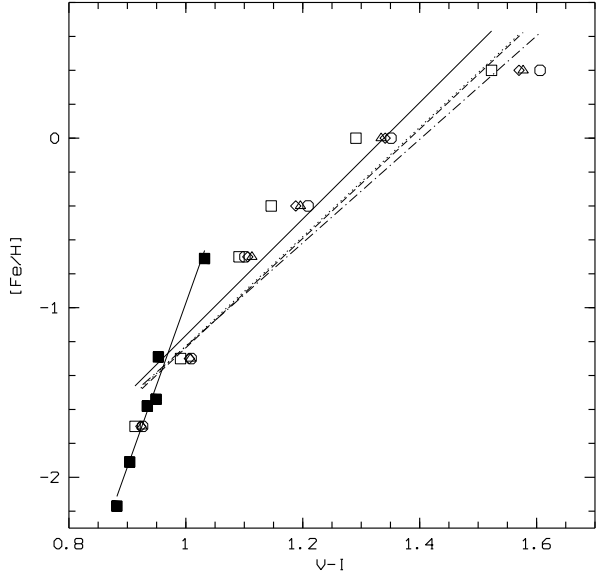}}
  \caption{Calibration of the linear $[\mbox{M}/\mbox{H}]-(V-I)_{0,g}$-relation according to Sarajedini 
           (filled symbols) in comparison to the recalibration for higher metallicities. The key for the 
           symbols is the same as used in Fig. \ref{diaCalibdV}, except for the filled symbols.
           \label{diaSarVI}}
\end{figure}

\begin{figure}
  \resizebox{\hsize}{!}{\includegraphics{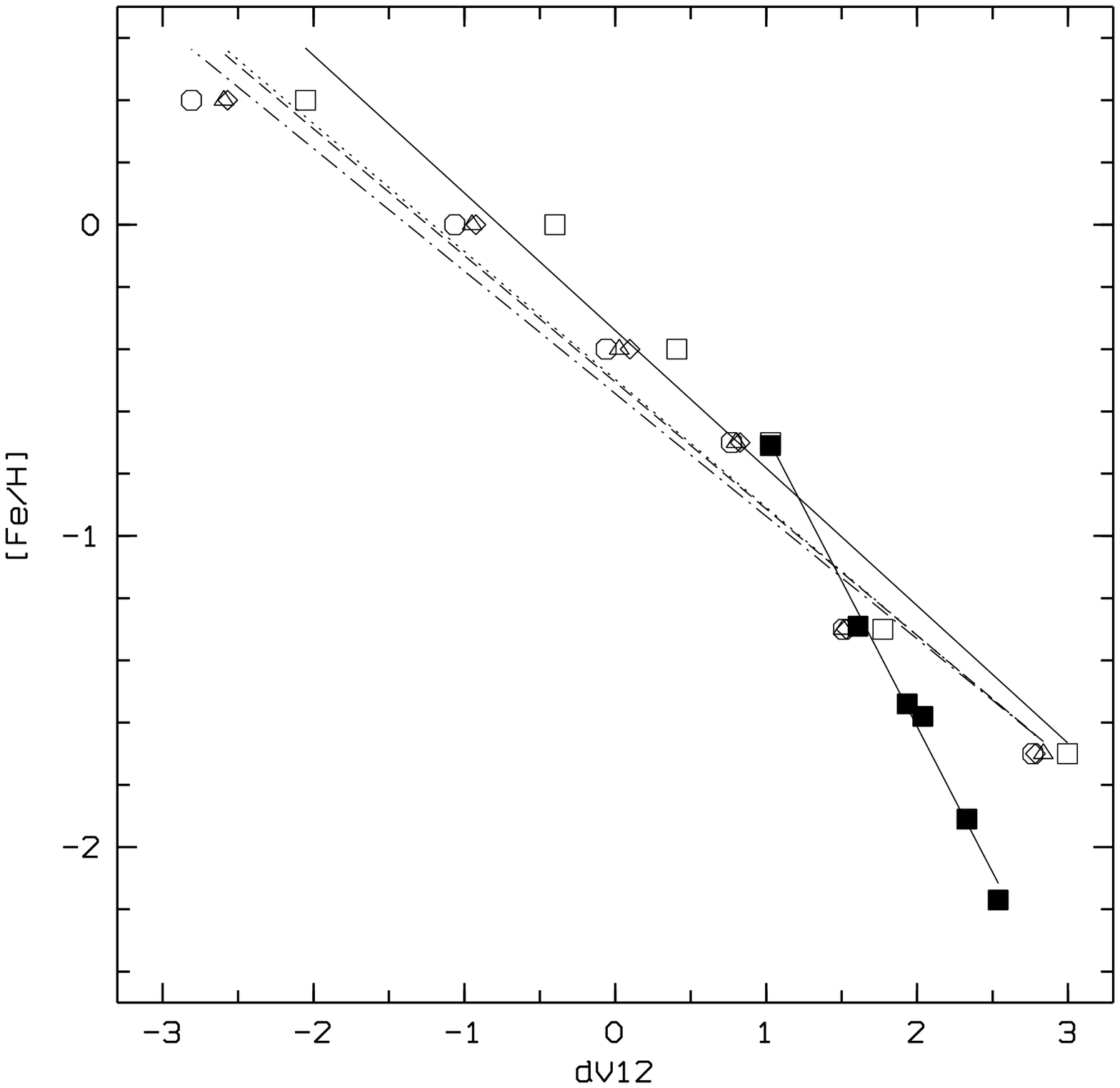}}
  \caption{Calibration of the linear $[\mbox{M}/\mbox{H}]-\Delta V_{1.2}$-relation according to Sarajedini 
           (filled symbols) in comparison to the recalibration for higher metallicities. The key for the
           symbols is the same as used in Fig. \ref{diaCalibdV}, except for the filled symbols. 
           \label{diaSar12}}
\end{figure}

Sarajedini (\cite{SAR94}) proposed a method to simultaneously determine metallicity and reddening. 
For this, he used the (linear) $[\mbox{M}/\mbox{H}]-(V-I)_{0,g}$-relation and the dependence of 
metallicity on the luminosity of the RGB at the absolute colour of $V-I=1.2$ mag, in linear form as well.
He calibrated both relations for a metallicity range of $(-2.2 \le [\mbox{M}/\mbox{H}] \le -0.70)$ dex. 
We recalibrated these relations in order to use them for our clusters. Using NGC 6791 and the 
Padova-tracks.  The graphic results are shown in Figs. \ref{diaSarVI} and \ref{diaSar12}; the calibration 
coefficients are given in Table \ref{tabSarCalib}. 
\begin{table*}\
  \begin{center}
  \begin{tabular}{lcccc}
  \hline
                    & $a_{V-I}$ & $b_{V-I}$   & $a_{1.2}$   & $b_{1.2}$ \\
  \hline
  Padova  7.9 Gyr & $3.432$     & $-4.595$    & $-0.442$    & $-0.340$  \\
  Padova 13.2 Gyr & $3.225$     & $-4.451$    & $-0.441$    & $-0.497$  \\
  Padova 14.5 Gyr & $3.222$     & $-4.458$    & $-0.407$    & $-0.506$  \\
  Padova 15.8 Gyr & $3.052$     & $-4.280$    & $-0.394$    & $-0.543$  \\
  Sarajedini        & $9.668$   & $10.64$     & $-0.9367$   & $0.2606$  \\
  \hline
  \end{tabular}
  \end{center}
  \caption{Coefficients for Sarajedini- and Padova-relations. The equations have the form
           $[\mbox{M}/\mbox{H}]=a_{V-I}(V-I)_{0,g} + b_{V-I}$ and 
           $[\mbox{M}/\mbox{H}]=a_{1.2}\Delta V_{1.2} + b_{1.2}$.
           The errors are $\Delta a_{V-I}=0.43$, $\Delta b_{V-I}=0.23$, 
           $\Delta a_{1.2}=0.04$ und $\Delta b_{1.2}=0.16$. \label{tabSarCalib}}
\end{table*}
For a discussion and new calibration of Sarajedini's method see Caretta \& Bragaglia (\cite{CAA98}).
We did not make use of this method, as the extrapolation of Sarajedini's calibration did not seem
to be advisable, with reference to Figs. \ref{diaSarVI} and \ref{diaSar12}.

\subsubsection{Distance}
The brightness $M_V^{HB}$ of the horizontal branch is the best distance indicator for GCs. However, 
there is a lively discussion on how this brightness depends on the metallicity of the cluster.
 
We take the LMC distance as the fundamental distance for calibrating
the zero point in the relation between metallicity and horizontal
branch/RR\,Lyrae brightness.  The third fundamental distance
determination beside trigonometric parallaxes and stellar stream
parallaxes is the method of Baade-Wesselink parallaxes. It had been
applied to the LMC in its modified form known as Barnes-Evans
parallaxes. So far, it has been applied to Cepheids in
\object{NGC~1866} (Gieren et~al. \cite{GIE94}), and the most
accurate LMC distance until now stems from the period-luminosity
relation of LMC Cepheids by Gieren et al. (\cite{GIE98}). We adopt the
distance modulus from the latter work, which is $18.46\pm0.06$ mag, and
which is in very good agreement with most other work (e.g. Tanvir
\cite{TAN96}).

If we adopt the apparent magnitude of RR\,Lyrae stars in the LMC from
Walker (\cite{WAL92}), $18.94\pm0.1$ mag for a metallicity of
$[\element{Fe}/\element{H}]=-1.9$ dex, and the metallicity dependence from
Caretta et~al.  (\cite{CAB98}), one gets
\begin{equation}
  M_V(RR) = (0.18\pm0.09)([\element{Fe}/\element{H}]+1.6)+0.53\pm0.12
  \label{equCaretta}
\end{equation}
This zero-point is in excellent agreement with the one derived from
HB-brightnesses of old LMC globular clusters, if the above metallicity
dependence is used (Olszewski et al. \cite{OLS91}).

With relation \ref{equCaretta}, with the reddenings (as shown in Table \ref{tabReddening}, 
column $E_{V-I}^{rel}$) and with the extinction $A_V=R_V^I E_{V-I}$ we can calculate the distance moduli 
\begin{equation}
(m-M)_0=V_{HB}-A_V-M_V^{HB}
\label{equDistModHB}
\end{equation}
The values for $M_V^{HB}$ and $[\mbox{M}/\mbox{H}]$ are listed in Table \ref{tabFeHdV}, and the results are
given in Table 
\ref{tabDistRel}. $R_V^I$ comes from Table \ref{tabDifRed}. 
\begin{table*}[hb]
  \begin{center}
  \begin{tabular}{lccccc}
  \hline
  NGC   &$A_V$          &$M_V^{HB}$      &$(m-M)_0$       &$r$           &$r_{Harris}$ \\ \hline
  5927  &$0.79\pm 0.12$ &$0.77\pm 0.12$  &$14.75\pm 0.17$ &$8.9\pm 0.7$  &$7.4$     \\ 
  6316  &$1.32\pm 0.18$ &$0.71\pm 0.13$  &$15.39\pm 0.23$ &$12.0\pm 1.2$ &$11.5$    \\
  6342  &$1.04\pm 0.15$ &$0.74\pm 0.13$  &$14.69\pm 0.20$ &$8.7\pm 0.8$  &$9.1$     \\
  6441  &$1.05\pm 0.16$ &$0.75\pm 0.13$  &$15.70\pm 0.21$ &$13.8\pm 1.3$ &$9.7$     \\
  6760  &$1.48\pm 0.13$ &$0.72\pm 0.13$  &$14.88\pm 0.19$ &$9.5\pm 0.8$  &$7.3$     \\
  6528  &$1.11\pm 0.16$ &$0.82\pm 0.12$  &$15.15\pm 0.24$ &$10.7\pm 1.1$ &$7.4$     \\
  6553  &$1.76\pm 0.13$ &$0.81\pm 0.12$  &$13.95\pm 0.24$ &$6.2\pm 0.7$  &$4.7$     \\
  \hline
  \end{tabular}
  \end{center}
  \caption{Distances of GCs via $M_V^{HB}-[\mbox{M}/\mbox{H}]$-relation. The given errors only account for
           the errors in $M_V^{HB}$. The values of column $r_{Harris}$ are taken from Harris (\cite{HAR96}).
           Distances in kpc, brightness in mag.\label{tabDistRel}}
\end{table*}

\subsection{Comparison\label{ssecComp}}
The distances determined via the $M_{HB}-[\mbox{M}/\mbox{H}]$-relation are larger than those determined by 
the isochrone-fitting (Table \ref{tabDistComp}). However, as the related reddenings do not show any 
significant differences, this effect is attributed to the $M_{HB}-[\mbox{M}/\mbox{H}]$-relation and the 
isochrone-fitting itself. 
As described above, the isochrone fitting is lacking the desired accuracy especially because the 
TOP cannot be resolved for most of the clusters. Moreover, the fact that AGB and RGB cannot be 
distinguished in our CMDs leads to a systematic error in the isochrone distances in the sense that
the isochrones tend to have been fitted with brightnesses which are too large. 
In the following, we discuss some possible explanations for differences between distances taken from the
literature and this work. It should be remembered that the distance errors amount to about 10\%.

\begin{enumerate}
  \item The distance to \object{NGC~6528} increases by nearly 30\% compared to Richtler et al. (\cite{RTL98}).
        Taking into account, 
        that the isochrone (Fig.\ref{dia6528dciso}) might have been fitted slightly too low, we still get a
        distance of about $8.1$ kpc. Moreover, Richtler et al. determine the absolute reddening
        via the differentially reddened CMD, which leads to larger values ($0.6 \le E_{V-I} \le 0.8$)
        compared to our $E_{V-I}=0.46$. Thus the distance modulus decreases by about $0.4$ mag, as
        equation \ref{equDistModHB} is corrected more strongly on reddening. Finally, the different
        slopes of the reddening vector have to be regarded. Richtler et al. assume $A_V/E_{V-I}=2.6$,
        our slope, which we determined via the slope of the HB, amounts to $A_V/E_{V-I}=2.4$.
        On the whole, we get a difference between Richtler et al. and this work of $0.7$ mag in the 
        distance modulus.
  \item In the CMDs of \object{NGC~5927} and 6760 the differential reddening becomes noticable especially along
        the steep part of the RGB, as this is running nearly perpendicularly to the reddening vector. 
        Around the turn over of the AGB/RGB and for its redder part, it leads to an elongation, but
        not to a broadening of the structures. Fitting an isochrone to the broadened RGB, one generally
        would use the middle of the RGB as an orientation, as one cannot distinguish between reddening
        effects and photometric errors in the outer regions. However, the red part of the AGB/RGB
        approximately keeps its unextinguished brightness. Thus the differential 
        reddening might be overestimated, which leads to decreasing distances. A similar point
        can be made for the determination of $E_{V-I}$ via the $(V-I)_{0,g}-[\mbox{M}/\mbox{H}]$-relation.
        Measuring the colour $(V-I)_g$ in the differentially reddened diagram is best done at the
        middle of the broadened RGB again. This leads to an increased reddening, i.e. the distance modulus
        will be corrected too strongly for extinction. Overestimating the colour by $0.1$ mag leads to
        a decrease in distance of about 10\%.
  \item For \object{NGC~6441}, Harris (\cite{HAR96}) cites a value of $V_{HB}=17.10$ mag. From our CMD we get
        $17.66$ mag. This lower brightness is supported by (V,B-V)-CMDs of Rich et al. (\cite{RIC97}),
        especially as Harris' value comes from a CMD by Hesser \& Hartwick (\cite{HES76}), whose lower
        limiting brightness is around $17.3$ mag.
  \item The distances as determined via the $[\mbox{M}/\mbox{H}]-\Delta V$- and the 
        $(V-I)_{0,g}-[\mbox{M}/\mbox{H}]$-relation relate
        to the differentially dereddened CMDs, i.e to the minimal absolute reddening. However, the papers
        we obtained the cited values from (Table \ref{tabReddening}), do not take differential reddening into
        account (e.g. Armandroff (\cite{ARM88}) for \object{NGC~6342} and 6760, Ortolani et al. (\cite{ORT90},
        \cite{ORT92}) for \object{NGC~6528} and 6553). So their absolute reddenings are systematically larger 
        and the distances smaller. Interestingly, the absolute reddening for \object{NGC~6316} of 
        $E_{V-I}=0.63$ mag as determined in this work fits very well the value of $E_{V-I}=0.61$ mag given 
        by Davidge et al. (\cite{DAV92}); \object{NGC~6316} shows the smallest differential reddening 
        ($\delta E_{V-I}=0.07$ mag) of our cluster sample. 
  \item Finally, the distances depend on the assumed extinction law. The value varies between
        $3.1 \le R_V^B \le 3.6$ (Savage \& Mathis \cite{SAV79}, Grebel \& Roberts \cite{GRE95}, see 
        Fig \ref{diaExtVar} and discussion). This effect should have the strongest influence on the
        distances as determined in this work. Taking an absolute reddening of $0.5$ mag, the variation 
        between the above cited values results in a difference of $0.25$ mag in the distance modulus. 
        This corresponds to about 25\%
        of the distance in kpc.
\end{enumerate} 
        
The distance error mostly depends on the absolute reddening used. The errors in the metallicites have
only a minor influence on the distances (see Table \ref{tabVIcalib}). They amount to around 3\%
of the total distance in kpc. 
In conclusion, the increased distances $r_{rel}$ (Table \ref{tabDistComp}) are due to the fact that we
determine the distance-relevant parameters using the differentially dereddened CMDs.  
\begin{table}[hb]
\begin{center}
\begin{tabular}{lccccc}
  \hline
  NGC  &$r_{Harris}$ &$r_{iso}$  &$r_{rel}$  &$r_{lit}$  \\ \hline
  5927 &$ 7.4$    &$ 8.4$     &$ 8.9$        &           \\
  6316 &$11.5$    &$12.3$     &$12.0$        &           \\
  6342 &$ 9.1$    &$ 7.4$     &$ 8.7$        &           \\
  6441 &$ 9.7$    &$11.8$     &$13.9$        &           \\
  6760 &$ 7.3$    &$ 8.9$     &$ 9.5$        &           \\
  6528 &$ 7.4$    &$ 9.3$     &$10.7$        &$6.6$      \\
  6553 &$ 4.7$    &$ 5.4$     &$ 6.2$        &$5.2$      \\
  \hline
  \end{tabular}
  \end{center}
  \caption{Comparison of the distances taken from Harris (\cite{HAR96}), $r_{Harris}$, with the values
           of this work. $r_{iso}$ and $r_{rel}$ contain the distances determined via isochrone-fitting
           and $M_{HB}-[\mbox{M}/\mbox{H}]$-relation. The last column shows recently determined distances
           for \object{NGC~6528} (Richtler et al. \cite{RTL98}) and for \object{NGC~6553} 
           (Guarnieri et al. \cite{GUA98}).
           \label{tabDistComp}}
\end{table}
\clearpage

\subsection{Masses}
To classify the clusters according to Burkert \& Smith (\cite{BUR97}), we have to determine the masses
from the total absolute brightnesses. Because we could not measure the apparent total brightness, we used the
values given by Harris (\cite{HAR96}). With the extinctions and distance moduli given above 
(Table \ref{tabDistRel}), we get the absolute total brightnesses via
\begin{equation}
M_V^{total}=V^{total}- A_V - (m-M)_0
\end{equation}
We determined the masses using a mass-to-light-ratio of $\left(\frac{M}{L}\right)_V=3$ (Chernoff \& 
Djorgovski \cite{CHE89}). Table \ref{tabMasses} shows the results. Thus, \object{NGC~6441} is one of the 
most massive clusters of the galaxy. $\omega$ Cen/NGC 5139 has $\log(M/M_{\odot})= 6.51$ (Harris 
\cite{HAR96}).
\begin{table}[hb]
  \begin{center}
  \begin{tabular}{lccc}
  \hline
  NGC     & $V^{total}$   & $M_V^{total}$    & $\log\left(\frac{M}{M_{\odot}}\right)$ \\ \hline
  5927    & $8.01$        & $-7.52$          & $5.42$                           \\
  6316    & $8.43$        & $-8.28$          & $5.72$                           \\
  6342    & $9.66$        & $-6.07$          & $4.85$                           \\
  6441    & $7.15$        & $-9.61$          & $6.26$                           \\
  6760    & $8.88$        & $-7.48$          & $5.41$                           \\
  6528    & $9.60$        & $-6.65$          & $5.07$                           \\
  6553    & $8.06$        & $-7.65$          & $5.47$                           \\
  \hline
  \end{tabular}
  \end{center}
  \caption{Absolute total brightnesses and masses for all clusters. The apparent brightnesses
           were taken from Harris (\cite{HAR96}). \label{tabMasses}}
\end{table}

\section{Classification and assignment}

After having determined the parameters of our cluster sample, we now discuss each cluster's possible
affiliations with the galactic structure components i.e. halo, disk or bulge for each  cluster.
The necessary  criteria are introduced in the following subsection.

\subsection{The assignment criteria }
\subsubsection{Disk and Halo:  Zinn (1985)}
Zinn (\cite{ZIN85}) divided the GC-system into a metal-poor ($[\mbox{M}/\mbox{H}] \le -0.8$ dex) halo- 
and a metal-rich ($[\mbox{M}/\mbox{H}] \ge -0.8$) disk-subsystem. This distinction also correlated 
with the kinematics and spatial distribution of their objects. The resulting criteria are listed in 
Table \ref{tabZinn}. Equation \ref{equVrad} gives the orbital velocity $v_c$ of a cluster depending 
on its observed radial velocity $v_{rad}$. $v_c$ can be compared to the net rotation as given in 
Table \ref{tabZinn}.
\begin{table}[hb]
  \begin{center}
  \begin{tabular}{l|ccc}
  \hline
  subsystem           &$[\mbox{M}/\mbox{H}]\le -0.8$dex &$[\mbox{M}/\mbox{H}]\ge -0.8$dex \\ \hline
  $v_{rot}$[km/s]     &$50 \pm 23$                      &$152 \pm 29$                     \\
  $\sigma_{rot}$[km/s]&$114$                            &$71$                             \\ 
  \hline
  \end{tabular}
  \end{center}
  \caption{Kinematics and spatial distribution of the metal-rich and -poor subsystems of GCs
           according to Zinn (1985) \label{tabZinn}}
\end{table}

\subsubsection{Bulge and (thick) disk: Minniti (1995,1996)}
Minniti (\cite{MIN95}, \cite{MIN96}) divided Zinn's metal-rich disk system further into GCs belonging
to the (thick) disk on the one hand and to the bulge on the other. Comparing the GCs with their corresponding
field population, he assigned the GCs with galactocentric distances $R_{gc} \le 3$ kpc to the bulge and
the ones with $R_{gc} \ge 3$ kpc to the thick disk. 

\subsubsection{Inner halo, bar and disk: Burkert \& Smith}
Burkert \& Smith (\cite{BUR97}) used the masses of the metal-rich GCs to distinguish between a group
belonging to the inner halo and a group which can be further divided into a bar- and a ring-system using
the kinematics and spatial distribution of the clusters (see Table \ref{tabBurkert}).
\begin{table}[hb]
  \begin{center}
  \begin{tabular}{l|ccc}
  \hline
  group                  & inner halo    & bar        & disk        \\ \hline
  $\log(M/M_s)$          & $\ge 5.55$    & $\le 5.55$ & $\le 5.55$  \\
  $v_{rot}$[km/s]        & $24\pm 23$    & $24\pm 23$ & $164\pm 6$  \\
  $v_{rot}/\sigma_{rot}$ & $0.3$         & $0.3$      & $6$         \\
  spatial                & concentrated  & bar-like   & $(4\le R_{gc}\le 6)$kpc \\
  distribution           & to center     & structure  &             \\
  \hline
  \end{tabular}
  \caption{Criteria for subgroups of the metal-rich GCs according to Burkert \& Smith (\cite{BUR97}).
           \label{tabBurkert}}
  \end{center}
\end{table} 

\subsubsection{Radial velocities}
Unfortunately, there do not exist any data on proper motions of our clusters. The only kinematic information
available are radial velocities, catalogued by Harris (\cite{HAR96}). Thus, we can only check, whether a
disk orbit is compatible with a given radial velocity.  This is possible by comparing the
measured radial velocity $v_{rad}$ with the expected one, calculated via equation \ref{equVrad} assuming 
that disk clusters move on circular orbits in the galactic plane.  
\begin{equation}
  v_{rad}=v_c\sin \left(l+\arctan\left(\frac{y}{R_s-x}\right)\right) -v_s\sin (l),
  \label{equVrad}
\end{equation}
where $l$ is the galactic longitude, and $x$ and $y$ are the heliocentric coordinates.
We used $R_s=8.0$ kpc and $v_s=220$ km/s. $v_c$ gives the velocities of the clusters in the plane, 
corresponding to the galactic rotational velocity $v_{rot}(R_s-x)$ with the values taken from
Fich \& Tremaine (\cite{FIC91}).

\subsubsection{Metallicity gradient}
The metallicity gradient of the disk is an uncertain criteria insofar, as it is defined for the outer
ranges of the galactic disk. We use a metallicity gradient referring to the population of old open clusters. 
The oldest of these objects have ages similar to the youngest GCs (Phelps et al. \cite{PHE94}). Their  
scale height is comparable to other thick disk objects. Assuming that they are related to a possible 
disk population of GCs (Friel \cite{FRI95}), we can use their metallicity gradient 
\begin{equation}
\frac{\partial [\mbox{Fe}/\mbox{H}]}{\partial R_{gc}}=-0.091 \pm 0.014
\end{equation}
(Friel \cite{FRI95}) as a criterion for whether our GCs belong to the galactic thick disk or not. 

\subsection{The assignment}
Using the above criteria, we assigned the clusters of our sample according to Table \ref{tabAssign}.
The values of the parameters necessary to decide on group membership are listed in Table
\ref{tabAssDat}.
\begin{table}[hb]
  \footnotesize
  \begin{center}
  \begin{tabular}{l|ccccc}
  NGC   & Zinn & Minniti & Burkert & $v_{rad}^{dsk}$ & $\partial_r[\mbox{M}/\mbox{H}]$ \\ \hline
  5927  & d    & (d)     & d       & +d              & -d     \\
  6316  & ?    & ?       & bu      & +d              & -d     \\
  6342  & ?    & bu      & ba      & -d              & -d     \\
  6441  & ?    & ?       & bu      & -d              & -d     \\
  6760  & d    & ?       & d       & -d              & -d     \\
  6528  & ?    & bu      & ba      & -d              & -d     \\
  6553  & (h)  & bu      & ba      & -d              & -d     \\
  \hline
  \end{tabular}
  \end{center}
  \normalsize  
  \caption{Assignment of the clusters to the systems  \textbf{d}isk, \textbf{h}alo, \textbf{bu}lge
           and \textbf{ba}r according to the criteria in the first column. \textbf{?} is used 
           if no assignment is possible,  the symbols d, h, bu and ba together with 
           \textbf{+} or \textbf{-} relate to criteria which only can decide whether an object
           belongs to a certain group or not. 
           Symbols in brackets denote uncertainties explained in the text. 
           \label{tabAssign}}
\end{table} 
\begin{table*}
\begin{center}
\begin{tabular}{l|c|c|c|c|c|c|c}
\hline
NGC                               &5927     &6316     &6342    &6441     &6760     &6528    &6553    \\ \hline
$[\mbox{M}/\mbox{H}]$[dex]        &$-0.29$  &$-0.60$  &$-0.42$ &$-0.39$  &$-0.57$  &$0.00$  &$-0.04$ \\
$R_{gc}$[kpc]                     &$5.0$    &$4.1$    &$1.7$   &$6.1$    &$5.6$    &$2.8$   &$2.0$   \\
$z$[kpc]                          &$0.8$    &$1.3$    &$1.6$   &$-1.3$   &$-0.7$   &$-0.8$  &$-0.3$  \\
$\log(M/M_{\odot})$               &$5.42$   &$5.72$   &$4.84$  &$6.26$   &$5.41$   &$5.07$  &$5.47$  \\
$v_{rot}$[km/s]                   &$220$    &$220$    &$210$   &$225$    &$220$    &$195$   &$210$   \\
$v_{rad}$[km/s]                   &$-116$   &$72$     &$81$    &$18$     &$-28$    &$185$   &$-7$    \\
$v_{rot}^{dsk}$[km/s]             &$265$    &$-308$   &$114$   &$18$     &$121$    &$1910$  &$37$    \\
$v_{rad}^{dsk}$[km/s]             &$-75$    &$-32$    &$164$   &$-58$    &$56$     &$15$    &$58$    \\   
$[\mbox{M}/\mbox{H}]_{oc}(r)$[dex]&$0.18$   &$0.20$   &$0.46$  &$0.01$   &$0.11$   &$0.33$  &$0.51$  \\
\hline
\end{tabular}
\end{center}
\caption{All relevant parameters for the assignment. $[\mbox{M}/\mbox{H}]$ gives the metallicity 
         according to 
         the $[\mbox{M}/\mbox{H}]-\Delta V$-relation, $R_S$ the heliocentric distance, $R_{gc}$ the 
         galactocentric distance and $z$ the distance to the disk.
         The M in $\log(M/M_{\odot})$ stands for cluster mass, $v_{rot}$ for the orbital  
         velocity, derived via the rotational velocity curve of Fich \& Tremaine (\cite{FIC91}), 
         $v_{rad}$ for the observed radial velocity and $v_{rot}^{dsk}$ for the orbital velocity 
         of the clusters, calculated using $v_{rad}$ and the assumption of circular cluster orbits
         (in order to compare with a net rotation).
         $v_{rad}^{dsk}$ contains the expected radial velocity assuming disk orbits and
         $[\mbox{M}/\mbox{H}]_{oc}(r)$ the values derived via a metallicity gradient of the old open clusters
         (Friel \cite{FRI95}).
         \label{tabAssDat}}
\end{table*}

As for the metallicities of our clusters, they all belong to the disk system according to Zinn, 
which is obvious as the sample had been selected in this way. Not so obvious is the comparison
with the net rotation of Zinn's disk group. Only \object{NGC~5927} shows a value of $v_{rot}$ which is not
totally off the net rotation as given in Table \ref{tabZinn}.

\begin{figure}
  \resizebox{\hsize}{!}{\includegraphics{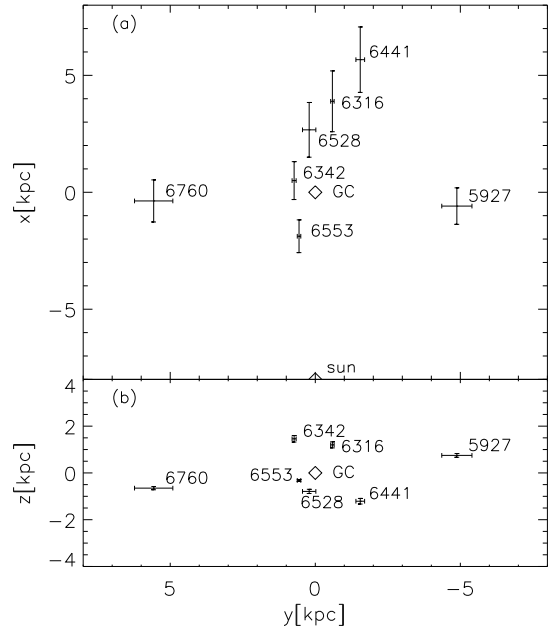}}
  \caption{Distribution of the cluster sample in the
           galactocentric x-y-plane (a) and y-z-plane (b). In
           the upper panel, the observer is located at $z \ge 0$. 
           The distribution in the y-z-plane is as seen from the sun. The
           distances used correspond to $r_{rel}$ in Table 
           \ref{tabDistComp}.\label{diaXYZ}}
\end{figure}

The clusters belonging to the bulge according to Minniti's criterion are members of the bar following 
the arguments of Burkert \& Smith (\cite{BUR97}). Binney et al. (\cite{BIN97}) quote a value of 
$20^{\circ}$ for the angle between x-axis in galactocentric coordinates and the major semiaxis of the 
bulge structure. Its end lying nearer to the sun is located at small galactic longitudes ($y \le 0$ in 
cartesian coordinates). 
Fig \ref{diaXYZ} shows the spatial distribution of our cluster sample. The coordinates
of the 'bar' clusters \object{NGC~6342}, 6528 and 6553 according to Burkert \& Smith seem to be consistent 
with a structure described by Binney et al. (\cite{BIN97}). However, as the referee pointed out, we do not
know how long-lived the Milky-Way bar is, and other tracers of old populations such as RR Lyrae do
not follow the bar (Alcock et al. \cite{ALC98}). Moreover, the distance between the 'bar' clusters
\object{NGC 6528} and \object{NGC 6553} is about $5$ kpc, which is much larger than the length of the
Milky-Way bar according to most authors (e.g. Binney et al. \cite{BIN97}). Also note in Fig \ref{diaXYZ},
that the errors in the x-coordinate are larger than those in y and z. 

There are only two 'disk' clusters remaining, assuming Burkert \& Smith's definition of
disk clusters: \object{NGC~5927} and \object{NGC~6760}. However, the radial velocities corroborate this result for 
\object{NGC~5927} only. For any other cluster, the radial velocities seem to exclude an assignment to the disk. 

The metallicity gradient of the old open clusters leads to the conclusion that none of our clusters
is to be assigned to the thick disk. Taken the whole sample of metal rich clusters (i.e. clusters with
$[\mbox{M}/\mbox{H}] \ge -0.8$ dex according to Zinn \cite{ZIN85}, see Tables 
\ref{tabRestAss}, \ref{tabRestDat}), we find only three objects, which could be disk clusters according
to the metallicity gradient criterium.
\begin{table*}
\footnotesize
\begin{center}
\begin{tabular}{lcccccccc}
\hline
Name     &$R_{gc}$&$z$  &$[\mbox{M}/\mbox{H}]$&$v_{rad}$&$v_{rad}^{dsk}$&$\log(M/M_{\odot})$&$v_{rot}^{dsk}$&$[\mbox{M}/\mbox{H}]_{oc}$\\
\hline
NGC 104  &$7.3$   &$-3.0$&$-0.76$ &$ -19$  &$ -26$         &$6.16$             &$202$         &$0.06$ \\
Lynga 7  &$4.2$   &$-0.3$&$-0.62$ &$   8$  &$ -95$         &                   &$106$         &$0.35$ \\
NGC 6256 &$2.3$   &$ 0.5$&$-0.70$ &        &               &$4.87$             &              &$0.52$ \\
NGC 6304 &$2.2$   &$ 0.6$&$-0.59$ &$-107$  &$ -38$         &$5.32$             &$478$         &$0.52$ \\
NGC 6356 &$7.0$   &$ 2.6$&$-0.50$ &$  27$  &$ -56$         &$5.81$             &$-361$        &$0.09$ \\
NGC 6352 &$3.3$   &$-0.7$&$-0.70$ &$-121$  &$ -93$         &$4.99$             &$244$         &$0.42$ \\
Ter 2    &$1.6$   &$ 0.4$&$-0.25$ &$ 109$  &$  79$         &$4.41$             &$305$         &$0.58$ \\
Liller 1 &$2.6$   &$ 0.0$&$ 0.22$ &$  53$  &$  73$         &$5.45$             &$127$         &$0.49$ \\
Ter 1    &$1.5$   &$ 0.1$&$-0.35$ &$  35$  &$ -40$         &$3.71$             &$-108$        &$0.59$ \\
Ton 2    &$1.4$   &$-0.5$&$-0.50$ &        &               &$4.84$             &              &$0.60$ \\
NGC 6388 &$4.4$   &$-1.4$&$-0.60$ &$  81$  &$ 155$         &$6.32$             &$ 54$         &$0.32$ \\
Pal 6    &$1.4$   &$ 0.2$&$-0.10$ &$ 201$  &$  31$         &$5.34$             &$1112$        &$0.60$ \\
Ter 5    &$0.6$   &$ 0.2$&$-0.28$ &$ -94$  &               &$5.56$             &              &$0.67$ \\
NGC 6440 &$1.2$   &$ 0.5$&$-0.34$ &$ -79$  &$ 180$         &$5.89$             &$-49$         &$0.62$ \\
Ter 6    &$0.6$   &$-0.3$&$-0.65$ &$ 126$  &$ -77$         &$5.14$             &$-305$        &$0.67$ \\
NGC 6496 &$4.4$   &$-2.0$&$-0.64$ &$ 113$  &$ 136$         &$5.28$             &$-364$        &$0.32$ \\
Ter 10   &$0.8$   &$-0.3$&$-0.70$ &        &               &$5.52$             &              &$0.66$ \\
NGC 6539 &$3.0$   &$ 0.9$&$-0.66$ &$ -45$  &$ 130$         &$5.71$             &$ 32$         &$0.46$ \\
NGC 6624 &$1.2$   &$-1.1$&$-0.42$ &$  53$  &$ 181$         &$5.39$             &$ 71$         &$0.62$ \\
NGC 6637 &$1.5$   &$-1.5$&$-0.71$ &$  40$  &$-191$         &$5.40$             &$-52$         &$0.59$ \\
Pal 8    &$5.2$   &$-1.5$&$-0.48$ &$ -43$  &$-134$         &$4.60$             &$-27$         &$0.26$ \\
Pal 10   &$6.4$   &$ 0.3$&$-0.10$ &        &               &$4.71$             &              &$0.14$ \\
Pal 11   &$7.6$   &$-3.4$&$-0.39$ &$ -68$  &$ -24$         &$5.11$             &$-72$         &$0.04$ \\
NGC 6838 &$6.7$   &$-0.3$&$-0.73$ &$ -23$  &$  25$         &$4.62$             &$161$         &$0.12$ \\ 
\hline
\end{tabular}
\normalsize
\end{center}
\caption{The relevant parameters of the remaining metal rich GCs according to Harris (\cite{HAR96}). 
         The columns are labeled as in Table \ref{tabAssDat}. Distances are given in kpc,
         metallicities in dex and velocities in km/s.
         \label{tabRestDat}}
\end{table*}

\begin{table}
\footnotesize
\begin{center}
\begin{tabular}{lcccccc}
\hline
Name     &Zinn & Minniti & Burkert &$v_{rad}$  &$[\mbox{M}/\mbox{H}]_{oc}$ \\\hline
NGC 104  &d    &-        &?        &+d         &-d    \\
Lynga 7  &d    &-        &-        &-d         &-d    \\
NGC 6256 &-    &bu       &ba       &-d         &-d    \\
NGC 6304 &?    &bu       &ba       &-d         &-d    \\
NGC 6356 &?    &-        &?        &-d         &-d    \\
NGC 6352 &?    &-        &ba       &+d         &-d    \\
Ter 2    &?    &bu       &ba       &+d         &-d    \\
Liller 1 &d    &bu       &ba(d)    &+d         &+d    \\
Ter 1    &?    &bu       &ba       &-d         &-d    \\
Ton 2    &?    &bu       &ba       &-          &-d    \\
NGC 6388 &d    &-        &?        &d?         &-d    \\
Pal 6    &?(bu)&bu       &ba       &-d         &-d    \\
Ter 5    &-    &bu       &bu       &-          &-d    \\
NGC 6440 &?    &bu       &bu       &-d         &-d    \\
Ter 6    &?    &bu       &bu       &-d         &-d    \\
NGC 6496 &?    &-        &ba       &+d         &-d    \\
Ter 10   &?    &bu       &ba       &-          &-d    \\
NGC 6539 &?    &bu?      &bu       &-d         &-d    \\
NGC 6624 &?    &bu       &ba       &-d         &-d    \\
NGC 6637 &?    &bu       &ba       &-d         &-d    \\
Pal 8    &?    &-        &d        &-d         &-d    \\
Pal 10   &-    &-(d)     &d        &-          &+d    \\
Pal 11   &?    &?        &d        &d?         &+d?   \\
NGC 6838 &d    &-        &d        &-d         &-d    \\ \hline
\end{tabular}
\normalsize
\end{center}
\caption{Suggested assignment according to the criteria discussed above for the remaining
         metal rich GCs. The columns are labeled as in Table \ref{tabAssign}.
         \label{tabRestAss}}
\end{table}
  
Some of the clusters do not meet any of the criteria. Interestingly, they are the most
massive, but metal-poorest objects of the sample. These objects are \object{NGC~6316}, 6760, and 6441 as
well as (in Table \ref{tabRestAss}) NGC 104, 6356 and 6388. Although NGC 104 seems to be 
a disk cluster and mostly is referred to as such, the large distance to the galactic plane (3 kpc)
does not support this assignment. Probably, the mentioned objects belong to the halo, being its 
metal-richest clusters. Zinn (\cite{ZIN85}) and Armandroff (\cite{ARM93}) point to the fact
that the division into metal-rich and poor clusters is by no means an exact one, but that
there is a metal-rich sample of halo clusters as well as a metal-poorer one of disk objects.  
Richtler et al. (\cite{RTL94}) discussed the existence of a subgroup of disk clusters according 
to Zinn (\cite{ZIN85}), based on an analysis  of the metallicities and the minimum 
inclination angles derived from $z/R_{gc}$-values for these
clusters. They conclude that the clusters NGC 6496, 6624 and 6637
might not be disk clusters after all, but belong to the halo. Adding
their argument to the above discussion, we end up with 3 probable
disk members (\object{NGC~5927}, additionally from Table \ref{tabRestAss}
Liller 1 and Pal 10. Pal 11 is excluded because of its large minimum
inclination angle.) and 9 clusters (NGC 104, 6316, 6356, 6388, 6441,
6496, 6624, 6637 and 6760) that more likely belong to the halo
than to the (thick) disk. The rest of the clusters (\object{NGC~6342}, 6528,
6553 and the remaining ones of Table \ref{tabRestAss}) fall in with
the bulge/bar-group of Minniti (\cite{MIN95}) and Burkert \& Smith
(\cite{BUR97}). 

\section{Conclusions}
We derived the parameters for five GCs near the galactic center in a uniform manner, employing 
a new calibration of methods which relate structures in the CMDs with the parameters.
Taking into account the differential reddening and correcting the CMDs for it leads to
more accurately determined parameters and a decreasing absolute reddening. There might be a
systematic effect on distances, if the differential reddening is not taken care of.

With the $[\mbox{M}/\mbox{H}]-\Delta V$- method we present a accurate way to estimate 
differentially metallicities of metal-rich GCs. Especially it might prove useful for surveys of
clusters in V, V-I, as their CMDs only need to contain the HB and turnover of the AGB/RGB.    

The metallicities of our program clusters all lie in the range of the clusters
constituting the classical disk-system of GCs in the Milky
Way. However, different criteria defining subgroups of the GC-system
partly lead to differing results. Most of the metal-rich GCs seem to belong
to a bar/bulge-structure, and only a minority could clearly be
addressed as 'disk'-clusters. So the classical disk-system is more
likely to be a mixture between a halo- and bulge-component.

\begin{acknowledgements}
We are indebted to E.K. Grebel for the most interesting and valuable 
discussions, especially on the technique of differential dereddening. 
We would like to thank the referee D. Minniti for helpful comments 
and criticism.
\end{acknowledgements}


\end{document}